\documentstyle[12pt]{article}
\newcommand{\pr}{\partial}

\newcommand{\nn}{\nonumber}

\newcommand{\be}{\begin{equation}}
\newcommand{\ee}{\end{equation}}
\newcommand{\bea}{\begin{eqnarray}}
\newcommand{\eea}{\end{eqnarray}}

\begin{document}

 \title{Dirac variables in gauge theories}

 \author{
  Victor Pervushin, \\
 {\normalsize\it Joint Institute for Nuclear Research},\\
 {\normalsize\it 141980, Dubna, Russia.}
 }


\maketitle

 \medskip
 \medskip
\vspace{-7cm}

 \hfill{   hep-th/0109218~~~~~~~}

\vspace{7cm}

 \hfill{\sf Lecture notes in DAAD Summerschool \phantom{aaaa} }

 \hfill{\sf on  Dense Matter in
 Particle - and Astrophysics,\phantom{aaaaa}}

 \hfill{\sf JINR, Dubna, Russia, August 20 - 31, 2001\phantom{aaaaa}}

 \begin {abstract}

 The review is devoted to  a relativistic formulation of the first
 Dirac quantization of QED (1927) and its generalization
 to the non-Abelian theories
 (Yang-Mills and QCD) with the topological degeneration of initial data.
 Using the Dirac variables we give a systematic description of
 relativistic nonlocal bound states  in QED with
 a choice of the time axis of quantization
 along the eigenvectors of their total momentum operator.

 We show that the Dirac variables of the non-Abelian fields are
 topologically degenerated, and there is a pure gauge
 Higgs effect in the sector of the zero winding number
 that leads to a nonperturbative physical vacuum in the form
 of the Wu-Yang monopole. Phases of the topological degeneration
 in the new perturbation theory are  determined by an equation of
 the Gribov ambiguity of the constraint-shell gauge
 defined as an integral of the Gauss equation with zero initial data.
 The constraint-shell non-Abelian dynamics
 includes  zero mode of the Gauss law differential operator,
 and a rising potential of the instantaneous interaction,
 that rearranges the perturbation series and changes the assymptotic
 freedom formula.

 The Dirac variables in QCD
 with the topological degeneration of initial data
 describe constituent gluon,  and quark masses,
 the spontaneous chiral symmetry breaking,
 color confinement in
 the form of quark-hadron duality as a consequence of summing
 over the Gribov copies. A solution of U(1)-problem is given
 by mixing the zero mode with $\eta_0$ - meson.
 We discuss reasons why all these physical effects disappear
 for  arbitrary gauges of physical sources in the standard
 Faddeev-Popov integral.

 \end{abstract}


 \hfill{\sf Is devoted to the memory of V.N. Gribov and I.V.
 Polubarinov \phantom{aaaaa}}

 \newpage
 {\bf \Large Content}
 \sf

 1. {\large \bf Historical Introduction} \hfill{3}

 2. {\large \bf Dirac variables in  QED} \hfill{4}

 2.1. Gauge-fixing method of 1967 \hfill{4}

 2.2. Dirac reduction and Dirac variables \hfill{5}

 2.3. Relativistic covariance             \hfill{8}

 2.4. Quantization and Feynman path integral \hfill{10}

 2.5. Gauge equivalence theorem and Faddeev-Popov integral \hfill{11}

 3. {\large \bf  QED of bound states: spectrum and S-matrix}             \hfill{13}

 3.1. Markov-Yukawa prescription                           \hfill{13}

 3.2. Effective Lagrangian of bilocal fields               \hfill{15}

 3.3. Quantization of bilocal fields                       \hfill{17}

 3.4. Schwinger-Dyson equation: the fermion spectrum       \hfill{18}

 3.5. Bethe-Salpeter equation: the bound-state spectrum    \hfill{19}

 3.6. Schr\"odinger equation                               \hfill{22}

 3.7. Spontaneous chiral symmetry breaking                 \hfill{23}

 3.8. Relativistic wave functions for multiparticle systems \hfill{24}

 3.9. Relativistic covariant unitary S-matrix for bound states \hfill{28}

 4. {\large \bf Dirac variables in   Yang-Mills theory\\
 with the topological degeneration of the physical states}      \hfill{29}

 4.1. {\it Constraint-shell} radiation variables
 in perturbation theory                                        \hfill{29}

 4.2. Topological degeneration of initial data                 \hfill{31}

 4.3. Physical vacuum and the gauge Higgs effect               \hfill{33}

 4.4. Dirac method and Gribov copies                           \hfill{34}

 4.5. Topological dynamics                                     \hfill{36}

 4.6. Zero mode of Gauss law and Dirac variables               \hfill{37}

 4.7. Constraining with zero mode                              \hfill{38}

 4.8. Feynman  path integral                                   \hfill{39}

 4.9. Rising potential induced by monopole                     \hfill{40}

 4.10. F-P path integral        \hfill{40}

 5. {\large \bf Dirac variables in  QCD: hadronization and confinement}     \hfill{41}

 5.1. Action and first principles of QCD                       \hfill{41}

 5.2. Topological degeneration of QCD                          \hfill{41}

 5.3. Superfluid dynamics of gluonic liquid                    \hfill{43}

 5.4. Feynman path integral                                    \hfill{45}

 5.5. A free rotator: topological confinement                  \hfill{46}

 5.6. Confinement as destructive interference  \hfill{47}

 5.7. Relativistic equations for gluonic systems               \hfill{48}

 5.8. Hadronization and chiral Lagrangian limit                \hfill{52}

 5.9. The low energy theorems                                   \hfill{55}

 5.10. U(1)- Problem                                            \hfill{60}

 6. {\large \bf Conclusion}                                                 \hfill{62}

\newpage
\section{Historical Introduction}

 The first papers by Dirac~\cite{cj},
 Heisenberg, Pauli~\cite{hp}, and Fermi~\cite{F}
 on the quantization of electrodynamics
 ran into difficulties of the determination of physical variables.
 The interpretation of all gauge components as independent variables
 contradicts the quantum principles; whereas excluding
  nonphysical variables contradicts  the relativistic
  principles.

 The consistent quantum description of gauge constrained systems
 was considered by Dirac, Schwinger, Feynman,
 Faddeev, Gribov, and other physicists~\cite{cj,sch,fey,f,g}
 as one of most fundamental problems of theoretical physics
 in the 20th century.

 The first quantization of electrodynamics belongs to Dirac~\cite{cj}
 who disregarded the relativistic principles and excluded
 nonphysical components  by the reduction of the initial action
 to the  solution of the Gauss law constraint,
 i.e. the equation for the time-like component of the gauge vector field.
 The Gauss law connects initial data of the time-like component
 with the data of all other fields.
 But this  {\it constraint-shell} method had a set of defects,
 including nonlocality, explicit noncovariance as the dependence
 on the external time axis of quantization, and complexity.
 Feynman-Schwinger-Tomonaga formulation of QED admitted a simpler
 method based on the  extended dynamics where all components
 were considered on  equal footing with fixing a relativistic gauge.

 At the beginning of the 60s, Feynman found
 that the naive generalization of his method of  construction
 of  QED did not work for the non-Abelian theory.
 The unitary S-matrix in the non-Abelian theory was obtained in
 the form of the Faddeev-Popov (FP) path integral~\cite{fp1} by the
 brilliant application of the theory of connections in vector bundle.
 There is an opinion that the FP path integral is
 the highest level of  quantum description of
 gauge relativistic constrained systems.  In any case, just this FP
 integral was the basis to prove renormalizability of the unified theory
 of electroweak interactions in papers by 't Hooft and Veltman
 awarded the Nobel prize in 1999.

 Nevertheless, in the context of the first Dirac quantization and its
 Hamiltonian generalizations~\cite{Dirac,gpk}, the intuitive status of
  the FP integral was so evident that  two years after
 the paper~\cite{fp1}, Faddeev gave the foundation of the FP integral by the
 construction of the unitary S-matrix~\cite{f} for an "equivalent non-Abelian
 unconstrained system" derived by resolving constraints in terms of the
 radiation variables  of the Hamiltonian description.

 Faddeev  showed that on the one hand, the constraint-shell dynamics
 is compatible with the simplest quantization by the
 standard Feynman path integral, on the other hand,
 this Feynman integral is equivalent to the FP integral in an
 arbitrary gauge. This equivalence was proved by the change of
 variables in the Feynman path integral that removed the time - like
 vector of the canonical quantization into the phase factors of
 physical source terms. These phase factors disappear for  S-matrix
 elements on the mass shell of elementary particles.
 In other words, Faddeev proved the equivalence of the {\it constraint-shell}
 approach with quantization of gauge theories by the gauge-fixing method
 only for  scattering amplitudes~\cite{f}
 where all color particle-like excitations of the fields are on their
 mass-shell. But the scattering amplitudes for color particles
 are nonobservable in QCD.
 The  observables are hadrons as colorless bound states
 where elementary particles are off mass-shell. Just for this case,
 the Faddeev  theorem of equivalence of different gauges becomes
 problematic even for QED in the sector of instantaneous bound
 states, as the FP integral in a relativistic gauge loses all
 propagators with  analytic properties
 that lead to instantaneous bound states identified with observable
 atoms.

 The Faddeev equivalence theorem~\cite{f} was proved before
 the revelation of nontrivial topological properties~\cite{ins,Fad,hooft}
 and the Gribov ambiguities~\cite{g}. These new facts were studied mainly
 on the level of the FP integral.  The topological degeneration
 of classical vacuum  was incorporated into the FP integral in the
 form of the instanton solutions~\cite{ins}.

 It seems more natural first, to study
 the topological degeneration of all non-Abelian
 physical states at the  more fundamental
 level of the  {\it constraint-shell dynamics}
 compatible with the
 Feynman path integral
 and, then, to obtain the corresponding non-Abelian  FP path integral
 analyzing  possible Gribov copies.

 The first quantization of electrodynamics was fulfilled
 just at the level of the  {\it constraint-shell dynamics}
 by Dirac~\cite{cj}.
 The Dirac reduction is based on the honest resolution
 of the Gauss law constraint and  introduction
  of the Dirac nonlocal gauge-invariant physical variables~\cite{cj,pol,lm},
 instead of gauge-fixing.
 The Dirac method is the way to distinguishe
 the unique (radiation) gauge of physical sources.

 The present review is devoted to the generalization
 of the Dirac variables~\cite{cj,pol,lm} to the non-Abelian theories
 with the topological degeneration of the initial data in QCD
 in the class of functions of  nontrivial topological
 transformations~\cite{vp1,p2,inp,in,kh}.

  We present here a set of arguments in favor of
 that the constructed {\it equivalent unconstrained system} contains
 the most interesting physical effects of hadronization and
 confinement in QCD that can be hidden in explicit solutions of constraints
 and equations of motion~\cite{wy,inp,in,bpr}.
 We show also that the FP integral in an arbitrary relativistic gauge
 loses all these effects. The relativistic covariant
 properties of the Dirac variables~\cite{z} allow us to use
 the Markov-Yukawa~\cite{ym} prescription of construction of the
 multilocal irreducible representations of the Poincare group and to
 formulate the generalized S-matrix formalism where the time
 axis of the Hamiltonian description is proportional to
 eigenvectors of the total momenta operator of any physical
 state~\cite{yaf,npbs,a20,a13,fb,fb1,a6}.

 \section{Dirac variables in   QED}

 \subsection{ Gauge-fixing method of 1967}

 We formulate the statement of the problem using QED.
 It is given by the action
 \begin{eqnarray}\label{w}
 W[A,\psi,\bar \psi] &=& \int d x \,\,\, \Bigl[ -\frac{1}{4} ( F_{\mu\nu}
 )^{2} + \bar {\psi} ( i \,\, \rlap/\nabla (A) - m^{0} ) \psi
 \Bigr]  \,\,\, ,
 \end{eqnarray}
 \begin{eqnarray}\label{vw}
 \nabla_{\mu}(A) &=& \partial_{\mu} -i e {A}_{\mu} \,\,\,\,\,
 \rlap/\nabla = \nabla_\mu \cdot \gamma^\mu \,\,\, ;
 \nonumber \\
 \,\,\,\,\,F_{\mu\nu} &=&
 \partial_{\mu}A_{\nu} - \partial_{\nu}A_{\mu}~.
 \end{eqnarray}

 This action  contains gauge fields more than independent degrees of freedom.
 First of all it is invariant with respect to gauge transformations
  \be\label{gauge}
  A_{\mu}^\Lambda=A_{\mu}+ \partial_{\mu}\Lambda~,\psi^\Lambda=
 \exp[ie\Lambda]\psi~.
 \ee
 One supposes that this invariance allows us to remove one field degree
 of freedom by the help of arbitrary gauges
 \be \label{gf}
 F(A_{\mu})=0,~~~~F(A^{u}_{\mu})=M_F u \not= 0~,
 \ee
 where the second equation means that the gauge fixes field unambiguously.

 The standard quantization of the gauge field system in the
 gauge $F(A_{\mu})=0$ is based on the FP path integral~\cite{fp1}
 \begin{eqnarray}\label{fpi11}
 Z^{FP}[s^F, {\bar s}^F, J^F]\;=\;\int \prod_{\mu}DA^F_\mu D
 \psi^F D{\bar \psi}^F \Delta _{FP}^F
 \delta (F(A^F)) e^{iW[ A^F,\psi^F {\bar \psi}^F ] + S^F}~,
 \end{eqnarray}
 where  $\Delta _{FP}^F=det M_F$ is FP determinant and
 \be\label{fpso1}
 S^F=\int d^4x \left( {\bar s}^F \psi^F +
 {\bar \psi}^F s^F
 + A^F_{\mu}J^{\mu}\right)~
 \ee
 are the sources. The foundation of the intuitive FP integral
 by the canonical quantization was made in Faddeev's paper~\cite{f}.

 After Faddeev papers of 1967-69 there dominates very popular
 opinion  that all physical results do not depend on a gauge
 $F(A_{\mu})=0$. A gauge is defined for reasons of simplicity
 and convenience, including the condition $det M_F \not= 0$;
 the opposite case $det M_F = 0$ is called the Gribov ambiguity~\cite{g}.
 All applications of quantum gauges theory after 1967-69,
 including investigation of the Gribov ambiguity~\cite{singer}, topological
 degeneration of initial data~\cite{ins}, hadronization~\cite{a22},
 parton models~\cite{efremov}, etc., were fulfilled at the level of
 the FP integral~(\ref{fpi11}).

 While, in Faddeev's paper~\cite{f} there was established the range of
 validity of the FP integral  and its gauge independence.
 Faddeev managed to prove the FP integral by the canonical quantization
 only in the sector
 of scattering amplitudes for elementary particles on their
 mass-shell. This proof  becomes doubtful for
 bound states where elementary particles are off mass-shell,
 the latter is crucial for QCD where only bound states are observable.

 The present review is devoted  to consideration of the problems of
 bound states, Gribov copies, and topological degenerations of initial
 data at the more fundamental level of the Dirac canonical
 quantization of 1927~\cite{cj}.

 \subsection{Dirac reduction and Dirac variables}

 The situation with canonical quantization of gauge theories before the FP
 revolution of 1967 was presented in the review by I.V.
 Polubarinov~\cite{pol}. Igor Vasil'evich intended to sent his
 manuscript to Usp. Fiz. Nauk. It is pity that this nice review
 was not published. After 1967, Polubarinov tried to include  the
 FP integral in his review, but he was not satisfied by the level
 of the physical foundation of the FP - scheme of quantization
 in the context of the first (constraint-shell) formulation
 of QED that belongs to Dirac~\cite{cj}. We reproduce here this
 formulation.

 The Dirac quantization was based on  a definite frame of
 reference distinguished by the time axis $\eta_\mu=(1,0,0,0)$,
 which allows us to establish boundary conditions and initial data.
 In special relativity, Einstein identified a frame of reference
 with a set of the watches, rulers, and other physical instruments
 for measurement of physical quantities (that include in our case a
 spectrum of the field excitations).
 The classical equations are split on the Gauss law constraint
 \be \label{var1}
 \frac{\delta W}{\delta A_0}=0
 ~\Rightarrow~~
 \Delta A_0= \partial_i  \partial_0 A_i +j_0
 ~~~~~~~(\Delta =\partial_i \partial_i,~j_{\mu}=e\bar\psi\gamma_{\mu}
 \psi)~,
 \ee
 and equations of motion
 \be \label{var2}
 \frac{\delta W}{\delta A_k}=0~
 ~\Rightarrow~~
 \partial_0^2 A_k- \partial_k \partial_0 A_0-(\delta_{ki}\Delta-
 \partial_k \partial_i ) A_i=j_k~,
 \ee
 \be \label{var3}
 \frac{\delta W}{\delta \psi}=0~
 ~\Rightarrow~~\bar \psi ( i \,\, \rlap/\nabla (A) + m^{0} )=0~,
 \ee
 $$
 \frac{\delta W}{\delta \bar \psi}=0~
 ~\Rightarrow~~( i \,\, \rlap/\nabla (A) - m^{0} )\psi=0~.
 $$
 The problem of canonical quantization meets with
 the non-dynamic status of the time component $A_0=A_\mu  \eta_\mu$.
 The non-dynamic status of $A_0$ is not compatible with
 quantization
 of this component as the fixation of $A_0$ (by the Gauss law) and
 its zero momentum $E_0=\partial {\cal L}/\partial (\partial_0
 A_0)=0$
 contradict to the commutation relation and uncertainty principle.
 To keep quantum principle, Dirac excluded the time component
 using the Gauss law constraint~(\ref{var1}):
 an explicit solution of this Gauss law
 \be \label{var1s}
 A_0(t,x)= a_0[A]+\frac{1}{\Delta}j_0(t,x)~,
 \ee
 where
 \be \label{var2s}
 a_0[A]=\frac{1}{\Delta}
 \partial_i  \partial_0 A_i(t,x)~,
 \ee
 connects initial data of $A_0(t_0,x)$ with a set of
 the initial data of the
 longitudinal component $\partial_i \partial_0 A_i(t,y),$
 and current $j_0(t,y)$ in the whole space, here
 \be \label{coulomb}
 \frac{1}{\Delta}f(x)=-\frac{1}{4\pi}\int d^3y\frac{f(y)}{|x-y|}
 \ee
 is the Coulomb kernel of the nonlocal distribution. One can
 substitute the solution (\ref{var1s}) into equation for spatial
 component (\ref{var2})
 \begin{eqnarray}  \label{vartr}
 \frac{\delta W}{\delta A_i}\;\Bigl\vert_{ \frac{\delta W}{\delta
 A_0}\;=\;0.}~\Rightarrow~[\delta_{ik}-\partial_i\frac{1}{\Delta}\partial_k]
 (\partial_0^2-\Delta) A_k=j_i-\partial_i\frac{1}{\Delta}\partial_0
 j_0.
 \end{eqnarray}
 We can see, that the constraint-shell equations~(\ref{vartr})
 contain only two transverse physical variables as the gauge - invariant
 functionals
 \be \label{**}
 A^*_i(t,\vec x) =
 [\delta_{ik}-\partial_i\frac{1}{\Delta}\partial_k]  A_k~.
 \ee
 Dirac generalized these gauge - invariant variables onto all fields
 by gauge transformations
 \begin{eqnarray}\label{vak}
 \sum_{a=1,2} e_k^aA_a^D=A_{k}^{D}[A]& =&
 v[A] ( A_{k} + i \frac{1}{e} \partial_{k} )
 ( v[A] )^{-1},\nn\\
 \psi^{D}[A,\psi]& =& v[A] \psi \,,\,\,
 \end{eqnarray}
 where the gauge factor~\cite{cj}  is defined by
 \be\label{va}
 v[A] =  \exp \bigl\{ ie\int_{t_0}^{t} dt' a_0 (t')\bigr\}~.
 \ee
 Using the gauge transformations~(\ref{gauge})
 \be\label{vgauge}
 a^{\Lambda}_0=a_{0}+\partial_0 \Lambda~\Rightarrow~
 v[A^{\Lambda}]= \exp[ie\Lambda(t_0,\vec x)]v[A]\exp[ie\Lambda(t,\vec x)]~,
 \ee
 we can find that  initial data of
 the gauge - invariant Dirac variables~(\ref{vak}) are
 degenerated with respect to stationary gauge transformations
  \be\label{gauge1}
 A_i^D[A^\Lambda]\;=\;A_i^D[A]+\partial_i \Lambda(t_0,\vec x),~~~~~
 \psi^{D}[A^\Lambda,\psi^\Lambda]
 =\exp[ie\Lambda(t_0,\vec x)]\psi^{D}[A,\psi]~.
 \ee
 The Dirac variables (\ref{vak}) as the functionals of initial fields
 satisfy the Gauss law constraint
 \be\label{gc2}
 \partial_0 \left(\partial_i A^D_i(t,\vec x) \right) \equiv 0~.
 \ee
 Thus, explicit resolving the Gauss law allows us
 to remove two degrees of freedom and to reduce
 the gauge group into the subgroup of the stationary gauge
 transformations~(\ref{gauge1}).

 We can fix a stationary phase $\Lambda(t_0,\vec x)=\Phi_0 (\vec x)$
 by  an additional constraint in the form of the
 time integral of the Gauss law constraint~(\ref{gc2})
 with zero initial data
 \be \label{ngauge}
 \partial_i A_i^D=0
 \end{equation}
 We call this equation the {\it constraint-shell "gauge"}. This "gauge"
 restricts initial data to a phase distinguished by
 the equations $\Delta \Phi_0 (\vec x)=0$.
 Nontrivial solutions of this equation we call the degeneration
 of initial data, and the Gribov copies of the constraint-shell "gauge".
 The degeneration of initial data is determined by topological
 properties of the manifold of stationary gauge transformations
 in the class of functions with a finite density of energy.
 In the case of the three-dimensional QED there is only a trivial
 solution $\Phi_0 (\vec x)=0$. In this case
 $$
 \psi^D=\psi^*,~~~~~~~ A^D= A_i^*~.
 $$
 In the one-dimensional QED and non-Abelian theory
 the degeneration of initial data is the evidence
 of a zero mode of the Gauss law constraint
 that describes the topological excitation of gauge field with the
 Coleman spectrum of the electric tension~\cite{p2,in,col,gip}.
 We consider this zero mode in the Section devoted to Yang-Mills theory.

 Dirac constructed  an {\it equivalent unconstrained system}
 the equation of which reproduce the equations of the
 initial theory~(\ref{w}).
 \be \label{red}
 W^*= W|_{\delta W/{\delta A_0}=0}=\int d^4x \frac{1}{2}
 [\sum_{a=1,2}(\partial_\mu
 A_a^*
 \partial^\mu A_a^*) + \frac{1}{2}j_{0}^{*} \frac{1}{\Delta}j_{0}^{*} -
 j_{i}^{*} A_{i}^{*} + {\bar {\psi}}^{*} (i{\hat \partial} -
 m){\psi}^{*}]\;.
  \ee
 To derive this {\it equivalent unconstrained system} that
 contains only physical variables,
 Dirac~\cite{cj,pol} proposed to change the order of
 constraining and varying. He substituted the solution of
 the Gauss law constraint~(\ref{var1s}) into the initial
 action~(\ref{w}) in the rest frame $\eta_\mu=(1,0,0,0)$
 and introduced the Dirac variables~(\ref{vak}).

 To combine the nonlocal physical variables $A_i^*$ and
 variational principle formulated for local fields, we can
 introduce three independent variables. In this case,
 the Dirac action
 should be added by the Lagrange multiplier \cite{pol}
 \be \label{red1}
 W_{\rm eff}=W^* +\int d^4x \lambda_L (x)\partial_i A_i^* \;.
  \ee
 The local equations of the {\it equivalent unconstrained
  system}~(\ref{red1})
 \be \label{varc}
 \frac{\delta W_{\rm eff}}{\delta A^*_i}=0~,~~~~~~
 \frac{\delta W_{\rm eff}}{\delta \lambda_L}=0~,~~~~~~
 \frac{\delta W^*}{\delta \psi^*}=0~,~~~~~~\frac{\delta W^*}{\delta \bar \psi^*}=0
 \ee
 completely coincides with the equations of the initial constrained
 action~(\ref{var1})-~(\ref{var3}).
 Eqs.~(\ref{varc}) reproduce the constraint~(\ref{ngauge}) and
 lead to the equation for the Lagrange multiplier
 \be \label{phase}
 \Delta \lambda_L (t,\vec x)=0~.
 \ee
 The latter coincides with the equation for the stationary phase.
 In three-dimensional QED both they $\lambda$, and $\Phi_0$ are
 equal to zero, in the class of functions where
 the physical fields are defined.

 In three-dimensional QED there are only  three subtle differences of the
 {\it equivalent unconstrained
  system}~(\ref{red1}) from the initial gauge theory~(\ref{w}).
 First of them is the origin of the current conservation law.
 In the initial constrained system~(\ref{w}),
 the current conservation law  $\partial_{0}j_{0}=\partial_{i}j_{i}$
 follows from the equations for the gauge fields;
  whereas the
 similar law $\partial_{0}j^*_{0}=\partial_{i}j^*_{i}$ in the
 {\it equivalent unconstrained  system}~(\ref{red})
  follows only from the classical equations for the fermion
  fields. This difference becomes essential  in quantum theory.
  In the  second case, we cannot use
 the current conservation law, if the quantum fermions are off mass-shell,
 in particular, in an atom.
 What we observe in an atom? The bare fermions, or {\it
 dressed} ones~(\ref{vak})? Dirac supposed~\cite{cj} that we can observe
 only {\it gauge invariant} quantities of the type of the {\it
 dressed} fields.
 Really, we can convince that  {\it dressed} fields~(\ref{vak})
  as nonlocal functionals  from initial gauge fields
  are   invariant with respect to the time dependent gauge transformations
  of these initial fields~(\ref{gauge}).

 The gauge
 invariance with respect to the time dependent gauge transformations
 is the second difference of the nonlocal Dirac
 variables~(\ref{vak}) from the initial fields
 of the constrained system~(\ref{w}) with usual
 transformational properties with respect to gauge and Lorentz
 transformations.

 The gauge constraint $ \partial_{i}A_i=0$, in the gauge-fixing method, is
 associated with the relativistic noncovariance.
 Whereas, the observable nonlocal variables~(\ref{vak}) depend on
 the time axis by the {\it relativistic covariant} manner.
 Polubarinov's review~\cite{pol}
 was mainly devoted to the relativistic covariant formulation
 of the Dirac quantization~\cite{cj}.

 The gauge-fixing method and its terminology "the Coulomb gauge"
 do not reflect these three properties of the Dirac observables in
 the  {\it constraint-shell} QED~(\ref{red}): the
 off-mass-shell
 nonconservation of the current, gauge invariance~(\ref{gauge}), and
 relativistic covariance.

 In fact,  the term {\it gauge}~(\ref{ngauge})
 means  a {\it choice of nonlocal
 variables}, or more exactly, a {\it gauge of physical sources}
 associated with these variables.

 This Dirac construction  of a relativistic covariant
 {\it equivalent unconstrained  system} can
 be generalized also on  massive vector theories~\cite{hpp}.
 The generalization of the first Dirac quantization on
 the non-Abelian theory, including QCD~\cite{vp1,p2,inp,kh,bpr},
  is the topic of the present review.

  \subsection{Relativistic covariance}

 Relativistic transformations of the Dirac variables are  discussed  in details in
 the Polubarinov review~\cite{pol} (see also~\cite{31,32,33}).
 If we make usual relativistic transformations of the initial
 fields $A_i, A_0, \psi  $ with the parameter $\epsilon_i$
 \begin{eqnarray}\label{ultf}
  \delta_{L}^{0} A_{k}  &=& \epsilon_{i} ( x_{i}^{\prime}
 \partial_{0^{\prime}} - x_{0}^{\prime} \partial_{i^{\prime}} )
 A_{k}(x^{\prime}) + \epsilon_{k} A_0~,
 \nonumber \\ \\
  \delta_{L}^{0} \psi &=& \epsilon_{i} ( x_{i}^{\prime}
 \partial_{0^{\prime}} - x_{0}^{\prime} \partial_{i^{\prime}} )
 \psi ( x^{\prime} ) + \frac{1}{4} \epsilon_{k} [\gamma_{i},
 \gamma_{j} ] \psi (x^{\prime} )~,  \nonumber
 \end{eqnarray}
 then the physical variables~(\ref{vak}) suffer the Heisenberg-Pauli
 transformations~\cite{hp}
 \begin{eqnarray}\label{ltf}
 A_{k}^{*} [ A_{i} &+& \delta_{L}^{0} A ] - A_{k}^{*} [ A ]  =
 \delta_{L}^{0} A_{k}^{*} + \partial_{k} \Lambda~, \nonumber \\ \\
 \psi^{*} [ A &+& \delta_{L}^{0} A , \psi + \delta_{L}^{0} \psi ] -
 \psi^{*} [ A, \psi ] = \delta_{L}^{0} \psi^{*} + i e \Lambda
 (x^{\prime}) \psi^{*}~,
 \end{eqnarray}
  were
 \begin{eqnarray} \label{lqed}
 \Lambda [A^*,j^*_0] = \epsilon_{k} \frac{1}{{\bf \partial}^{2}} ( \partial_{0}
 A_{k}^{*} + \partial_{k} \frac{1}{\Delta} j_{0}^{*} ) \,\,\, .
 \end{eqnarray}
 These transformation were interpreted by Heisenberg and Pauli~\cite{hp} (with
 reference to the unpublished note by von Neumann) as the
 transition from the Coulomb gauge with respect to the time axis in
 the rest frame \(\eta_{\mu}^0=(1,0,0,0)\) to the Coulomb gauge
 with respect to the time axis in the moving frame (see Fig.1)
 \[
 \eta'_{\mu}=\eta_{\mu}^0 + \delta_L {\eta_{\mu}}^0 \;=\;
 {(L\eta^0)}_{\mu}.\]

\unitlength=1mm \special{em:linewidth 0.4pt} \linethickness{0.4pt}
\begin{picture}(110.00,130.00)
\put(17.00,100.00){\vector(0,1){30.00}}
\put(50.00,100.00){\vector(0,1){30.00}}
\put(80.00,100.00){\vector(1,3){10.00}}
\put(114.00,100.00){\vector(1,3){10.00}}
\put(63.00,116.00){\line(1,0){12.00}}
\put(63.00,114.00){\line(1,0){12.00}}
\put(71.00,119.00){\line(5,-4){5.00}}
\put(76.00,115.00){\line(-5,-4){5.00}}
\put(32.00,114.00){\makebox(0,0)[cc]{${\eta_0}=(1,0,0,0)$}}
\put(99.00,114.00){\makebox(0,0)[cc]{${\eta}^{\prime}={\eta_0
}+\delta^0_L{\eta_0}$}}
\put(58.00,121.00){\makebox(0,0)[cc]{$A_0^T=0$}}
\put(138.00,121.00){\makebox(0,0)[cc]{$A_0^{T^{\prime}}=(\eta^{\prime}
\cdot A) =0$}}
\put(64.00,105.00){\makebox(0,0)[cc]{$\eta=(1,0,0,0)$}}
\put(128.00,105.00){\makebox(0,0)[cc]{$\eta^{\prime}= \eta
+\delta^0_L \eta $}} \put(20.00,90.00){\makebox(0,0)[cc]{ \tt
Lorentz }} \put(17.50,85.00){\makebox(0,0)[cc]{\tt Frame}}
\put(52.00,90.00){\makebox(0,0)[cc]{\sl Gauge}}
\put(82.00,95.00){\makebox(0,0)[cc]{\underline {New} }}
\put(85.00,90.00){\makebox(0,0)[cc]{\tt Lorentz }}
\put(81.95,85.00){\makebox(0,0)[cc]{\tt Frame}}
\put(116.00,95.00){\makebox(0,0)[cc]{\underline {New} }}
 \put(117.00,90.00){\makebox(0,0)[cc]{\sl Gauge}}
 \end{picture}
 \vspace{-8.4cm}
 \begin{center}
{\bf Figure 1.}
 \end{center}

 These transformations correspond the "change of variables"
  \begin{eqnarray} \label{cgauge}
 \psi^*(\eta), A^*(\eta) \rightarrow \psi^*(\eta^{\prime}),
 A^*(\eta^{\prime}) \,\,\, ,
 \end{eqnarray}
 so that they became transverse with respect to the new time axis $\eta'$
 (or, from the point of view of the "gauge-fixing" method of reduction,
 the transformations~(\ref{ultf}),~(\ref{ltf})) correspond to the
 "change of gauge".

 In result  we got the relativistic covariant
 separation of the interaction on the Coulomb potential
 (instantaneous with respect to the time- axis $\eta_\mu$) and on
 the retardation.

 The Coulomb interaction  has the covariant form
 \begin{eqnarray} \label{rc}
 { W}_{C} = \int d^4 x d^4 y \frac{1}{2} j_{\eta}^{*}(x) V_C(z^{\perp})
 j_{\eta}^{*}(y) \delta(\eta \cdot z) \,\,\, ,
 \end{eqnarray}
 Here
 \begin{eqnarray} \label{ccar}
 j_{\eta}^{*} = e \bar {\psi}^* \rlap/\eta \psi^* \,\, , \,\,
 z_{\mu}^{\perp} = z_{\mu} - \eta_{\mu}(z \cdot \eta) \,\, , \,\,
 z_\mu = (x - y)_\mu  \,\, , \,\,
 \end{eqnarray}
 \begin{eqnarray} \label{cf}
 V_C(r) = - \frac{1}{4\pi r} \,\,\, , \,\,\, r = \vert {\bf z} \vert~.
 \end{eqnarray}
 Finite Lorentz transformations from the time axis $\eta^{(1)}$ to the
 time axis $\eta^{(2)}$ where constructed in paper~\cite{pol} using
 the gauge transformations
 \be
 ieA^{(2)}=U_{(2,1)}[ieA^{(1)}+\partial ]U_{(2,1)}^{-1}~,~~~
 \psi^{(2)}=U_{(2,1)} \psi^{(1)}~,
 \ee
  where $U_{2,1}=v_{(2)}v_{(1)}^{-1}$, and $v_{(2)},v_{(1)}$ the
 Dirac gauge factors~(\ref{va}) for the time axes $\eta^{(2)}$
 and $\eta^{(1)}$ respectively.

 \subsection{Quantization and Feynman path integral}

  The initial action~(\ref{w}) is not compatible with
 quantum principles.
 The Dirac formulation of the {\it equivalent unconstrained system}
 keeps the quantum  principles by the value
 of excluding the nonphysical components.
 We quantize the {\it equivalent unconstrained system}
 with gauge- invariant physical variables~(\ref{vak}).
 The corresponding commutation relations
 \begin{eqnarray*}
 i \bigl[ \partial_{0} A^{*}_i ( {\bf x}, t ) \,\, , \,\, A^{*}_j (
 {\bf y}, t ) \bigr]  &=& ( \delta_{ij} -
 \partial_{i}
 \frac{1}{ \Delta } \partial_{j} )
  \delta^3 ( {\bf x} - {\bf y} ) \,\,\, ,     \\  \nonumber  \\
 \bigl\{  {\hat \psi}^{* +} ({\bf x},t ) , \hat {\psi}^{*} ( {\bf
 y},t ) \bigr\} &=& \delta^3 ( {\bf x} - {\bf y} )
 \end{eqnarray*}
 lead to the generating functional for Green's function of the
 obtained unconstrained system in the form of the Feynman path
 integral
 \be \label{fi}
 Z_{\eta}^{*}[ s, {\bar {s}}^*, J^* ]\;=\;\int \prod_{j} DA^*_j
 D\psi^*D{\bar \psi}^*
 e^{iW^{*}[A^*,\psi^*, {\bar \psi}^*] + i S^* },
 \ee
 with external source terms
 \be \label{si}
 S^*\;=\;\int d^4x \left({\bar s}^* \psi^* + {\bar \psi}^* s^*
 +J^*_i A^*_i \right).
 \ee
 By the construction of the {\it unconstrained system} this generating
 functional is  {\it gauge - invariant} and {\it relativistic covariant}.
 Relativistic transformation properties of the
 quantum fields should repeat the ones of the Dirac
 variables~(\ref{vak}) as non-local functionals
 of the initial fields.
 As it was shown in papers~\cite{sch,pol,z,31,32,33}
 that quantum theory with the gauge - invariant
 Belinfante energy-momentum tensor on the constraint
 \begin{eqnarray} \label{beli}
 T_{ \mu \nu}  & = & F_{\mu \lambda }F_{\nu}^{\lambda} + {\bar
 {\psi}} \gamma_{\mu} [ i\partial_{\nu} + e A_{\nu} ] \psi
  -  g_{\mu \nu} L + \nonumber \\
 & + & { i \over 4 }\partial_{\lambda} [ \bar {\psi}
 \Gamma_{\mu\nu}^{\lambda} \psi ]\; ,
 \end{eqnarray}
 \begin{eqnarray*}
 \Gamma_{\mu\nu}^{\lambda} = {1\over
 2}[\gamma^{\lambda}\gamma_{\mu}]\gamma_{\nu} - g_{\mu\nu }
 \gamma^{\lambda} - g_{\nu}^{\lambda}\gamma_{\mu} \;
 \end{eqnarray*}
 completely reproduced the symmetry properties of the "classical"
 theory (\ref{ultf}), (\ref{ltf}), (\ref{lqed})
 \be \label{ql}
 i\epsilon_k[M_{0k},\psi^*]=\delta^0_L\psi^*+ie\Lambda[A^*,j^*_0]\psi^*;
 ~~~~M_{0k}=\int d^3x[x_kT_{00}-tT_{0k}]~.
 \ee
  This Lorentz transformation of the operator quantization means
 on the level of the Feynman path integral the change of the
 time axis
 \be\label{rfi}
 Z_{L\eta}^{*}[ s^*, {\bar {s}}^*,J^* ]\;=\; Z_{\eta}^{*}[
 Ls^*, L{\bar {s}}^*,LJ^* ].
 \ee
 This scheme of quantization depends on a choice of
 the time axis.
 If one choose a definite frame of reference with
 the initial time axis, any Lorentz transformation
 turns this time axis by the relativistic covariant manner.
 In this meaning, the {\it constraint dynamics}
 is {\it relativistic covariant}.
 Another problem is to find conditions when measurable physical
 quantities and results of theoretical calculations
 do not depend on the time axis (identified with a physical device).
 This independence exists only for scattering amplitudes
 of particles onto their mass-shell~\cite{f}. In this case, one
 can say about the {\it relativistic invariance}
 of the scattering amplitudes of the local degrees of freedom.
 But, it is well-known that the Green functions (in particular,
 one-particle Green function) and instantaneous bound states depend on
 the time axis. In general case,  measurable  quantities in electrodynamics
 depend on the time axis and other parameters of a physical device
 including its size and energy resolution~\cite{p2}.

 If a nonlocal process depends on the time axis, one should
 to establish a principle of the choice of this time axis.

 We shall use the generalization of the Markov-Yukawa principle~\cite{ym}
 supposed in~\cite{inp,yaf,mpl,nsh}:
 in concrete calculations the time-axis is chosen to be parallel
  to a total momentum of any state.

  In particular, this choice and the nonlocal relativistic
  transformations (\ref{ql}) remove all infrared divergences from
  the one particle Green function in the radiation
 variables~\cite{inp,mpl,nsh}
  \be \label{green1}
 i (2\pi)^4 \delta^4(p-q) G(p-q)=\int\limits_{ }^{ } d^4d^4y
 \exp(ipx-iqy)<0|T\bar \psi^*(x)\psi^*(y)|0>~,
 \ee
 $$
 G(p)=G_0(p)+ G_0(p) \Sigma (p) G_0(p) + O(\alpha^4)~,~~~
 G_0(p)=[\not p-m]^{-1}
 $$
 $$
 \Sigma (p)=\frac{\alpha}{8\pi^3 i}\int\limits_{ }^{ }
 \frac{d^4q }{q^2+i\epsilon}\left[  \left(\delta_{ij}-q_i
 \frac{1}{\vec q^2} q_j  \right)
 \gamma_i G_0(p+q)\gamma_j
   + \gamma_0  G_0(p+q)\gamma_0\frac{1}{\vec q^2}\right]=
 \frac{\alpha}{4\pi}\Pi(p),
 $$
 where $\Pi(p)$ is
 $$
  m(3D+4)-D (\not p-m)+\frac{1}{2}
 (\not p-m)^2\left[\frac{(\not p+m)}{p^2} \left(ln\frac{m^2-p^2}{m^2} \right)
 \left(1+\frac{\not p(\not p-m)}{2p^2}\right)-
 \frac{\not p}{2p^2}\right],
 $$
 and $D$ is a ultra-violet divergence.
 The transition to another Lorentz frame $p'_{\mu}=(p'_0,\vec p'\not= 0)$
 is accompanied by the additional diagrams
 $<0|T\bar \psi^*(x)\delta_L\psi^*(y)|0>$ induced by the transformation
 (\ref{ql}). As result in another frame we get
 the same relativistic covariant expression depending on the new
 momentum $p'$ (see details in~\cite{mpl,nsh}).

 \subsection{Gauge equivalence theorem and FP integral}

 Thus, the constraint-shell generational
 functional~(\ref{fi}) is relativistic covariant~(\ref{rfi})
 and gauge-invariant by the construction.
 The main difference of this functional from the
 intuitive FP integral~(\ref{fpi11}) is the information contained
 in the solution of the constraint, i.e., the electrostatic phenomena
 of the instantaneous interaction, including the Coulomb-like bound states.

 The Faddeev-Popov integral for the
 generating functional of Green functions in the gauge \( F(A)\;
 =\; 0\) can be obtained from the Feynman integral ~\cite{f}
 by two steps: *) a change of variables, and **) a change of sources.

 *) {\bf the change of variables} is fulfilled by the Dirac
 factors~(\ref{vak}), (\ref{va})
 \begin{eqnarray}\label{vakf}
 A_{k}^{*}[A^F]& =& v[A^F] ( A^F_{k} + i \frac{1}{e} \partial_{k} )
 ( v[A^F] )^{-1},\nn\\
 \psi^{*}[A^F]& =& v[A^F] \psi \,,\,\,
 \end{eqnarray}
 \begin{eqnarray}\label{vaf}
 v[A^F] =  \exp
 \bigl\{ i\, e \frac{1}{{\Delta}}
 \partial^{j} A^F_{j} \bigr\} \,\,\, .
 \end{eqnarray}
 This change introduces additional degrees of freedom
 and the FP determinant $\Delta _{FP}^F$ of the transition to new
 variables of integration.
 These degrees are removed by additional constraints $F(A)=0$.
 Finally, the {\it  constraint-shell} functional
 $Z^*$ (\ref{fi}) takes the equivalent form of the FP path integral
 \begin{eqnarray}\label{fpi}
 Z^{*}[s^*, {\bar s}^*, J^*]\;=\;\int \prod_{\mu}DA^F_\mu D
 \psi^F D{\bar \psi}^F \Delta _{FP}^F
 \delta (F(A^F)) e^{iW[ A^F,\psi^F {\bar \psi}^F ] + S^*},
 \end{eqnarray}
 all electrostatic monopole physical phenomena that
 depend on the time axis are
 concentrated in the Dirac gauge factor $v(A)$ that
 accompanies the physical sources ${\bar s}^*,s^*,J^*$.
 \be\label{fpis}
 S^*\;=\;\int d^4x \left( ( v[A^F] )^{-1}{\bar s}^* \psi^F +
 {\bar \psi}^F  ( v[A^F] )^{-1}s^*
 +J^*_i A^*_i[A^F]\right)
 \ee
 Thus, this change of variables corresponds to the rearrangement of
 the Feynman diagrams. In particular,
 after the change of variables
 the sum of the  Coulomb  kernel and  transverse photon propagator
 \begin{eqnarray*}
 {\cal K}^{R}(J)&=&J_{0}^{(1)}{1 \over {\bf q}^{2}}J_{0}^{(2)}+ \nonumber \\
 &+&J_{i}^{(1)}(\delta_{ij}-q_{i}{1 \over {\bf q}^{2}}q_{j}) {1
 \over q_{0}^{2}-{\bf q}^{2}}J_{j}^{(2)}
 \end{eqnarray*}
 converts  into the identically equivalent sum of
 the Feynman gauge propagator ${\cal K}^{F}$ and
 the longitudinal term $ {\cal K}^{L} $ :
 \be \label{sds}
 {\cal K}^{R}(J)  \equiv
 {\cal K}^{F}(J)+{\cal K}^{L}(J)~,
 \ee
 where
 $$
 {\cal K}^{F}(J)  =
 -[J_{0}^{(1)}J_{0}^{(2)}-J_{i}^{(1)}J_{i}^{(2)}]{1\over
 q_{0}^{2}-{\bf q}^{2}}~,
 $$
 $$
 {\cal K}^{L}(J)  =
 \left[(q_{0}J_{0}^{(1)})(q_{0}J_{0}^{(2)})
  -  (q_{i}J_{i}^{(1)}) (q_{j}J_{j}^{(2)})\right]{1\over {\bf
 q}^{2}(q_{0}^{2}-{\bf q}^{2})}~.
 $$

 **) The next step is {\bf the change of sources}
 \be\label{fpso}
 \label{ps}  S^* ~\Rightarrow~S^F=\int d^4x \left( {\bar s}^F \psi^F +
 {\bar \psi}^F s^F
 + A^F_{\mu}J^{\mu}\right)~.
 \ee
 In the result we get the original FP integral~(\ref{fpi11})
 \begin{eqnarray}\label{fpi1}
 Z^{FP}[s^F, {\bar s}^F, J^F]\;=\;\int \prod_{\mu}DA^F_\mu D
 \psi^F D{\bar \psi}^F \Delta _{FP}^F
 \delta (F(A^F)) e^{iW[ A^F,\psi^F {\bar \psi}^F ] + S^F}~.
 \end{eqnarray}
 without the Dirac factors.
 Changing the sources we lose the
 Dirac factor together with the  class of the {\bf spurious diagrams}
 that remembered the electrostatic
 phenomena and instantaneous bound states.
 One of these spurious diagrams
 is the longitudinal term $ {\cal K}^{L} $ in equation~(\ref{sds}).
 This longitudinal term disappears only
 on the mass-shell (because of the
 current conservation law
  $J_{0}^{(1,2)}q_{0}=J_{i}^{(1,2)}q_{i}$).
 As we have seen before, the Dirac nonlocal gauge-invariant observable
  currents do not satisfy the conservation law for
 particles off the mass-shell, in particular, in  bound states.

 Really, the FP perturbation theory in the relativistic gauge~(\ref{fpi1})
 contains only photon propagators with the light-cone singularities
 forming the Wick-Cutkosky bound states with the spectrum differing
 from the observed one which corresponds to the instantaneous Coulomb
 interaction~\cite{kum}. The Wick-Cutkosky bound states have the
 problem of a tachyon and the probability interpretation.
  These problems where solved by the quasipotential method
  ~\cite{kad} that introduces instantaneous bound states, and
  the corresponding  time axis as the Markov-Yukawa
 prescription~\cite{ym}.


 Thus, in QED, the fundamental
 constraint-shell functional~(\ref{fi})
 coincides with the FP integral~(\ref{fpi}) in the sector of scattering
 amplitudes for elementary particles on their mass-shell~\cite{f}
 that is not actual for solution of the problem of hadronization in QCD.
 For description of the bound state sector in gauge theories,
 including QCD,
 we have three alternatives:\\
 1) the FP integral~(\ref{fpi}),\\
 2) its quasipotential approach to bound states at the level of the
 FP integral~\cite{kad}, and\\
 3) the constraint-shell functional~(\ref{fi}) added by the
 Markov-Yukawa prescription of the choice of the time axis.

 After Faddeev papers of 1967-69
 solutions of all problems of gauge theories, including the description of bound states,
 the Gribov ambiguity~\cite{g}, topological degeneration~\cite{ins,Fad},
  were considered only at the level of the FP integral~(\ref{fpi}).

 The task of the present paper is to consider these problems at the most
 fundamental level of the Hamiltonian approach to quantization of
 gauge theories. It is based on a definite frame of reference, that
 includes a choice of the time axis,
 initial data, boundary conditions, normalization of wave functions,
 bound states, etc.. It is useful to recall the following
 words by Max Born about quantum theory (\cite{born}, p. 108):
 " The clue is the point ..., that quantum mechanics does not describe
 a situation in an objective external world, but
 a definite experimental
 arrangement for observing a section of the external word.
 Without this
 idea even the formulation of a dynamical problem in quantum theory is
 impossible. But if it is acceptable, the fundamental indeterminacy in
 the  physical predictions becomes natural as no experimental
 arrangement can ever be absolute precise."
 Following to Max Born one can say that the Hamiltonian
 description of any quantum system is determined by
 "a definite experimental
 arrangement for observing a section of the external word".
 If it is acceptable, the dependence of the
  quantum description of nonlocal processes on a frame
 of reference becomes natural as any experimental
 arrangement is included in this frame.
 The Hamiltonian method determines the energy
 spectrum of physical states. Therefore, a direction of the
 total momentum of any physical state is distinguished.
 In the case of bound states, the time axis is chosen
 along the total momentum of any bound state in the context
 of the Markov-Yukawa solution~\cite{ym,yaf,ed} of the problem
 of the relativistic covariance in QED.

 \section{QED of bound states: spectrum and S-matrix}

 \subsection{Markov-Yukawa prescription}

 One of the first definitions of the physical bound states in QED
 belongs to Lord
 Eddington: "A proton yesterday and electron today do not make an
 atom"~\cite{ed}. It is clear that we can observe experimentally
 two particles as a bound state ${\cal M}(x,y)  $ at one and the
 same time
 \begin{eqnarray} \label{3-1}
 {\cal M}(x,y) = e^{iMX_0} \psi(z_i) \delta(z_0) \,\,\, ,
 \end{eqnarray}
 where $X_\mu$ and $z_\mu$ are the total and relative coordinates
 \begin{eqnarray} \label{3-2}
 X_\mu = \frac{(x+y)_{\mu}}{2} \,\,\, , \,\,\, z_\mu = (x-y)_\mu \,\,\, .
 \end{eqnarray}

 This principle of the simultaneity has more deep mathematical
 meaning~\cite{yaf,a14} as the constraint of irreducible nonlocal
 representations of the Poincare group for arbitrary bilocal field
 $ {\cal M}(x,y) = {\cal M}(z \vert X)   $
 \begin{eqnarray}\label{3-3}
 z_\mu \frac{\partial}{\partial X_\mu} {\cal M}(z \vert X) = 0~,
 ~~~~~~~~~~~~~~{\cal M}(z|X)\equiv {\cal M}(x,y)
 \,\,\, .
 \end{eqnarray}
 This constraint is not connected with the dynamics of interaction
 and realized the Eddington simultaneity.

 The general solution of the irreducibility constraint~(\ref{3-3})  can be
 written in the form of the expansion of the bilocal field ${\cal
  M}(z \vert X)$ with respect to "in" and "out" plane waves
 \begin{eqnarray} \label{3-4}
  {\cal M}(z \vert X) = \sum_{A} \int
  \frac{ d^3 {\cal P}_A }{ \sqrt{(2\pi)^3 \omega_A }}
 \left[ e^{ i{\cal P}_A\cdot X} \Phi_{{\cal P}_A}(z_A^{\perp})
  a^{(+)}_{{\cal P}_A} + e^{- i{\cal P}_A\cdot X} \bar{\Phi}_{{\cal
  P}_A}(z_A^{\perp}) a^{(-)}_{{\cal P}_A} \right] \delta \Bigl(\frac{{\cal
 P}_A \cdot z}{ \sqrt{ {\cal P}^2 }   }\Bigr)
 \end{eqnarray}
 where $a^{(\pm)}_{{\cal P}_A}$ are coefficients of the expansion.
 $\Phi_{{\cal P}_A}(z_A^{\perp}) $,~$ \bar{\Phi}_{{\cal
 P}_A}(z_A^{\perp})$  are the normalized amplitudes in the space of
 relative coordinates orthogonal to the total momentum of an atom
  ${\cal P}_A   $
 \begin{eqnarray}  \label{3-5}
 (z_A^{\perp})_\mu = z_\mu - {\cal P}_{A\mu}
 \bigl( \frac{{\cal P}_A \cdot z}{{\cal P}^2 } \bigr) \,\,\, .
 \end{eqnarray}
 It is clear, that at the point of the existence of the bound state
 with the definite total momentum $ {\cal P}_{A\mu} $ any
 instantaneous interaction~(\ref{rc})  with the time- axis $\eta_\mu$
 parallel to this momentum $\eta_\mu \sim {\cal P}_{A\mu} $
 \begin{eqnarray} \label{3-6}
 \eta_\mu  {\cal M} (z \vert X)   \sim {\cal P}_{A\mu} {\cal M} (z
 \vert X) = \frac{1}{i} \frac{\partial}{\partial X_\mu} {\cal M} (X
 \vert z)
 \end{eqnarray}
 is much greater then any "retardation" interaction~\cite{love}. It is just
 our principle of the choice of the
 time- axis of the Dirac quantization
 of a gauge theory.  A time axis is chosen to be parallel to the total
 momentum of a considered state. In particular, for bound states
 this choice means that the coordinate of the potential coincides
 with the space of the relative coordinates of the bound state wave
 function in the accordance with the Markov-Yukawa prescription~\cite{ym}
 and the Eddington concept of simultaneity~\cite{ed}.
 In this case, we get the relativistic covariant
 dispersion law and invariant mass spectrum. The  relativistic
 generalization of the Coulomb potential  is not only the change of
 the form of the potential, but also the change of a direction of
 its motion in four-dimensional space to lie along the total
 momentum of the bound state. The relativistic covariant unitary
 perturbation theory in terms of such the relativistic
 instantaneous bound states has been constructed in~\cite{yaf}. In
 this perturbation theory, each instantaneous bound state in QED
 has a proper equivalent unconstrained system of the Dirac quantization.
 The manifold of
 frames corresponds to the manifold of "equivalent unconstrained
 systems".  In this case, the bilocal fields~(\ref{eta2})
 automatically belong to the irreducible representation of the
 Poincare group \cite{a14}.

 By analogy, we introduce for the $N$- local field
 the total and relative coordinates
 \begin{eqnarray}
 X_{\mu} = {1 \over N} \sum_{i=1}^{N} x_{i\mu} \,\, , \,\,
 z_{\mu}^{(i)} = x_{i\mu} - X_{\mu}
 \end{eqnarray}
 which are connected by the identity
 \begin{eqnarray*}
 \sum_{i=1}^{N} z_{\mu}^{(i)} = 0 .
 \end{eqnarray*}
 Then, the generalization of the Markov - Yukawa condition takes the form
 \begin{eqnarray}\label{myn}
 z_{\mu}^{(i)} { \partial \over {\partial X_{\mu}} }
 \Phi ( z_{\mu}^{(1)}, z_{\mu}^{(2)}, \, ... \, , z_{\mu}^{(N)} ) = 0
 \, \, \, \,  (i=1,2, \, ... \, N) .
 \end{eqnarray}
 Let $ {\cal P}_{\mu} $ be the eigenvalue of the operator for the total
 4-momentum, and $ \eta_{\mu} $ be the unit vector in the direction
 $ {\cal P} (  \eta_{\mu} \sim {\cal P}_{\mu} ) .$
 Owing to the condition~(\ref{myn})
 the $N$- local function
 $$
 \underline {\Phi} ( p_{\mu}^{\perp (1)},
 p_{\mu}^{\perp (2)} \,\, ... \,\, ,p_{\mu}^{\perp (N)} \vert {\cal P} )~,
 $$
 being the Fourier transform of $ \Phi(z_{\mu}^{(1)},X_{\mu})$ with
 respect to all coordinates, depends only on the transverse relative momenta
 \begin{eqnarray}
 p_{\mu}^{(i)\perp } =
 p_{\mu}^{(i)} - \eta_{\mu} ( p^{(i)} \cdot \eta ), ~~~~\,\,\,
 \sum_{i=1}^{N} p_{\mu}^{(i) \perp} = 0 .
 \end{eqnarray}

 See also the generalization of the Markov-Yukawa condition for
 three-local~\cite{a30} and N-local~\cite{a31} cases.

 \subsection{Effective Lagrangian of bilocal fields}

 The  constraint-shell QED allows us to construct the
 "bound state" relativistic covariant
 perturbation theory with respect to "retardation"
 ~\cite{yaf,love}.
 Our solution of the problem of relativistic invariance
 of the nonlocal objects is the choice of the time axis as a
 vector operator with eigenvalues proportional to total momenta
 of  bound states~(\ref{3-6})~\cite{yaf}.
 In this case, the relativistic covariant
 unitary S-matrix can be defined as the Feynman path integral
 \be \label{fi2}
 Z_{\hat \eta}^{*}[ s, {\bar {s}}^*, J^* ]\;=<*|\int
 D\psi^*D{\bar \psi}^*
 e^{iW_C^{*}[\psi^*, {\bar \psi}^*] + i S^* }|*>~,
 \ee
 where
 \be
  <*|F|*>=\;\int \prod_{j} DA^*_je^{iW_0^{*}[A^*]}F
 \ee
 is the averaging over transverse photons
 \begin{eqnarray} \label{act2}
 W_{C} [ \psi , \bar{\psi} ] &=&
 \int d^{4}x [ \bar{\psi}(x) ( i \rlap/{\partial}- ie\rlap/A^* - m^{0} ) \psi(x) +   \nonumber  \\
 &+& { 1 \over 2 } \int d^{4}y ( \psi(y) \bar{\psi}(x) ) {\cal
 K}^{(\eta)} ( z^{\bot} \mid X ) ( \psi(x) \bar{\psi}(y) ) ] .
 \end{eqnarray}
 Here $ \rlap/\partial = \partial^{\mu} \gamma_{\mu},\,\, {\cal
 K}^{(\eta)}$ is the kernel
 \begin{eqnarray} \label{3-7}
 {\cal K}^{(\eta)}( z^{\perp} \mid X ) = \rlap/\eta V(z^{\perp})
  \delta(z \cdot \eta ) \rlap/\eta , \, \, ~~~~~~~~( \rlap/\eta = \eta^{\mu}
 \gamma_{\mu}=\gamma \cdot \eta,~~~~ z_{\mu}^{\perp} = z_{\mu} - \eta_{\mu}
 ( z \cdot \eta ))~ ,
 \end{eqnarray}
  and $z$ and $X$ are the relative and total coordinates defined
  in equation~(\ref{3-2}),
 and $ V( z^{\perp} ) $ is the potential depending on the
 transverse  component of the relative coordinate
 with respect to the time axis $\eta$
 \begin{eqnarray}\label{eta2}
 \eta_{\mu} \sim i { \partial \over { \partial X_{\mu} }} ~.
 \end{eqnarray}

In the context of this perturbation theory the bound state total
 momentum operators (23) can form the "new quantum numbers" of the
 type the Isgur- Wise ones~\cite{a1,a2}.

 It seems that a most straightforward way for constructing a theory of
 bound states is the redefinition of action (\ref{act2}) in terms of bilocal
 fields by means of the Legendre transformation \cite{a15,pre}
 \begin{eqnarray}  \label{3-8}
 {1 \over 2} &\int &d^{4}x d^{4}y  ( \psi(y) \bar{\psi}(x) ) {\cal
 K}(x,y) ( \psi(x) \bar{\psi}(y) )  = \nonumber \\  = -{1 \over 2}
 &\int &d^{4}x d^{4}y  {\cal M}(x,y) {\cal K}^{-1}(x,y) {\cal
 M}(x,y) + \\  +  &\int &d^{4}x d^{4}y ( \psi(x) \bar{\psi}(y) )
 {\cal M}(x,y) \nonumber
 \end{eqnarray}
 where $ {\cal K}^{-1} $ is the inverse of the kernel~(\ref{3-7}).
 Following ref. \cite{pre}, we introduce the short - hand notation
 \begin{eqnarray} \label{short}
 \int d^{4}x d^{4}y \psi (y) \bar{\psi}(x) ( i
 \rlap/\partial -ie\rlap/A^*- m^{0} ) \delta^{(4)} (x-y)
 &=&  ( \psi \bar{\psi} , - G_{A}^{-1} ) \, \, ,  \\
 \int d^{4}x d^{4}y ( \psi(x) \bar{\psi}(y) ) {\cal M}(x,y) &=& (
 \psi \bar{\psi}, {\cal M} ) .
 \end{eqnarray}

 After quantization ( or integration ) over $N_{c}$ fermion fields
 and normal ordering, the functional~(\ref{fi2}) takes the form
 \be
 \label{fi3}
 Z_{\hat \eta}^{*}[ s, {\bar {s}}^*, J^* ]\;=<*|\int \prod D{\cal M}
 e^{iW_{eff}[{\cal M}] + i S_{eff}[{\cal M}]}|*>
 \ee
 where
 \begin{eqnarray} \label{3-9}
 W_{eff}[{\cal M}] = ( \psi \bar{\psi}, ( - G_{A}^{-1} + {\cal M})
 ) - { 1 \over 2} ( {\cal M}, {\cal K}^{-1} {\cal M} )
 \end{eqnarray}
 is the effective action, and
\begin{eqnarray} \label{3-10}
S_{eff}[{\cal M}] = ( s^* \bar{s^*}, (  G_{A}^{-1} - {\cal
M})^{-1} )
 \end{eqnarray}
is the source term. The effective action can be decomposed in the
form
 \begin{eqnarray} \label{3-11}
W_{eff}[{\cal M}] = - {1 \over 2} N_{c} ( {\cal M}, {\cal K}^{-1}
{\cal M} ) + i N_{c} \sum_{n=1}^{\infty} {1 \over n} \Phi^{n} .
\end{eqnarray}
Here $ \Phi \equiv G_{A} {\cal M} , \Phi^{2} , \Phi^{3} $ etc.
mean the following expressions
\begin{eqnarray}   \label{3-12}
\Phi (x,y) \equiv G_{A} {\cal M} = \int d^{4}z G_{A} (x,z) {\cal M}(z,y), \nonumber \\
\Phi^{2} = \int d^{4} x d^{4}y \Phi(x,y) \Phi(y,x) , \\
\Phi^{3} = \int d^{4} x d^{4}y d^{4}z \Phi(x,y) \Phi(y,z)
\Phi(z,x) \,\,  , etc \nonumber
\end{eqnarray}

As a result of such quantization, only the contributions with
inner fermionic lines ( but no scattering and dissociation channel
contribution ) are included in the effective action since we are
interested only in the bound states.

 The requirement for the choice of the time axis (\ref{eta2}) in
 bilocal dynamics is equivalent to Markov - Yukawa
 condition \cite{ym}~(\ref{ym})
 \begin{eqnarray}\label{ym}
 z_{\mu} \cdot i{ \partial {\cal M}(z|X) \over { \partial X_{\mu}}
 } = 0~,
~~~~~~~~~~~~~~{\cal M}(z|X)\equiv {\cal M}(x,y)
 \end{eqnarray}
 where $ z_{\mu} = ( x-y)_{\mu} $ and $ X_{\mu} = (1/2)(x+y)_{\mu}
 $ are relative and total coordinates.

 \subsection{Quantization of bilocal fields}

 The first step to the quantization of the effective action  is the
 determination of its minimum
 \begin{eqnarray}\label{8}
 N_{c}^{-1}{ \delta W_{eff} ({\cal M}) \over { \delta {\cal M}} } =
 =- {\cal K}^{-1} {\cal M} + i \sum_{n=1}^{\infty} G_{A}( {\cal M}
 G_{A})^{n} \equiv - {\cal K}^{-1} {\cal M} + {i \over { G_{A}^{-1}
 - {\cal M} } } = 0 .
 \end{eqnarray}
  We denote the corresponding classical solution for the bilocal
 field by $ \Sigma (x-y) $. It depends only on the difference
 $x-y$ because of translation invariance of vacuum solutions.

 The next step is the expansion of the effective action around the
 point of minimum $ {\cal M} = \Sigma + {\cal M}^{\prime} $ ,
 \begin{eqnarray}\label{9}
 W_{eff} ( \Sigma + {\cal M}^{\prime} ) &= & W_{eff}^{(2)}+W_{int}; \nonumber \\
 W_{eff}^{(2)} ({\cal M}^{\prime} ) &= & W_{Q}(\Sigma) + N_{c}[ -
  {1\over2} {\cal M}^{\prime} {\cal K}^{-1} {\cal M}^{\prime}
 + { i \over 2} ( G_{\Sigma} {\cal M}^{\prime} )^{2} ] ; \nonumber \\
 W_{int}=\sum_{n=3}^{\infty}W^{(n)}& = & i N_{c} \sum_{n=3}^{\infty} {1\over n} ( G_{\Sigma} {\cal
 M}^{\prime} )^{n}, \, \, \,~~~~~~~~~  ( G_{\Sigma} = ( G_{A}^{-1} -
 \Sigma)^{-1} ),
 \end{eqnarray}
  and the representation of the small fluctuations $ {\cal
 M}^{\prime} $  as a sum over the complete set of orthonormalized
 solutions $ \Gamma $, of the classical equation
 \begin{eqnarray}\label{10}
 { \delta^{2}W_{eff} ( \Sigma + {\cal M}^{\prime} ) \over { \delta
 {\cal M}^{2}} } \vert_{ {\cal M}^{\prime} = 0 }  \cdot \Gamma = 0.
 \end{eqnarray}
 with a set of quantum numbers ($H$) including masses $M_H=\sqrt{{\cal
 P}_\mu^2}$ and energies $\omega_H=\sqrt{{\vec {\cal P}}^2+M_H^2}$
 \be\label{set}
 {\cal M}^{\prime}(z|X)=\sum\limits_H\int\frac{d^3\vec {\cal
 P}}{(2\pi)^{3/2}\sqrt{2\omega_H}}\int\frac{d^4q}{(2\pi)^4}
 \{ e^{i\vec {\cal P}\vec{X}} \Gamma_H(q^{\bot}|{\cal P})a^+_H({\cal P})
 +e^{-i\vec {\cal P}\vec{X}} \bar{\Gamma}_H(q^{\bot}|-{\cal P})a^-_H({\cal P})\}
 \ee
 The bound state creation and annihilation operators obey
 the commutation relations
 \begin{eqnarray} \label{comrel}
 \biggl[
 a^{-}_{H'}( \vec{\cal P'} )     ,
 a^{+}_{H} ( \vec{\cal P}  )
 \biggr]   = \delta_{H'H}
 \delta^3 (
 \vec{\cal P'} - {\cal P} ) \,\,\, .
 \end{eqnarray}
 \be \nonumber
 \biggl[ a^{\pm}_H({\cal P}),a^{\pm}_{H'}({\cal P}')\biggl]=0~.
 \ee
 The  corresponding Green function takes form
 \be\label{green}
 {\cal G}(q^{\bot},p^{\bot}|{\cal P})=\sum\limits_H
 \left\{
 \frac{\Gamma_H(q^{\bot}|{\cal P})\bar{\Gamma}_H(p^{\bot}|-{\cal P})}
 {({\cal P}_0-\omega_H-i\varepsilon)2\omega_H}-
 \frac{\Gamma_H(p^{\bot}|{\cal P}))}
 {({\cal P}_0-\omega_H-i\varepsilon)2\omega_H}~.
 \right\}~,
 \ee

 To normalize vertex functions we can use the
 "free" part of effective action (\ref{9})
 for the quantum bilocal meson {\cal M}
 with the commutation relations~(\ref{comrel}).
 The substitution of the off - shell
 $\sqrt{{\cal P}^2} \neq M_H$ decomposition
 (\ref{set}) into the "free" part of effective
 action defines the reverse Green function of the bilocal field
 ${\cal G}({\cal P}_0)$
 \begin{eqnarray}
 W_{eff}^{(0)} [{\cal M}] = 2\pi
 \delta ( {\cal P}_0 - {\cal P'}_0 )
 \sum_H \int \frac{d \vec{\cal P}}{\sqrt{2 \omega }}
 a^{+}_{H}( \vec{\cal P} )~
 a^{-}_{H} ( - \vec{\cal P}  )
    {\cal G}^{-1}_H ({\cal P}_0) \,\,\, ,
 \end{eqnarray}
 where ${\cal G}^{-1}_H ({\cal P}_0) $ is the reverse
 Green function which can be represented as difference of two terms
 \begin{eqnarray}
 {\cal P}_H^{-1}({\cal P}_0) =
 I( \sqrt{ {\cal P}^2 } ) -I(M_H^{ab}(\omega))
 \end{eqnarray}
 where $M_H^{ab}(\omega)$ is the eigenvalue of the equation
 for small fluctuations~(\ref{10}) and
 \begin{eqnarray}
 I(\sqrt{{\cal P}^2}) &=&
 N_c \int \frac{d^3 q^\perp}{(2\pi)^3} \cdot \nonumber \\ \\
 &\cdot &
 \biggl\{
 \frac{i}{2\pi}
 \int dq_0 \mbox{tr}
 \bigl[
 G_{\Sigma_b}(q-\frac{{\cal P}}{2} )
    \bar{\Gamma}_{ba}^H (q^\perp | -{\cal P})
 G_{\Sigma_a}( q+ \frac{{\cal P}}{2} )
    {\Gamma}_{ab}^H (q^\perp | {\cal P})
 \bigr]
 \biggr\}        \,\,\, .  \nonumber
 \end{eqnarray}
 According to quantum field theory, the normalization condition
 is defined by formula
 \begin{equation}
 2 \omega =
 \frac{\partial{\cal G}^{-1}({\cal P}_0)}{\partial{\cal P}_0}|_{{\cal P}_0=\omega({\cal P}_1)}
 = \frac{dM({\cal P}_0)}{d{\cal P}_0}\frac{dI(M)}{dM}|_{{\cal P}_0=\omega}~.
 \end{equation}
 Finally, we get that
  solutions of equation (\ref{10}) satisfy the normalization
 condition~\cite{nakanishi}
 \be\label{nakanishi}
 iN_c\frac{d}{d {\cal P}_0}\int \frac{d^4q}{(2\pi)^{4}}
 tr \left[  \underline{G}_\Sigma(q-\frac{{\cal
 P}}{2})\bar{\Gamma}_H(q^{\bot}|-{\cal P}) \underline{G}_\Sigma(q+\frac{{\cal
 P}}{2}){\Gamma}_H(q^{\bot}|{\cal P})
 \right]=2\omega_H
 \ee
 and
 \be\label{nak}
 \underline{G}_\Sigma(q)=\frac{1}{\not q-\underline{\Sigma}(q^{\bot})},~~~~~~
 \underline{\Sigma}(q) = \int d^{4}x \Sigma(x) e^{iqx}
 \ee
 is the fermion Green function.

 \subsection{Schwinger-Dyson equation: the fermion spectrum}

 The equation  of stationarity (\ref{8})   can be rewritten form of
  Schwinger-Dyson (SD) equation
 \begin{eqnarray}\label{sd}
 \Sigma(x-y) = m^{0} \delta^{(4)} (x-y) + i {\cal K}(x,y)
 G_{\Sigma}(x-y)~.
 \end{eqnarray}
 It describes the spectrum of Dirac
  particles in bound states.
 In the momentum space  with
 $$
 \underline{\Sigma}(k) = \int d^{4}x \Sigma(x) e^{ikx}
 $$
  for the Coulomb type kernel we obtain the following equation for
 the mass operator (
 $ \underline {\Sigma} $ )
 \begin{eqnarray} \label{sdm}
 \underline{\Sigma}(k) = m^{0} + i \int { d^{4}q \over { (2\pi)^{4}
 }} \underline {V} ( k^{\perp} - q^{\perp} ) \rlap/\eta
 \underline{G}_{\Sigma}(q) \rlap/\eta ,
 \end{eqnarray}
 where $ G_{\Sigma}(q) = ( \rlap/q - \underline{\Sigma}(q))^{-1} ,
 \underline {V} ( k^{\perp} ) $ means the Fourier transform of the
 potential, $ k^{\perp}_{\mu} = k_{\mu} - \eta_{\mu} ( k \cdot
 \eta) $
 is the transverse with respect to $ \eta_{\mu} $ relative
 momentum.
 The quantity $ \underline{\Sigma} $
 depends only on the transverse momentum
 \begin{eqnarray*} \label{sd11}
 \underline{\Sigma}(k) = \underline{\Sigma}(k^{\perp}),
 \end{eqnarray*}
 because of the instantaneous form of the potential $
 \underline{V}(k^{\perp} )$ in any frame.
 The fermion spectrum can be obtained by solving the Schwinger-Dyson
 equation (\ref{sd}).
 We may integrate it  over the longitudinal momentum $ q_{0}= (q \cdot
 \eta) $ using the representation
 \begin{eqnarray} \label{sd21}
 \underline{\Sigma}_{a}(q) = \rlap/q^{\perp} + E_{a} ( q^{\perp})
 S^{-2}_{a} (q^{\perp})
 \end{eqnarray}
 for the self - energy with
 \begin{eqnarray} \label{sd3}
 S^{-2}_{a}(q^{\perp})= \exp \{ -  \hat{\rlap/q}^{\perp} 2
 {\upsilon}_{a}(q^{\perp}) \} , \,\,\, \hat{q}^{\perp}_{\mu} =
 q^{\perp}_{\mu} / \vert q^{\perp} \vert
 \end{eqnarray}
 where $ S_{a} $ is the Foldy - Wouthuysen type transformation
 matrix with the parameter $ {\upsilon}_{a} $.

 Then, one has
 \begin{eqnarray} \label{sd4}
 \underline{G}_{\Sigma_{a}} &=&
 [ q_{0} \rlap/\eta - E_{a}(q^{\perp}) S_{a}^{-2}(q^{\perp})]^{-1} = \nonumber \\
 & = & [ { \Lambda^{(\eta)}_{(+)a} (q^{\perp}) \over { q_{0} -
  E_{a}(q^{\perp}) +i \epsilon} } +
 { \Lambda^{(\eta)}_{(-)a} (q^{\perp}) \over
 { q_{0} + E_{a}(q^{\perp}) +i \epsilon} } ] \rlap/\eta
 \end{eqnarray}
 where
 \begin{eqnarray} \label{ope1}
  \Lambda^{(\eta)}_{(\pm)a}(q^{\perp})= S_{a}(q^{\perp})
 \Lambda^{(\eta)}_{(\pm)}(0) S_{a}^{-1}(q^{\perp}) , \,\,\,
 \Lambda^{(\eta)}_{(\pm)}(0)= ( 1 \pm \rlap/\eta ) / 2
  \end{eqnarray}
 are the operators separating the states with positive ( $ + E_{a}
 $ ) and negative ( $ - E_{a} $ ) energies.
 As a result, we obtain the following equations for the one -
 particles energy $ E $ and the angle $ {\upsilon} $ :
 \begin{eqnarray}\label{sd1}
 E_{a}(k^{\perp}) \cos 2 {\upsilon}(k^{\perp}) = m^{0}_{a} +
 {1\over2}\int { d^{3}q^{\perp}\over (2\pi)^{3} }
 \underline{V}(k^{\perp}-q^{\perp}) \cos 2{\upsilon}(q^{\perp})
 \end{eqnarray}
 \begin{eqnarray} \label{sd2}
 E_{a}(k^{\perp}) \sin 2 {\upsilon}(k^{\perp}) = \vert k^{\perp}
 \vert + {1\over2}\int { d^{3}q^{\perp}\over (2\pi)^{3} }
 \underline{V}(k^{\perp}-q^{\perp}) \vert k^{\perp} \cdot q^{\perp}
 \vert \sin 2{\upsilon}(q^{\perp})
 \end{eqnarray}

 \subsection{Bethe - Salpeter equation: bound-state spectrum}

 Equations for the spectrum of the bound states (\ref{10})  can be
 rewritten in the form of the
 Bethe-Salpeter (BS) one~\cite{a18}
 \begin{eqnarray}\label{bs}
 \Gamma = i {\cal K}(x,y) \int d^{4}z_{1} d^{4}z_{2}
 G_{\Sigma}(x-z_{1}) \Gamma(z_{1},z_{2}) G_{\Sigma}(z_{2}-y)~.
 \end{eqnarray}
In the momentum space we obtain with
 $$\underline{\Gamma}(q \vert {\cal P}) = \int d^{4}x d^{4}y
 e^{i{x+y \over2} {\cal P}} e^{i(x-y)q} \Gamma(x,y)
 $$
 for the Coulomb type kernel  the following equation for
 and the vertex function ( $ \underline
 {\Gamma} $ )
 \begin{eqnarray} \label{bs0}
 \underline{\Gamma}(k, {\cal P}) = i \int {d^{4}q \over (2\pi)^{4}}
 \underline {V} ( k^{\perp} - q^{\perp} ) \rlap/\eta [
 \underline{G}_{\Sigma}(q+{ {\cal P} \over 2 }) \Gamma(q \vert
 {\cal P} ) \underline{G}_{\Sigma}(q-{ {\cal P} \over 2 }) ]
 \rlap/\eta
 \end{eqnarray}
 where $
 \underline {V} ( k^{\perp} ) $ means the Fourier transform of the
 potential, $ k^{\perp}_{\mu} = k_{\mu} - \eta_{\mu} ( k \cdot
 \eta) $ is the transversal with respect to $ \eta_{\mu} $ relative
 momentum, ${\cal P}_{\mu}$ is the total momentum.

 The quantity  $ \underline {\Gamma} $
 depends only on the transversal momentum
 \begin{eqnarray*}
 \underline{\Gamma}(k \vert {\cal P}) =
 \underline{\Gamma}(k^{\perp} \vert {\cal P}) ,
 \end{eqnarray*}
 because of the instantaneous form of the potential $
 \underline{V}(k^{\perp} )$ in any frame.

 We consider the Bethe - Salpeter equation~(\ref{bs})
 after integration over the longitudinal momentum
 $ q_{0} $. The vertex function takes the form
 \begin{eqnarray} \label{bs1}
 \Gamma_{ab}(k^{\perp} \vert {\cal P}) = \int { d^{3}q^{\perp}
 \over (2\pi)^{3} } \underline{V} ( k^{\perp}-q^{\perp} )
 \rlap/\eta \psi_{ab}(q^{\perp}) \rlap/\eta,
 \end{eqnarray}
 where the bound state wave function $ \psi_{ab} $ is given by
 \begin{eqnarray} \label{bs3}
 \psi_{ab}(q^{\perp})= \rlap/\eta [ {
 \bar{\Lambda}_{(+)a}(q^{\perp} \Gamma_{ab}(q^{\perp} \vert {\cal
 P} ) {\Lambda}_{(-)b}(q^{\perp}) \over { E_{T} - \sqrt{ {\cal
 P}^{2} } + i \epsilon } } + { \bar{\Lambda}_{(-)a}(q^{\perp}
 \Gamma_{ab}(q^{\perp} \vert {\cal P} ) {\Lambda}_{(+)b}(q^{\perp})
 \over { E_{T} + \sqrt{ {\cal P}^{2} } - i \epsilon } } ]
 \rlap/\eta
 \end{eqnarray}
 $ E_{T} = E_{a} + E_{b} $ means
 the sum of one - particles energies of the two particles ($ a$)
 and ( $ b$ ) defined by~(\ref{sd1}),~(\ref{sd2})  and the notation~(\ref{ope1})
 \begin{eqnarray}  \label{ope2}
 \bar{\Lambda}_{(\pm)}(q^{\perp}) = S^{-1}(q^{\perp})
 \Lambda_{(\pm)}(0) S(q^{\perp}) = {\Lambda}_{(\pm)}( - q^{\perp})
 \end{eqnarray}
 has been introduced.

 Acting with the operators (\ref{ope1}) and (\ref{ope2}) on
 equation (\ref{bs1}) one gets
 the equations for the wave function $ \psi $ in an arbitrary moving
 reference frame
 \begin{eqnarray} \label{bs2}
 &(& E_{T}(k^{\perp}  ) \mp \sqrt { {\cal P}^{2} } )
 \Lambda^{(\eta)}_{(\pm)a}(k^{\perp}) \psi_{ab}(k^{\perp})
 {\Lambda}^{(\eta)}_{(\mp)b}( - k^{\perp}) = \nonumber \\  &=&
 \Lambda^{(\eta)}_{(\pm)a}(k^{\perp}) [ \int { d^{3}q^{\perp} \over
 (2\pi)^{3} } \underline{V} (k^{\perp}-q^{\perp})
 \psi_{ab}(q^{\perp}) ] {\Lambda}^{(\eta)}_{(\mp)b}( - k^{\perp}) .
 \end{eqnarray}

 All these equations ~(\ref{bs1}) and ~(\ref{bs2})
 have been derived without any
 assumption about the smallness of the relative momentum $ \vert
 k^{\perp} \vert $ and for an arbitrary total momentum
 $$ {\cal
 P}_{\mu} = ( \sqrt { M_{A}^{2} + {\vec {\cal P}}^{2} } , {\vec
 {\cal P}} \neq 0 )~.
 $$
 We expand the function $\Psi$ on the projection operators
 \be  \label{bs4}
 \Psi=\Psi_++\Psi_-,~~~~~\Psi_{\pm}=
 \Lambda^{(\eta)}_{\pm}\Psi
 \Lambda^{(\eta)}_{\mp}~.
 \ee
 According to Eq.~(\ref{bs3}), $\Psi$ satisfies the identities
 \be  \label{bs5}
 \Lambda^{(\eta)}_{+}\Psi\Lambda^{(\eta)}_{+}=
 \Lambda^{(\eta)}_{-}\Psi\Lambda^{(\eta)}_{-}\equiv 0~,
 \ee
 which permit the determination of an unambiguous expansion of $\Psi$
 in terms of the Lorentz structures:
 \be \label{bs6}
  \Psi_{\pm(a,b)}=S^{-1}_{(a)}\left\{ \gamma_5 L_{\pm(a,b)}(q^{\bot})+
  (\gamma_{\mu}-\eta_{\mu}\not \eta)N^{\mu}_{\pm(a,b)} \right\}
 \Lambda^{(\eta)}_{\mp}(0)S^{-1}_{(b)}~,
 \ee
 where $L_{\pm}=L_1\pm L_2$, $N_{\pm}=N_1\pm N_2$. In the rest
 frame $\eta_{\mu}=(1,0,0,0)$ we get
 $$
 N^{\mu}=(0,N^i)~;
~~~~~~N^i(q)=\sum\limits_{a=1,2 }^{ }N_{\alpha}(q)e^i_{\alpha}(q)+
 \Sigma(q) \hat q^i~.
 $$
 The wave functions $L,N^{\alpha},\Sigma$  satisfy equations

\begin{center}
\underline {1. Pseudoscalar particles.}
\end{center}

\begin{eqnarray}\label{pse}
 M  \stackrel{0}{L}_{2} &=&
E  \stackrel{0}{L}_{1}
+
\int \frac{d {\bf q}}{(2\pi)^{3}}
V({\bf p}-{\bf q})
(  {\tt c}^{-}_{p} {\tt c}^{-}_{q}
+ \xi {\tt s}^{-}_{p} {\tt s}^{-}_{q})
\stackrel{0}{L}_{1} \,\,\, ;
\nonumber \\  \\
M  \stackrel{0}{L}_{1} &=&
E \stackrel{0}{L}_{2}
+
\int \frac{d {\bf q}}{(2\pi)^{3}}
V({\bf p}-{\bf q})
(  {\tt c}^{+}_{p} {\tt c}^{+}_{q}
+ \xi {\tt s}^{+}_{p} {\tt s}^{+}_{q})
\stackrel{0}{L}_{2}      \,\,\, .
\nonumber
\end{eqnarray}
\begin{eqnarray*}
\xi = {\hat p}_{i} \cdot {\hat q}_{i} \,\,\, .
\end{eqnarray*}

\begin{center}
\underline {2.Vector particles.}
\end{center}

\begin{eqnarray} \label{vecpar}
M \stackrel{0}{N}_{2}{}^{\alpha} &=&
E \stackrel{0}{N}_{1}{}^{\alpha}  + \nonumber \\
&+&
\int \frac{d {\bf q}}{(2\pi)^{3}}
V({\bf p}-{\bf q})
\{
(  {\tt c}^{-}_{p} {\tt c}^{-}_{q}
{\underline \delta}^{\alpha\beta}
+ {\tt s}^{-}_{p} {\tt s}^{-}_{q}
( {\underline \delta}^{\alpha\beta} \xi -
\eta^{\alpha}{\underline \eta}^{\beta} ) )
\stackrel{0}{N}_{1}^{\beta}
+
(\eta^{\alpha} {\tt c}^{-}_{p} {\tt c}^{+}_{q})
\stackrel{0}{\Sigma}_{1} \}   \,\,\,  ; \nonumber \\  \\
M \stackrel{0}{N}_{1}{}^{\alpha} &=&
E \stackrel{0}{N}_{2}{}^{\alpha}  +  \nonumber \\
&+&
\int \frac{d {\bf q}}{(2\pi)^{3}}
V({\bf p}-{\bf q})
\{
(  {\tt c}^{+}_{p} {\tt c}^{+}_{q}
{\underline \delta}^{\alpha\beta}
+ {\tt s}^{+}_{p} {\tt s}^{+}_{q}
( {\underline \delta}^{\alpha\beta} \xi -
\eta^{\alpha}{\underline \eta}^{\beta} ) )
\stackrel{0}{N}_{2}^{\beta}
+
(\eta^{\alpha} {\tt c}^{+}_{p} {\tt c}^{-}_{q})
\stackrel{0}{\Sigma}_{2}  \} \,\,\, .
\nonumber
\end{eqnarray}
\begin{eqnarray*}
\eta^{\alpha} = {\hat q}_{i} {\hat e}^{\alpha}_{i}(p) \,\,\, , \,\,\,
{\underline \eta}^{\alpha} = {\hat p}_{i} {\hat e}^{\alpha}_{i}(q) \,\,\, , \,\,\,
{\underline \delta}^{\alpha\beta}
= {\hat e}^{\alpha}_{i}(q){\hat e}^{\beta}_{i}(p) \,\,\, .
\end{eqnarray*}

\begin{center}
\underline {3. Scalar particles.}
\end{center}

\begin{eqnarray}\label{sca}
M \stackrel{0}{\Sigma}_{2} & = &
E \stackrel{0}{\Sigma}_{1}  +  \nonumber \\
&+&
\int \frac{d {\bf q}}{(2\pi)^{3}}
V({\bf p}-{\bf q})
\{
( \xi {\tt c}^{+}_{p} {\tt c}^{+}_{q} + {\tt s}^{+}_{p} {\tt s}^{+}_{q})
\stackrel{0}{\Sigma}_{1}
+
({\underline \eta}^{\beta}
{\tt c}^{-}_{p} {\tt c}^{+}_{q}) \stackrel{0}{N}_{1}{}^{\beta}  \} \,\,\, ; \nonumber \\  \\
M \stackrel{0}{\Sigma}_{1} & = &
E \stackrel{0}{\Sigma}_{2}   + \nonumber \\
&+&
\int \frac{d {\bf q}}{(2\pi)^{3}}
V({\bf p}-{\bf q})
\{
( \xi {\tt c}^{-}_{p} {\tt c}^{-}_{q} + {\tt s}^{-}_{p} {\tt s}^{-}_{q})
\stackrel{0}{\Sigma}_{2}
+                          ( {\underline \eta}^{\beta}
{\tt c}^{+}_{p} {\tt c}^{-}_{q}) \stackrel{0}{N_{2}}^{\beta} \} \,\,\, .  \nonumber
\end{eqnarray}

Here, in all equations,
\be
{\tt c}^{\pm}(p)=\cos[v_a(p) \pm v_b(p)]~,~~~~~~~~
{\tt s}^{\pm}(p)=\sin[v_a(p) \pm v_b(p)]
\ee
$v_a, v_b$ are the Foldy-Wouthuysen matrices of particles (a,b),
and $E=E^a+E^b$ is the sum of one-particle energies.

The normalization of these solutions is uniquely determined
by equation~(\ref{nakanishi})
\be\label{nakap}
\frac{2N_c}{M_L}\sum\limits_{ }^{ } \frac{d^3q}{(2\pi)^3}\left\{
L_1(q)L_2^+(q)+L_2(q)L_1^+(q)\right\}=1~,
\ee
\be\label{nakav}
\frac{2N_c}{M_N}\sum\limits_{ }^{ } \frac{d^3q}{(2\pi)^3}\left\{
N^{\mu}_1(q)N^{\mu+}_2(q)+N^{\mu}_2(q)N^{\mu+}_1(q)\right\}=1~,
\ee
\be\label{nakas}
\frac{2N_c}{M_{\Sigma}}\sum\limits_{ }^{ } \frac{d^3q}{(2\pi)^3}\left\{
\Sigma_1(q)\Sigma_2^+(q)+\Sigma_2(q)\Sigma_1^+(q)\right\}=1~.
\ee
 The description of instantaneous relativistic bound states
 in a hot and dense medium can be found in~\cite{bkm,puz}.

 \subsection{Schr\"odinger equation }

 If the atom is at rest ( $ {\cal P}_{\mu} = ( M_{A},0,0,0 ) $ )
 equation (\ref{bs2}) coincides with the Salpeter equation \cite{a18}. If
 one assumes that the current mass $ m^{0} $ is much larger than
 the relative momentum $ \vert q^{\perp} \vert $ then the coupled
 equations ~(\ref{bs1}) and ~(\ref{bs2}) turn into the Schr\"o\-dinger
 equation. In the rest frame ( $ {\cal P}_{0} = M_{A} $ ) equations
 ~(\ref{sd1}) and~(\ref{sd2})  for a large mass  $ ( m^{0} / \vert q^{\perp} \vert
 \rightarrow \infty ) $ describe a nonrelativistic particle
 \begin{eqnarray*}
 E_{a}( {\bf k} ) = \sqrt { ( m_{a}^{0})^{2} + {\bf k}^{2} } \simeq
 m^{0}_{a} + {1 \over 2} { {\bf k}^{2} \over m^{0}_{a} }, \\
 \tan 2 \upsilon = { k \over m^{0} } \rightarrow 0 ; \,\, S({\bf
 k}) \simeq 1 ; \,\, \Lambda_{ (\pm)} \simeq { {1 \pm \gamma_{0} }
 \over 2 } .
 \end{eqnarray*}
 Then, in equation (\ref{bs2}) only the state with positive energy remains
 \begin{eqnarray*}
  \psi  \simeq \psi_{(+)}= \Lambda_{(+)}\gamma_5\sqrt{4\mu}\psi_{Sch}, \,\,\,
 \Lambda_{(-)} \psi \Lambda_{(+)} \simeq 0 ,
 \end{eqnarray*}
 where $ \mu = m_{a} \cdot m_{b} / ( m_{a}+m_{b}) $.
 And finally the Schr\"odinger equation results in
 \begin{eqnarray} \label{dinge}
 [ {1 \over 2\mu} {\bf k}^{-2} + ( m^{0}_{a} + m^{0}_{b} - M_{A} )
 ] \psi_{Sch}({\bf k}) = \int { d{\bf q} \over (2\pi)^{3} }
 \underline{V} ({\bf k}-{\bf q}) \psi_{Sch}({\bf q}),
 \end{eqnarray}
 with the normalization $\int d^3q|\psi_{Sch}|^2/(2\pi)^3=1$.

 For an arbitrary total momentum $ {\cal P}_{\mu} $ equation~(\ref{dinge})
 takes the form
 \begin{eqnarray} \label{dinger}
 [ - {1 \over 2\mu}
 (k_{\nu}^{\perp})^{-2} + ( m^{0}_{a} + m^{0}_{b} - \sqrt{ {\cal
 P}^{2}}  ) ] \psi_{Sch}( k^{\perp})  = \int { d^{3} q^{\perp}
 \over (2\pi)^{3} } \underline{V} ( k^{\perp}- q^{\perp})
 \psi_{Sch}( q^{\perp}),
 \end{eqnarray}
 and describes a
 relativistic atom with nonrelativistic relative momentum $ \vert
 k^{\perp} \vert << m^{0}_{a,b} $. In the framework of such a
 derivation of the Schr\"odinger equation it is sufficient to
 define the total coordinate as  $ X=(x+y)/2$,
 independently of the magnitude of the masses of the two particles
 forming an atom.

 In particular, the Coulomb interaction leads to a positronium
 at rest with the bilocal wave function
 \be\label{posi}
 \Phi_P^{\alpha\beta}(t|\vec z)=\eta_P(t)\left(\frac{1+\gamma_0}{2}
 \gamma_5\right)^{\alpha\beta}\underline{\psi}_{Sch}(\vec z)
 \sqrt{\frac {m_e} 2}~,~~~~
 \underline{\psi}_{Sch}(\vec z)=
 \int\limits_{ }^{ }\frac{d^3p}{(2\pi)^3}e^{(i\vec p \vec z)} \psi_{Sch}(\vec p);
 \ee
 where $\underline{\psi}_{sch}(\vec z)$ is the Schr\"odinger
 normalizable wave function of the relative motion
 \be \label{relative}
 \left(-\frac 1{m_e}\frac {d^2}{d{\vec z}^2}
 -\frac {\alpha}{|\vec z|}\right)\underline{\psi}_{Sch}(\vec z)=
 \epsilon\underline{\psi}_{Sch}(\vec z);
  ~~~~~~~~~~~\left(~\int d^3z{\parallel\underline{\psi}_{Sch}(z)\parallel}^2=1\right)
 \ee
 $M_P=(2m_e-\epsilon)$ is the mass of a positronium,
 $\frac{1+\gamma_0} 2$ is the
 projection operator on the state with positive energies of an
 electron and positron.
 We have chosen the total-motion variable $\eta_P(t)$
 in ~(\ref{posi}) so that
 the effective action for the
 total motion of the positronium with anomaly term have the form
 similar to the $ \eta_0 $ -meson ones~\cite{2}
 \be \label{s}
 W_{eff}=\int dt \left\{\frac 1 2 \left({\dot\eta_P}^2-M_P^2{\eta_P}^2\right)
 V_{(3)} +
 C_P\eta_P \dot X[A] \right\}~,
 \ee
 where
 \be \label{t}
 C_P=\frac{\sqrt{2}}{m_e}8{\pi}^2\left(\frac{\underline{\psi}_{Sch}(0)}{m_e^{3/2}}\right)~,
 \ee
 and
 \be\label{wnqed}
 \frac d{dt}X[A]=\int d^3x F_{\mu\nu}{}^*F^{\mu\nu~,}
 ~~~~~~~~~~X[A]=\frac{e^2}{16\pi}\int d^3x(\epsilon_{ijk}A_i\pr_jA_k)
 \ee
 is the "winding number" functional (i.e., anomalous term) that
   describes the two $\gamma$ decay of  a positronium.

 \subsection{Spontaneous chiral symmetry breaking }

 The solution of the set of SD and Salpeter
 equations~(\ref{sd1}),~(\ref{sd2}),~(\ref{bs2})
 was considered in the numerous papers~\cite{fin,a3,a4} (see also
 review~\cite{puz}) for different potentials.
 One of the main results of these paper was the pure
 quantum effect of spontaneous
 chiral symmetry breaking for non-Coulomb potentials.
 The instantaneous interaction, in this case, leads to
 rearrangement of the perturbation series and strongly changes
 the spectrum of elementary excitations and bound states in contrast
 to the naive perturbation theory.

 To demonstrate this effect
 and estimate possibility of the considered
 relativistic equations, we consider
 the opposite case of massless particles, $ m^{0}_{a}=m^{0}_{b}
 \rightarrow 0 $ for an arbitrary potential. Suppose that in this case
 equations~(\ref{sd1}),~(\ref{sd2})
 \begin{eqnarray} \label{sdg1}
 2E_{a}(k^{\perp}) \cos 2 {\upsilon}(k^{\perp}) = \int {
 d^{3}q^{\perp}\over (2\pi)^{3} }
 \underline{V}(k^{\perp}-q^{\perp}) \cos 2{\upsilon}(q^{\perp})
 \end{eqnarray}
 \begin{eqnarray} \label{sdg2}
 2E_{a}(k^{\perp}) \sin 2 {\upsilon}(k^{\perp}) = \vert k^{\perp}
 \vert + \int { d^{3}q^{\perp}\over (2\pi)^{3} }
 \underline{V}(k^{\perp}-q^{\perp}) \vert \hat{k}^{\perp} \cdot
 \hat{q}^{\perp} \vert \sin 2{\upsilon}(q^{\perp})
 \end{eqnarray}
 have a nontrivial solution $ \upsilon(k^{\perp}) \neq 0 $. This
 solution describes the spontaneous breakdown of chiral symmetry
  \cite{yaf,a6,puz,fin,a3,a4}.

 It can easily be seen that equations ~(\ref{sdg1}) and~(\ref{sdg2})
 are identical with~(\ref{bs2})
 for the bound state wave function with zero eigenvalue, ${\cal
 P}_{\mu}^{2} = 0 $ and
 \begin{eqnarray}  \label{bsg2}
 \Lambda_{(+)} \psi \bar{\Lambda}_{(-)} & = &
 \Lambda_{(-)} \psi \bar{\Lambda}_{(+)} \equiv \psi \nonumber \\
 2E_{a}(k^{\perp}) \psi (k^{\perp}) &=& \int { d^{3}q^{\perp}\over
 (2\pi)^{3} } \underline{V}(k^{\perp}-q^{\perp}) \psi (q^{\perp}).
 \end{eqnarray}
 Therefore,
 \begin{eqnarray}
 \psi = \cos 2 \upsilon (k^{\perp}) / F
 \end{eqnarray}
 where $ F$ is a proportionality constant determined by the
 normalization~(\ref{nakap})~\cite{yaf}
 \be \label{1fpi}
 F=\frac{4 N_c}{M_H} \int\limits_{ }^{ } \frac{d^3q}{(2\pi)^3} L_2 \cos(2v(q))
 \ee
 In this way, the coupled equations~(\ref{sd1}),~(\ref{sd2}), and~(\ref{bs2})
 contain the Goldstone mode that accompanies
 spontaneous breakdown of chiral symmetry. Thus, in  the
 framework of instantaneous action  we get the proof of the
 Goldstone theorem in the bilocal variant.

 Just this example represents a model for the construction of a low
 - energy theory of light mesons, in which the pion is considered
 in two different ways, as a quark - antiquark bound state and as a
 Goldstone particle. So it turns out that our relativistic
 instantaneous model for bound states can, in the lowest order of
 radiative corrections, also describes mesons.
 It was shown that the spontaneous symmetry breaking
 is absent for the pure Coulomb potential in QED~\cite{a3}.
 The spontaneous symmetry breaking
 and the Goldstone meson in QCD are realized for
 the potential of hadronization.
 What is an origin of this potential in QCD?

 Recall that in QED the Coulomb potential is
 the consequence of the resolving the Gauss law constraint.
 This Coulomb potential is lost in any relativistic invariant
 gauge of sources in FP representation of the path integral.
 The defect of the pure relativistic artificial gauges
 is problem of a tachion and other non-physical states in
 the bound state spectrum~\cite{nakanishi}.
 The incorporation of the simultaneity for these gauges
 by the quasipotential approach~\cite{kad} could not
 describe both
 the spontaneous chiral symmetry breaking with the Goldstone meson
 and S-matrix elements of interactions of some instantaneous bound
 states (papers on this topic are absent in literature).

 \subsection{\bf The relativistic equation for multiparticle systems.}

  To derive  the relativistic covariant equations for many - particle,
 we use the operator approach \cite{fb1,fin,a3} with
 the Hamiltonian given by
 \begin{eqnarray} \label{38-1}
 {\cal H} =
 \int d{\bf x}
 \bar{\psi} (i \partial_{i} \gamma_{i} + m^{0} )\psi + {1 \over 2}
 \int d{\bf x} d{\bf y}
 ( \psi^{+}_{i} ( {\bf x} )
 \psi_{j} ( {\bf x} ) )
 V( {\bf x} - {\bf y} )
 ( \psi^{+}_{k} ( {\bf y} )
 \psi_{l} ( {\bf y} ) )  ~.
 \end{eqnarray}
 The first step for constructing the physical states consists in the definition
 of the one - quasi - particle creation ( $ a^{+},b^{+} $ ) and annihilation
 ($a,b$) operators with the help of the Bogoliubov fermion expansion \cite{a9}
 \begin{eqnarray}  \label{38-2}
 \psi_{\alpha}( {\bf x} ) =
 \sum_{s} \int { d{\bf q} \over (2\pi)^{3/2} } e^{i {\bf q}{\bf x} }
 [ a_{s}( {\bf q})
 \mu_{\alpha}( {\bf q},s) +
 b^{+}_{s}( - {\bf q})
 \nu_{\alpha}( - {\bf q},s) ]~ .
 \end{eqnarray}
 Here $
 \mu_{\alpha}( {\bf q},s) $ and $
 \nu_{\alpha}( - {\bf q},s) $
 are the coefficients determined from the Schr\"odinger equation for
 the one - particles energy
 \begin{eqnarray} \label{38-3}
 < a_{s}({\bf q}) \vert \hat{H} \vert a^{+}_{s} ( {\bf q}^{\prime} ) > =
 E({\bf q})
 < 0 \vert a_{s}({\bf q})  a^{+}_{s} ( {\bf q}^{\prime} ) \vert 0 > .
 \end{eqnarray}
 They can be represented via the Foldy - Wouthuysen matrix~(\ref{sd3})  as
 \begin{eqnarray*}
 \mu_{\alpha}( {\bf q},s) =
 S( {\bf q})_{\alpha \beta}
 \mu_{\beta}( 0,s) ; \,\,\,
 \nu_{\alpha}( - {\bf q},s) =
 S( - {\bf q})_{\alpha \beta}
 \nu_{\beta}( 0 ,s)
 \end{eqnarray*}
 with
 \begin{eqnarray*}
 S_{\alpha \alpha^{\prime}}( {\bf q})
 [ \sum_{s}
 \mu_{\alpha^{\prime}}( 0,s)
 \mu^{+}_{\beta^{\prime}}( 0,s) ]
 S^{-1}_{\beta^{\prime} \beta} ( {\bf q}) =
 { (S { 1+\gamma_{0} \over 2} S^{-1})}_{\alpha \beta}  \equiv {( \Lambda^{0}_{+}({\bf q}))}_{\alpha \beta}, \\
 S_{\alpha \alpha^{\prime}}( - {\bf q})
 [ \sum_{s}
 \nu_{\alpha^{\prime}}( 0,s)
 \nu^{+}_{\beta^{\prime}}( 0,s) ]
 S^{-1}_{\beta^{\prime} \beta} ( - {\bf q}) =
 { (S { 1- \gamma_{0} \over 2} S^{-1})}_{\alpha \beta}
 \equiv {( \Lambda^{0}_{-}( - {\bf q}))}_{\alpha \beta} .
 \end{eqnarray*}
 $ \Lambda_{+}^{0} $ and $ \Lambda _{-}^{0} $ are projection operators on
 states with positive, resp., negative energy. Then, equation~(\ref{38-3})
 takes the form
 of the Schwinger - Dyson equation~(\ref{sd1}),~(\ref{sd2})
 which can compactly be written as
 \begin{eqnarray} \label{38-4}
 E(p) S^{-2}({\bf p}) = m^{0} + p_{i}\gamma_{i} +{1\over2}
 \int { d{\bf q} \over (2 \pi)^{3} } \underline {V} ( {\bf p} - {\bf q} )
  S^{-2}({\bf q}) ~,
 \end{eqnarray}

 After inserting~(\ref{38-2}) into~(\ref{38-1}) the Hamiltonian can be given in the following
 manner :

 \begin{eqnarray} \label{38-5}
 {\cal H} &=& E_{0} + H_{1} + :H_{4}:, \nonumber \\
 E_{0} & = & <0\vert {\cal H} \vert 0>, \nonumber \\
 H_{1} & = & \sum_{(1)} E({\bf p}_{1}) ( a_{1}^{+} a_{1} + b_{1}^{+} b_{1} ) , \\
 :H_{4}: &=& {2\over3} \sum_{1,2,3,4,} \delta^{(4)}(p_{1}-p_{2}+p_{3}-p_{4})
 \underline{V} ({\bf p}_{1}-{\bf p}_{3}) \nonumber \\
 & \{ &
 a_{1}^{+} b_{ \hat {2} }^{+} a_{3}^{+} b_{ \hat {4} }^{+}
 \mu_{1}^{*} \nu_{ \hat {2} }^{*} \mu_{3}^{*} \nu_{ \hat {4} }^{*} +
 a_{1}^{+} b_{ \hat {2} }^{+} b_{ \hat {3} } a_{1}
 \mu_{1}^{*} \nu_{ \hat {2} } \nu_{ \hat {3} }^{*} \mu_{4} +  \nonumber \\ &+&
 b_{ \hat {1} } a_{2} a_{3}^{+} b_{ \hat {4} }^{+}
 \nu_{ \hat {1} }^{*} \mu_{2} \mu_{3}^{*} \nu_{ \hat {4} } +
 b_{ \hat {1} } a_{2} b_{ \hat {3} } a_{4}
 \nu_{ \hat {1} }^{*} \mu_{2} \nu_{ \hat {3} }^{*} \mu_{4} + \,\,\, ... \} + \,\,\, ... \nonumber
 \end{eqnarray}
 The following abbreviations have been
 used in~(\ref{38-5}):
 \begin{eqnarray*}
 \sum_{I} = \sum_{s_{I}} \int { d{\bf p}_{I} \over (2\pi)^{3/2} } , \,
 \{ I \} = \{ p_{I} , s_{I} \} , \,
 \{ \hat {I} \} = \{ - p_{I} , - s_{I} \} , \, I=1,2,3,4 .
 \end{eqnarray*}

 For diagonalizing the Hamiltonian~(\ref{38-5})
 with respect to pair correlations
 ($ a^{+}_{1} b^{+}_{ \hat {2} }$), ($ b_{ \hat {3} }a_{4}$)
 one defines a new vacuum as the coherent
 state
 \begin{eqnarray}  \label{38-6}
 \vert 0 >>_{\alpha} = \exp \{ \sum_{1,2,3,4} \alpha(1, \hat {2} , \hat {3} ,4)
 [ ( a_{1}^{+i_{1}} b_{ \hat {2} }^{+i_{1}} )( b_{ \hat {3} }^{+j_{1}} a_{4}^{+j_{1}}) ] \}
 \vert 0 >
 \end{eqnarray}
 and the creation operator for the bound state (of pair correlation)
 \begin{eqnarray} \label{38-7}
 B^{+}(n) = \sum_{1,2} \delta({\bf p}_{1}-{\bf p}_{2})
 [ X_{+}(1, \hat {2} ) a^{+i}(1) b^{+i}( \hat {2} ) - X_{-}(
 \hat {1} ,2) b^{j}( \hat {1} ) a^{j}(1) ].
 \end{eqnarray}
 The coefficient $X_{+}$ and $ X_{-}$ are determined from the Schr{\"o}dinger
 equation for the two - particle energy $ M_{B} $ ,
 \begin{eqnarray}  \label{38-8}
 { }_{\alpha}<< 0 \vert B(n) (H_{1}+H_{4}) B^{+}(n) \vert 0 >>_{\alpha} =
 M_{B} \,\, { }_{\alpha} << 0 \vert B(n) B^{+}(n) \vert 0 >>_{\alpha} ,
 \end{eqnarray}
 and the parameter $ \alpha $ in~(\ref{38-8}) is given with the help of the
 definition of the annihilation operator
 $ B(n) $ for the pair correlation
 \begin{eqnarray}  \label{38-9}
 B(n) \vert 0 >>_{\alpha} = 0~ .
 \end{eqnarray}
 Equation~(\ref{38-8}) coincides with equation~(\ref{bs2})
 in the rest frame (the Salpeter
 equation) for the meson spectrum
 \begin{eqnarray}  \label{38-10}
 (E_{1}({\bf p})+E_{2}({\bf p}) \mp M_{B} ) \psi_{\pm\pm}({\bf p}) =
 \Lambda_{\pm}({\bf p})
 [ \hat{I}_{ {\bf p}{\bf q} }\times (
 \psi_{++}({\bf q}) +
 \psi_{--}({\bf q}) )]
 \Lambda_{\pm}({ - \bf p})
 \end{eqnarray}
 up to the notation
 \begin{eqnarray}  \label{38-11}
 & & \psi = \psi_{++} + \psi_{- - } ; \,\,\, \psi_{\pm\pm}
 = \Lambda_{\pm} \psi \Lambda_{\mp} ; \nonumber \\
 & & \psi_{++}({\bf p})_{\alpha\beta} = \sum_{s_{1},s_{2}}
 X_{+} ( {\bf p} , {\bf p} , s_{1},s_{2} ) \mu_{\alpha}^{+} ({\bf p} ,s_{1})
 \nu_{\beta}({\bf p} ,s_{2}) ,  \\
 & & \psi_{- -}({\bf p})_{\alpha\beta} = \sum_{s_{1},s_{2}}
 X_{-} ( {\bf p} , {\bf p} , s_{1},s_{2} ) \nu_{\alpha}^{+} ({\bf p} ,s_{1})
 \mu_{\beta}({\bf p} ,s_{2}) ,    \nonumber  \\
 & & {\hat I}_{ {\bf p}{\bf q} } \times \psi(q) = \int\frac{d^3q}{(2\pi)^3}
 V({\bf p-q})\psi({\bf q})
 \nonumber
 \end{eqnarray}
 The one - particle energies
 $ E_{1}({\bf p}) $,
 $ E_{2}({\bf p}) $ in~(\ref{38-11}) are defined via the
 Schwinger - Dyson equation~(\ref{38-4}).

 Notice that equations of the type~(\ref{38-4}),~(\ref{38-11})
 are well - known from the
 nonrelativistic many - body theory (Landau's theory of fermi liquids
 \cite{fb24}, Random Phase Approximation \cite{fb25}) and play an essential
 role in the description of elementary excitation in atomic nuclei \cite{fb26}.
 Their relativistic analogies describing the Goldstone pion
 and the constituent masses of the light quarks are
 equations~(\ref{sd1}),~(\ref{sd2}),~(\ref{bs2}).

 Thus the Green function method discussed in sects.3.2.-3.5.,
 and the operator approach lead
 to one and the same equations and complement each other. The first allows
 one to make
 easily the relativistic generalization and to construct the effective
 bound state
 interaction Lagrangian, whereas the second yields an adequate interpretation
 of quantum states and enables one to describe more complicated system,
 (in QCD,  baryons and other many - quark states \cite{fb1}).

 Let us construct by means of the quasiparticle operator method
 the relativistic equation for a three - particle system.
 In the "coherent" vacuum~(\ref{38-7}) the
  creation operator of a three - particle system consists not only
 of creation operators for particles ($ a^{+}$) but also of annihilation
 operators for antiparticles ($b$) with the same quantum numbers
 \begin{eqnarray} \label{38-12}
 \underline{B}^{+} &=& \sum_{1,2,3} \delta({\bf p}_{1}+{\bf p}_{2}+{\bf p}_{3} )
 [
 X_{+++}(1,2,3) a^{i(+)} (1) a^{j(+)} (2) a^{l(+)} (3) + \nonumber \\ &+&
 X_{--+}(1,2,3) b^{i(+)}(1) b^{i(+)}(2) b^{i(+)} (3) +
 interchange \,\,\, of \,\,\, (1,2,3) ]  \epsilon^{ijk} .
 \end{eqnarray}

 The "baryon" functions are as follows:
 \begin{eqnarray*}
 \psi_{+++}(1,2,3)_{\alpha\beta\gamma} &=&
 \sum_{s_{1}s_{2}s_{3}} \mu^{+}_{\alpha}(1) \mu^{+}_{\beta}(2) \mu^{+}_{\gamma}(3)
 X_{+++}(1,2,3),  \\
 \psi_{--+}(1,2,3)_{\alpha\beta\gamma} &=&
 \sum_{s_{1}s_{2}s_{3}} \nu^{+}_{\alpha}(1) \nu^{+}_{\beta}(2) \nu^{+}_{\gamma}(3)
 X_{--+}(1,2,3),
 \end{eqnarray*}
 etc. \\
 Then, the eigenvalue equation for the Hamiltonian operator
 \begin{eqnarray}   \label{38-13}
 { }_{\alpha}<< 0 \vert \underline{B} H \underline{B}^{+} \vert 0 >>_{\alpha},
 \end{eqnarray}
 is equivalent to the following system for the "baryons" wave functions \\
 $ \psi_{+++}, \psi_{--+}, \psi_{-+-},  \psi_{+--}. $
 \begin{eqnarray} \label{38-14}
 &[&
 \left( \begin{array}{llll} +\\+\\+\\- \end{array} \right) E(1)
 \left( \begin{array}{llll} +\\+\\-\\+ \end{array} \right) E(2)
 \left( \begin{array}{llll} +\\-\\+\\+ \end{array} \right) E(3)
 \left( \begin{array}{llll} -\\+\\+\\+ \end{array} \right) M_{B} ]
 \psi _{ \left( \begin{array} {llll} +++\\--+\\-+-\\+-- \end{array} \right)}
 (1,2,3) = \nonumber \\
 &=& {2\over3}
 \Lambda_{ \left( \begin{array}{llll} +\\-\\-\\+ \end{array} \right)} (1)
 \Lambda_{ \left( \begin{array}{llll} +\\-\\+\\- \end{array} \right)} (2)
 \Lambda_{ \left( \begin{array}{llll} +\\+\\-\\- \end{array} \right)} (3)
 \nonumber \\
 & \{ &
 \hat{I}_{ \underline{1},\underline{2} } [
 \psi _{ \left( \begin{array}{llll} +++\\--+\\-+-\\+-- \end{array} \right)}
 (\underline{1},\underline{2},3) +
 \psi _{ \left( \begin{array}{llll} --+\\+++\\+--\\-+- \end{array} \right)}
 (\underline{1},\underline{2},3)  ] + \nonumber \\
 &+&
 \hat{I}_{ \underline{2},\underline{3} } [
 \psi _{ \left( \begin{array}{llll} +++\\--+\\-+-\\+-- \end{array} \right)}
 (1,\underline{2},\underline{3}) +
 \psi _{ \left( \begin{array}{llll} +--\\-+-\\--+\\+++ \end{array} \right)}
 (1,\underline{2},\underline{3})  ] + \\
 &+&
 \hat{I}_{ \underline{1},\underline{3} } [
 \psi _{ \left( \begin{array}{llll} +++\\--+\\-+-\\+-- \end{array} \right)}
 (\underline{1},2,\underline{3}) +
 \psi _{ \left( \begin{array}{llll} -+-\\+--\\+++\\--+ \end{array} \right)}
 (\underline{1},2,\underline{3})  ]  \} \nonumber
 \end{eqnarray}
 where
 \begin{eqnarray} \label{38-15}
 I_{
 \underline{1},
 \underline{2}}
 \psi(\underline{1},\underline{2},3) &=&
 \int { d{\bf q} \over (2\pi)^{3} }
 \underline{V}({\bf q}) \psi( {\bf p}_{1} - {\bf q},{\bf p}_{2} + {\bf q},{\bf p}_{3} )
 \nonumber \\
 {\bf p}_{1} &+& {\bf p}_{2} \, + \, {\bf p}_{3} = 0 .
 \end{eqnarray}
 Equation~(\ref{38-15}) is the analogue of the Salpeter equation~(\ref{38-10})
 for a bound state consisting of three particles.
 In the same notations eq.~(\ref{38-11}) has the form
 \begin{eqnarray*}
 &[& \left( \begin{array}{ll} +\\+ \end{array} \right)E_{1}(1)
 \left( \begin{array}{ll} +\\+ \end{array} \right) E_{2}( \hat {2} )
 \left( \begin{array}{ll} -\\+ \end{array} \right) M_{B} ]
 {\bar {\psi} }_{ \pm \pm } (1, \hat{2} ) = \\  \nonumber
 & = &  {4 \over 3} \Lambda_{\pm}(1) \{ {\hat {I}}_{
 \underline{1}, \hat{\underline {2} } }
 [ {\psi}_{++} (\underline{1}, \hat{\underline{2}} ) +
  {\psi}_{--} (\underline{1}, \hat{\underline{2}} ) ] \}
 \Lambda_{\pm}( \hat{2} )
 \end{eqnarray*}
 with the condition $ p_{1} = p_{2} = p $ and with the taking into account
 the identities
 \begin{eqnarray*}
 &\int& d{\bf q} \underline{V} ( {\bf p} -{\bf q} ) \psi ({\bf q})= \\ \nonumber
 = &\int& d{\bf q} \underline{V} ( {\bf q} ) \psi ({\bf q} + {\bf p} )= \\ \nonumber
 = &\int& d{\bf q} \underline{V} ( {\bf q} )
 {\psi} ( {\bf p}_{1} + {\bf q} , - {\bf p}_{2} - {\bf q} )
 \vert_ { p_{1} = p_{2} = p } \\  \nonumber
 & {\psi} &(p_{1}, -p_{2}) = \psi (1, \hat{2})
 \end{eqnarray*}
 The nonrelativistic
 reduction \cite{a18} from the Salpeter eq.~(\ref{bs2})
 to the Schr\" odinger equation,
 \begin{eqnarray*} \label{38-16}
 E_{a}({\bf p}) &\simeq& \sqrt { m_{a}^{2} + {\bf p}^{2} } \simeq
 m_{a} + {1\over2}{ {\bf p}^{2} \over m_{a} }, \\
 S_{a}({\bf p}) &\simeq&  1, \,\,\,
 \psi_{+++} \equiv \psi >> \psi_{ \left( \begin{array}{lll} +--\\-+-\\--+ \end{array}
 \right) },
 \end{eqnarray*}
 leads in our case to the well - known nonrelativistic equation for
 the wave function of three particle bound states
 \begin{eqnarray}
 &[&
 { {\bf p}_{1}^{2} \over 2m_{1} } +
 { {\bf p}_{2}^{2} \over 2m_{2} } +
 { {\bf p}_{3}^{2} \over 2m_{3} } -
 (M_{\underline{B}} - m_{1} -m_{2} - m_{3} ) ]
 \psi({\bf p}_{1},{\bf p}_{2},{\bf p}_{3} ) = \nonumber \\ &=& {2\over3} [
 \hat{I}_{ \underline{1},\underline{2} }
 \psi ( {\underline{\bf p}_{1}}, {\underline{\bf p}_{2}}, {\bf p}_{3} ) +
 \hat{I}_{ \underline{2},\underline{3} }
 \psi ( {\bf p}_{1}, {\underline{\bf p}_{2}}, {\underline{\bf p}_{3}} ) +
 \hat{I}_{ \underline{1},\underline{3} }
 \psi ( {\underline{\bf p}_{1}}, {\bf p}_{2}, {\underline{\bf p}_{3}} )  ] .
 \end{eqnarray}
 Here, the condition~(\ref{38-16}), which means the choice of the rest frame
 $ {\cal P}_{\mu} = ( M_{\underline{B}},0,0,0 ) $ , has to
 be fulfilled.

 Notice that the Jacobi coordinates, which allow to write the Hamiltonian
 in the term of two relative momenta, have sense only in the nonrelativistic
 limit.

 To describe the three-particle bound system
 in an arbitrary reference frame it is sufficient
 to substitute in~(\ref{38-14}) all relative momenta $ {\bf p}_{i} $ by the
 transversal ones, $ p_{\mu}^{\perp (i)}$, and the projection operators
 $ \Lambda_{\pm}({\bf p}) $ by the operators
 \begin{eqnarray*}
 \Lambda_{\pm} ({\bf p}^{\perp}) =
 S({\bf p}^{\perp})
 { {M_{\underline{B}} \pm \rlap/{\cal P}} \over 2M_{\underline{B}} }
 S({\bf p}^{\perp})^{-1} .
 \end{eqnarray*}
 In the same way one can generalize the equation~(\ref{38-14})
 and its relativization for an arbitrary
 $ N $- particle state.

 The method for constructing relativistic wave functions of many - quark
 system explained above unambiguously enables one to build from
 the nonrelativistic bound state wave function
 \begin{eqnarray*}
 {\chi}_{ \alpha_{1}, \alpha_{2}, \,\, ... \,\, \alpha_{N} } \cdot
 e^{i M X_{0}} \cdot \Phi_{
 \alpha_{1}, \alpha_{2}, \,\, ... \,\, \alpha_{N}
  }
 ({\bf {p}}^{1},{\bf {p}}^{2}, \,\, ... \,\, , {\bf {p}}^{N} ) , \,\,\,\,\,
 \sum_{i} {\bf {p}}(i) = 0
 \end{eqnarray*}
 relativistic wave functions for the same bound states with total
 momentum \\
 $ {\cal P}_{\mu} = ( \omega = \sqrt { {\cal {\bf P}}^{2} + M^{2} } ,
 {\cal {\bf P}} ) $,
 \begin{eqnarray*}
 {\chi}_{ \alpha_{1}, \alpha_{2}, \,\, ... \,\, \alpha_{N} } & \cdot &
  e^{i {\cal P} X } \cdot
 \Lambda_{+ \alpha_{1}^{ } \alpha_{1}^{\prime}} ( {p}^{(1) \perp }
 \Lambda_{+ \alpha_{2}^{ } \alpha_{2}^{\prime}} ( {p}^{(2) \perp } ) \,\, ... \,\,
 \Lambda_{+ \alpha_{N}^{ } \alpha_{N}^{\prime}} ( {p}^{(N) \perp } ) \cdot \\ &\cdot&
 \Phi_{ \alpha_{1}^{\prime}, \alpha_{2}^{\prime}, \,\, ... \,\,
 \alpha_{N}^{\prime} }
 ({p}^{(1) \perp},{p}^{(2) \perp}\,\, ... \,\, {p}^{(N) \perp}) , \\
 &(& \sum_{i} p_{\mu}^{(i) \perp} = 0 ).
 \end{eqnarray*}
 Here $ {\chi}_{ \alpha_{1}, \alpha_{2}, \,\, ... \,\, \alpha_{N} } $ is the
 matrix selecting one or another representation of the Lorentz group
 with a definite spin. ( A representation of the Poincare group that
 preserves
 the one - time dependence of wave functions see in ref. \cite{a29} ).

 \subsection{Relativistic covariant unitary S-matrix for bound states}

 The achievement of the relativistic covariant constraint-shell
 quantization of gauge theories is the description of both
 the spectrum of bound states and their S-matrix elements.

 It is convenient to write the matrix elements for the
 action~(\ref{3-9}),~(\ref{9}) in terms of the field operator
 $$
 \Phi^{\prime}(x,y)=\int d^4x_1 G_{\Sigma}(x-x_1){\cal M}'(x_1,y)
 =\Phi^{\prime}(z|X)
 $$
 Using the decomposition over the bound state quantum numbers $(H)$
 \be\label{set1}
 {\Phi}^{\prime}(z|X)=\sum\limits_H\int\frac{d^3\vec {\cal
 P}}{(2\pi)^{3/2}\sqrt{2\omega_H}}\int\frac{d^4q}{(2\pi)^4}
 \{ e^{i\vec {\cal P}\vec{X}} \Phi_H(q^{\bot}|{\cal P})a^+_H({\cal P})
 +e^{-i\vec {\cal P}\vec{X}} \bar{\Phi}_H(q^{\bot}|-
 {\cal P})a^-_H({\cal P})\}~,
 \ee
 where
 \be
 \Phi_{H(ab)}(q^{\bot}|{\cal P})=G_{\Sigma a}(q+{\cal P}/2)
 \Gamma_{H(ab)}(q^{\bot}|{\cal P})~,
 \ee
 we can write the matrix elements for the interaction $W^{(n)}$~(\ref{9})
 between the vacuum and the n-bound state
 \be \label{S-matrix}
 <H_1{\cal P}_1, ...,H_n{\cal P}_n|iW^{(n)}|0>=-i(2\pi)^4 \delta^4
 \left( \sum\limits_{i=1 }^{n }{\cal P}_i\right) \prod\limits_{j=1 }^{n }
 \left[\frac{1}{(2\pi)^32\omega_j}\right]^{1/2} M^{(n)}({\cal P}_1,...,{\cal P}_n)
 \ee
 $$
 M^{(n)}=\int \frac{id^4q}{(2\pi)^4 n} \sum\limits_{\{i_k\} }^{ }
 \Phi_{H_{i_{1}}}^{a_1,a_2}(q| {\cal P}_{i_1})
 \Phi_{H_{i_{2}}}^{a_2,a_3}(q-\frac{{\cal P}_{i_1}+{\cal P}_{i_2}}{2}| {\cal P}_{i_2})
 \Phi_{H_{i_{3}}}^{a_3,a_4}\left(q-\frac{2{\cal P}_{i_2}+
 {\cal P}_{i_1}+{\cal P}_{i_3}}{2}| {\cal P}_{i_3}\right)
 $$
 $$
 ...\Phi_{H_{i_{n}}}^{a_n,a_1}
 \left(q-\frac{2({\cal P}_{i_2}+...+{\cal P}_{i_{n-1}})+{\cal P}_{i_1}+
 {\cal P}_{i_n}
 }{2}| {\cal P}_{i_n}\right)
 $$
 ($\{i_k\}$ denotes permutations over $i_k$).

 Expressions~(\ref{set}),~(\ref{green}),~(\ref{set1}),~(\ref{S-matrix})
 represents Feynman rules for the construction of a quantum field theory
 with the action~(\ref{9}) in terms of bilocal fields.

 It was shown~\cite{a20} that the separable approximation
 of the constraint-shell gauge theory of bound states leads to
 the well-known Nambu Jona-Lasinio model~~\cite{a22,a21} and the
 phenomenological chiral Lagrangians~\cite{b4,b29} used for
 the description of the low-energy meson physics.
 Thus, the constraint-shell gauge theory of bound states
 is sufficient for describing the spectrum and interaction of
 hadrons as extended objects (without introducing the ideology of bags
 and string).

 In the context of the constraint-shell gauge theory,
 to solve the problem of hadronization in QCD, one needs
 to answer the questions:\\
 i) What is the origin of the potential of hadronization
 in the non-Abelian theory?\\
 ii) How to combine the Schr\"odinger equation for
 heavy quarkonia (that is derived by the residuum of poles
 of the quark Green functions) with the quark confinement~\cite{pn}?\\
 iii) What is the origin of the additional mass of
 the ninth pseudoscalar meson~\cite{w1}?

 \section{Dirac variables in Yang-Mills theory
 with the topological degeneration of physical states}

 \subsection{{\it Constraint-shell} radiation variables in perturbation theory}


 We consider the Yang-Mills theory with the local
 SU(2) group in four - dimensional Minkowskian space - time
 \begin{equation}\label{YM}
 W [\;A_\mu\;]\; =\;- \frac{1}{4} \int d^4x {F^{a}_{\mu\nu}{F_{a}^{\mu\nu}}}%
 \;= \frac{1}{2} \int d^4x \left( {F^{a}_{0i}}^2 - {B^{a}_{i}}^2
 \right)~,
 \end{equation}
 where the standard definitions of non-Abelian  electric tension
 $F^a_{0i} $
 $$
 F_{0i}\;=\;\partial_0 A_i^a - {D (A)} ^{ab}_i A_0^b, \;\;\;\;
 D ^{ab}_i \;=\; \left(\delta^{ab}\partial_i\ + g
 \epsilon^{acb} \,A^c_i\right)~
 $$
  and magnetic one $B^a_i$
 $$
 B^a_i = \epsilon_{ijk}\; \left( \partial_j A^a_k + \frac{g}{2}\,
 \epsilon^{abc}\,A^b_j\,A^c_k \right)
 $$
 are used.
 The action (\ref{YM}) is invariant with respect to gauge
 transformations $u(t,\vec x)$
 \be \label{gauge12} {\hat A}_{i}^u :=
 u(t,\vec x)\left({\hat A}_{i} + \partial_i \right)u^{-1}(t,\vec
 x),~~~~~~ \psi^u := u(t,\vec x)\psi~,
 \ee where ${\hat A_\mu}=g\frac{\tau^a }{2i} A_\mu^a~$.

 Solutions of the non-Abelian equations
 \begin{equation}\label{ym0}
 \frac{\delta W}{\delta A^a_0}= 0 \quad \Longrightarrow \quad
 \left[D^2\, (A)\right]^{ac} A^c_0 = D^{ac}\,_i\, (A)
 \partial_0 A^c_i
 \end{equation}

 \begin{equation}\label{ymi}
 \frac{\delta W}{\delta A^a_i}= 0 \quad \Longrightarrow \quad
 \left[\delta_{ij}D^2_k\, (A) -
 D_j\, (A)~D_i\, (A)\right]^{ac} A^c_j
 = D^{ac}\,_0\, (A)\left[\partial_0 A_i^c - {D (A)} ^{cb}_i
 A_0^b\right]
 \end{equation}
 are determined by boundary conditions and  initial data.

 The first Gauss equation~(\ref{ym0}) is the constraint. It
 connects initial data of $A^a_0$ with the ones of the
 spatial components $A^a_i$.
 To remove nonphysical variables,
 we can honestly solve this constraint in the
 form of the naive perturbation series
 \be\label{ymgps}
 A_0^c=a_0^c[A_i]=\frac{1}{\Delta}\partial_0 \partial_iA_i^c +
 ...
 \ee
 The resolving of the constraint and the substitution of
 this solution into the equations of motion distinguishes the gauge -
 invariant nonlocal (radiation) variables.
 After the substitution of this solution into ~(\ref{ymi})
 the lowest order of the  equation~(\ref{ymi}) in the coupling
 constant contains only transverse fields
 \be\label{ymps}
 (\partial_0^2-\Delta) A^{cT}_k+...=0.~~~~~
  A^{cT}_i=[\delta_{ik}-\partial_i\frac{1}{\Delta}\partial_k]A^{c}_k+...
 \ee
 This perturbation theory is well-known as the
 radiation~\cite{sch} (or Coulomb~\cite{f}) "gauge" with the generational
 functional of the Green functions in the form of the Feynman integral
 \bea
 \label{ymfi}
 Z_F[l^{(0)},J^{aT}]&=&\int\limits_{ }^{ }
 \int\limits_{ }^{ }\prod\limits_{c=1 }^{c=3 }
 [d^2A^{cT} d^2 E^{cT}]\nonumber\\
 &&\times\exp\left\{iW^T_{l^{(0)}} [A^T,E^T]-i\int\limits_{ }^{
 }d^4x [J^{cT}_{k} \cdot A^{cT}_{k}]\right\}~
 \eea
 with the constraint-shell action~(\ref{YM})
 \be\label{wT}
 W^T_{l^{(0)}} [A^T,E^T]=W^I\;\Bigl\vert_{ \frac{\delta W^I}{\delta
 A_0}\;=\;0}
 \ee
 given in the first order formalism
 \be \label{wI}
 W^I=\int dt \int d^3x \{ F^c_{0i}E^c_i-\frac{1}{2}[E^c_iE^c_i+B^c_iB^c_i]\}~.
 \ee
 The constraint
 \begin{eqnarray}\label{ymfic}
 \frac{\delta W^I}{\delta A_0}\;=\;0~\Rightarrow~
 D^{cd}_i\, (A)E^d_i=0~
 \end{eqnarray}
 is solved in terms of the radiation variables
 \begin{eqnarray}\label{ymrade}
 E^c_i=E^{Tc}_i +\partial_i \sigma~,
 ~~\partial_iE^{Tc}_i=0.
 \end{eqnarray}
 where the functions $\sigma^a$ take the form~\cite{f}
 \be
 \sigma^a[A^T,E^T]=\left( \frac{1}{D\,_i\,
 (A)\partial_i}\right)^{ac}\epsilon^{cbd}A_k^{Tb}E_k^{Td}
 \ee
 The operator quantization of the Yang-Mills theory in terms of the radiation
 variables belongs to Schwinger~\cite{sch} who proved the relativistic
 covariance of the radiation variables~(\ref{ymps}).
 This means that the radiation
 fields are transformed as the nonlocal functional
 \begin{eqnarray}\label{ymfin}
 \hat A_{k}^{T}[A]& =& v^{T} [A] ( \hat A_{k} + \partial_{k} )
 ( v^{T} [A] )^{-1},\nn\\
 {\hat A_{k}^{T}}\;&=&\;e\frac{A_{k}^{Ta} \tau^a}{2i}\,,
 \end{eqnarray}
 where the matrix \( v^{T} [A]\) is found from the condition of
 transversallity \(\partial_{k}\hat A_{k}^{T}=0\).
 At the level of the Feynman integral, as we have seen in QED,
 the relativistic covariance means the relativistic transformation
 of sources~\cite{mpl}.

 The definition~(\ref{ymfin}) can be treated as the
 transition to the new variables that
 allows us to rewrite the Feynman integral in the form of the FP
 integral~\cite{f,fs,ft}
 \bea
 \label{ymfpi}
 Z_F[l^{(0)},J^{aT}]&=&\int\limits_{ }^{ }
 \int\limits_{ }^{ }\prod\limits_{c=1 }^{c=3 }
 [d^4A^{c} ]\delta(\partial_i A^c_i) Det[D_i(A)\partial_i] \nonumber\\
 &&\times\exp\left\{i W [A]-i\int\limits_{ }^{
 }d^4x (J^{cT}_{k} \cdot A^{cT}_{k}[A])\right\}~.
 \eea
 It was proved~\cite{f,fs,ft} that on the mass-shell of the radiation
 fields the scattering amplitudes does not depends on the factor \(v^{T}
 [A]\). It remained only to reply question: Why does nobody observe
 these scattering amplitudes of mass-shell non-Abelian radiation fields?
 There are some possible answers on this question:
 the infrared unstability of the naive perturbation
 theory~\cite{ni,sh},
 the Gribov ambiguity, or zero of the FP determinant~\cite{g},
 the topological degeneration of the physical states~\cite{p2,in,pn}.
 In any case, in the non-Abelian theory there are not
 observable physical processes
 for which the gauge-equivalence theorem is valid.

 \subsection{Topological degeneration of initial data}

 One can find a lot of solutions of equations of classical electrodynamics.
 The nature chooses two types of functions:
 the monopole that determines nonlocal electrostatic phenomena
 (including instantaneous bound states), and multipoles that determine
 the spatial components of gauge fields with a nonzero magnetic tension.

 Spatial components of the non-Abelian fields considered above
 as radiation variables~(\ref{ymps}) in the naive perturbation
 theory are  also defined as multipoles.
 In the non-Abelian theory, there is the reason to count that
 {\it the spatial components of the non-Abelian fields belong to the
 monopole class of functions} like the time component of the Abelian
 fields.

 This fact was revealed by the authors of instantons~\cite{ins}. Instantons
 satisfy the duality equation in the Euclidean space, so that the instanton
 action coincides with the Chern-Simons functional (Pontryagin index)
  \be \label{1pont1}
  \nu[A]=\frac{g^2}{16\pi^2}\int\limits_{t_{in} }^{t_{out} }dt
 \int d^3x F^a_{\mu\nu} \widetilde{F}^{a\mu \nu} = X[A_{out}]- X[A_{in}]
 =n(t_{out}) - n(t_{in})~,
 \ee
 where
 \be \label{e1}
 X[A]=-\frac {1}{8\pi^2}\int\limits_V d^3x
 \epsilon^{ijk}Tr \left[{\hat A}_i\partial_j{\hat A}_k -
 \frac 2 3 {\hat A}_i{\hat A}_j{\hat A}_k\right]~,~~~~~~
 A_{in,out}=A(t_{in,out},x)
 \ee
 is the topological winding number functional of the gauge fields
 and $n$ is a value of this functional for a classical vacuum
 \be \label{clvac}
 \hat A_i=L^n_i=v^{(n)}(\vec{x})\partial_i{v^{(n)}(\vec{x})}^{-1}~.
 \ee
 The manifold of all classical vacua~(\ref{clvac})
 in the non-Abelian theory
 represents the group of three-dimensional paths lying  on the
 three-dimensional space of the $SU_c(2)$-manifold with the
 homotopy group $\pi_{(3)}(SU_c(2))=Z$. The whole group of
 stationary matrices is split into topological classes
  marked by integer numbers (the degree of the map)
 defined by the expression
 \be \label{1gn2}
 {\cal N}[n] =-\frac
 {1}{24\pi^2}\int d^3x ~\epsilon^{ijk}~ Tr[L^n_iL^n_jL^n_k]=n
 \ee
 which counts how many times a three-dimensional path $v(\vec{x})$
 turns around the $SU(2)$-manifold when the coordinate $x_i$ runs
 over the space where it is defined.

 Gribov in 1976 suggested to treat instantons as Euclidean solutions
 interpolating between classical vacua with different degrees of map.

 The degree of a map~(\ref{1gn2}) can be considered
 as the condition for normalization that determines a class of
 functions where the classical vacua $L^n_i$~(\ref{clvac}) are given.
 In particular, to obtain Eq.~(\ref{1gn2}) we should choose a classical
 vacuum in the  form
 \be \label{class0}
 v^{(n)}(\vec{x})=\exp(n \hat \Phi_0(\vec{x})),~~~~~ \hat \Phi_0=-
 i \pi\frac{\tau^a x^a}{r} f_0(r)~~~~~(r=|\vec x|)~,
 \ee
 where the function $f_0(r)$ satisfies the boundary conditions
 \be \label{bcf0}
 f_0(0)=0,~~~~~~~~~~~~~~
 f_0(\infty)=1~.
 \ee
 The normalization~(\ref{1gn2}) points out  that the vacuum values
 $L_i^n$  of spatial components $A_i$ belong to monopole-type
 class of functions. To show that these classical values are not
 sufficient to describe a physical vacuum in the non-Abelian theory,
 we consider
 a quantum instanton, i.e.  the corresponding zero vacuum solution
 of the Schr\"odinger equation
 \be \label{qvacuum}
  \hat H \Psi_0[A]=0~~~~~~(\hat H=\int d^3x [\hat E^2+B^2],~~\hat E=
 \frac{\delta}{i\delta A})~.
 \ee
 It can be constructed
 using the winding number functional~(\ref{e1}) and its derivative
 \be
 \frac{\delta}{\delta A_i^c} X[A]=\frac{g^2}{8\pi^2}B_i^c(A)
 \ee
 The vacuum wave functional in terms of the winding number (\ref{e1})
 takes the form of a plane wave~\cite{vp1}
 \be \label{twv}
 \Psi_0 [A] =\exp(i P_N X[A])
 \ee
 for nonphysical values of the
 topological momentum $P_N=\pm i 8 \pi^2/g^2$~\cite{vp1,kh}.
 We would like to drew attention
 of a reader to that in QED this type of the wave functional
 belongs to nonphysical part of a spectrum like the wave function of an
 oscillator $(\hat p^2+q^2)\psi_0=0$.
 The value of this nonphysical
 plane wave functional for classical vacuum~(\ref{clvac})
 coincides with quasiclassical instanton wave function
 \be\label{qins}
 \exp(iW[A_{\rm instanton}])=\Psi_0[A=L_{out}] \times \Psi^*_0[A=L_{in}]
 =\exp(-\frac{8\pi^2}{g^2}[n_{\rm out}-n_{\rm in}])~.
 \ee
 This exact relation between a classical instanton and its
 quantum version~(\ref{qvacuum})  points out
 that classical instantons are also nonphysical solutions,
 they permanently tunneling in Euclidean space-time between classical vacua
 with the zero energy that does not belong to physical spectrum
 \footnote{
 The author is
 grateful to V.N. Gribov for the discussion of the problem of
 instantons during a visit in  Budapest, May 1996.}.

 \subsection{Physical vacuum and the gauge Higgs effect}

 The next step is the assertion~\cite{jac} about
 the topological degeneration of initial data of not only
 classical vacuum but all physical fields
 with respect to stationary gauge transformations
 \be \label{top2}
 \label{1gnl} {\hat A}_i^{(n)}(t_0,\vec x)=v^{(n)}(\vec{x}){\hat
 A}^{(0)}_i(t_0,\vec{x}) {v^{(n)}(\vec{x})}^{-1}
 +L^n_i~,~~~~L^n_i=v^{(n)}(\vec{x})\partial_i{v^{(n)}(\vec{x})}^{-1}~.
 \ee
 The stationary transformations $v^n(\vec{x})$
 with $n=0$ are called the small ones; and those with $n \neq 0$
 the large ones~\cite{jac}.

 The group of transformations~(\ref{1gnl}) means that
 {\it the spatial components of the non-Abelian fields
 with a nonzero magnetic tension $B(A)\not= 0$ belong to the
 monopole class of functions} like the time component of the Abelian fields.
 In this case, non-Abelian fields with a nonzero magnetic tension contain
 a nonperturbative monopole-type term, and
 spatial components can be decomposed
 in a form of a sum of the monopole ${\Phi}^{(0)}_i(\vec{x})$ and multipoles
 $\bar A_i$
 \be
 \label{bar}
 \hat  A^{(0)}_i(t_0,\vec{x}) =
 \hat {\Phi}^{(0)}_i(\vec{x}) + \hat {\bar A}^{(0)}_i(t_0,\vec{x})~.
 \ee
 The multipole is considered as a weak perturbative part
 with asymptotics at the spatial infinity
 \be \label{ass1}
 \bar A_i(t_0,\vec{x})|_{\rm asymptotics} =
 O(\frac{1}{r^{1+l}})~~~~(l > 1)~.
 \ee

 Nielsen and Olesen~\cite{ni}, and  Matinyan and Savidy~\cite{sh}
 introduced a vacuum magnetic tension, using the fact that
 all asymptotically free theories are unstable, and the perturbation
 vacuum is not the lowest stable state.

 The extension of the topological classification of
 classical vacua  to all initial data of spatial components
 helps us to choose a vacuum monopole
 with a zero value of the winding number functional~(\ref{e1})
 \be\label{e10}
  X[A=\Phi_i^{c(0)}]=0~,~~~~~~~~~
 \frac{\delta X[A]}{\delta A_i^c}|_{A=\Phi^{(0)}}\not= 0.
 \ee
 The zero value of the winding number, transversality,
 and spherical symmetry fix a class of initial data for
 spatial components
 \be \label{wy}
 \hat \Phi_i= - i
 \frac{\tau^a}{2}\epsilon_{iak}\frac{x^k}{r^2}f(r)~.
 \ee
 They contain only one function $f(r)$.
 The classical equation for this function takes the form
 \be\label{higgs}
 D_k^{ab}(\Phi_i) F_{kj}^b(\Phi_i)=0~~\Rightarrow~~\frac{d^2 f}{dr^2}+
 \frac{f(f^2-1)}{r^2}=0~.
 \ee
 We can see three solutions of this equation.
 \be
 f_1^{PT}=0,~~~~~~~ f^{WY}_{1}=\pm 1~~~~~~~(r\not= 0)~ .
  \ee
 The first solution corresponds to the naive unstable perturbation
 theory with the assymptotic freedom formula.

 Two nontrivial solutions are well-known.  They are the Wu-Yang monopoles
 applied  for the construction of physical variables
 in the current literature~\cite{fn}. As it was shown in paper~\cite{bpr}
 the Wu-Yang monopole leads to rising potentials of the instantaneous
 interaction of the quasiparticle current.
 This interaction rearranges the perturbation series, leads to
 the gluon constituent mass, and removes the
 assymptotic freedom formula~\cite{fb,a6} as the origin of
 unstability.

 The Wu-Yang monopole is a solution of classical
 equations everywhere besides the origin of coordinates $r=0$.
 The corresponding magnetic field is
 \be
 \label{sb} B_i^a(\Phi_k)=\frac{x^a x^i}{gr^4}~.
 \ee
 Following Wu and Yang \cite{wy}, we consider the whole finite
 space volume, excluding  an $\epsilon$-region around the singular
 point. To remove
 a singularity at the origin of coordinates and regularize its
 energy, the Wu-Yang monopole is considered as the
 limit of zero size $\epsilon~\rightarrow~ 0$ for the
 Bogomol'nyi-Prasad-Sommerfield (BPS) monopole~\cite{BPS}
 \be\label{152}
 \label{bps} f^{WY}_{1}~\Rightarrow~ f^{BPS}_{1}= \left[1 -
 \frac{r}{\epsilon \sinh(r/\epsilon)}\right]
 \ee
 with the finite energy
 \be\label{v1}
  \int\limits_{
 }^{ }d^3x [B^a_i(\Phi_k)]^2 \equiv V <B^2>
 =\frac{4\pi}{g^2 \epsilon}\equiv \frac{1}{\alpha_s \epsilon}~.
 \ee
 In this case, the BPS-regularization of the Wu-Yang monopole
 is the analogue of the infrared regularization
 in QED by the introduction of the photon mass that also violates
 the initial equations of motion.
 The size of the BPS monopole  is chosen so that
 the parameter $\epsilon$  disappears in the infinite volume
 limit
 \be  \label{ve1}
  \epsilon=\frac{1}{\alpha_s<B^2>V}
 \ee
 and the vacuum energy-density of the monopole  solution $<B^2>$
 is removed by a finite counter-term in the Lagrangian
  $$
  \bar {\cal L}={\cal L}-\frac{<B^2>}{2} ~.
  $$
 This vacuum magnetic tension is the crucial difference of
 the topological degeneration of fields
 in Minkowski space from the topological degeneration of
 classical vacua of the instantons
 in the Euclidean one.

 The  problem is to formulate the Dirac quantization
 of weak perturbations of the non-Abelian fields in the
 presence of the nonperturbative monopole taking into account
 the topological degeneration of all initial data.

 \subsection{ Dirac method and Gribov copies}

 Instead of artificial equations (\ref{gf}), (\ref{fpi11})
 of the gauge-fixing method~\cite{fp1}
 \be \label{gf1}
 F(A_{\mu})=0,~~~F(A^{u}_{\mu})=M_F u \not=
 0~\Rightarrow~~Z^{FP}=\int \prod_{\mu}DA_\mu
 det M_F\delta (F(A)) e^{iW}
 \ee
 we repeat the Dirac {\it constraint-shell} formulation
 resolving the constraint (\ref{ym0})
 \begin{equation}\label{ym01}
 \frac{\delta W}{\delta A^a_0}= 0 \quad \Longrightarrow \quad
 \left[D^2\, (A)\right]^{ac} A^c_0 = D^{ac}\,_i\, (A)
 \partial_0 A^c_i~
 \end{equation}
 with  nonzero initial data
 \be
 \partial_0 A^c_i=0~\quad \Longrightarrow \quad
 A^c_i(t,\vec x)=\Phi_i^{c(0)}(\vec x)~.
 \ee
 The vacuum
 magnetostatic field $\Phi_i^{c(0)}$ has a zero value of the
 winding number (\ref{e1}) $X[\Phi_i^{c(0)}]=0$ and satisfies the classical
 equations everywhere besides of the small region near the origin
 of coordinates of the size of
 \be
 \epsilon\sim \frac{1}{\int d^3x B^2(\Phi)}\equiv\frac{1}{<B^2>V}
 \ee
 that disappears in the infinite volume limit.

 The second step is the consideration of the perturbation theory
 (\ref{bar}) where the constraint (\ref{ym01}) takes the form
 \be\label{ym01p}
 \left[D^2\, (\Phi^{(0)})\right]^{ac} A^{c(0)}_0=
 \partial_0[D^{ac}\,_i\, (\Phi^{(0)}) A^{c(0)}_i]~.
 \ee
 Dirac proposed~\cite{cj} that the time component $A_0$
 (the quantization of which contradicts to quantum principles)
 can be removed by gauge transformation, so that
 the constraint~(\ref{ym01p}) takes the form
 \be\label{ym01pd}
 \partial_0[D^{ac}\,_i\, (\Phi^{(0)}) A^{c(0)}_i]=0~.
 \ee
 We defined the {\it constraint-shell gauge}
 \be\label{pg}
 [D^{ac}\,_i\, (\Phi^{(0)}) A^{c(0)}_i]=0
 \ee
 as the zero initial data of this constraint (\ref{ym01pd}).

 The topological degeneration of initial data means
 that not only classical vacua but also
 all fields $A_i^{(0)}= {\Phi}_i^{(0)}+ {\bar A}_i^{(0)}$
 in the gauge (\ref{pg}) are degenerated
 \be
 \label{gnlp}
 \hat {A}_i^{(n)}=v^{(n)}(\vec{x})
 (\hat { A}_i^{(0)}
 +\partial_i) {v^{(n)}(\vec{x})}^{-1}~,~~~~~~~~~
 v^{(n)}(\vec{x})=\exp[n\Phi_0(\vec{x})]~.
 \ee
 The winding number functional~(\ref{e1}) after
 the transformation~(\ref{top2}) takes the form
 \be\label{X1}
 X[A_i^{(n)}]=
 X[A_i^{(0)}]+
 {\cal N}(n)+\frac{1}{8\pi^2}\int d^3x
 ~\epsilon^{ijk}~ Tr[\partial_i (\hat A^{(0)}_j L^n_k)]~,
 \ee
 where ${\cal N}(n)=n$ is given by eq.~(\ref{1gn2}).

 The  constraint-shell  gauge~(\ref{pg}) keeps its form
 in each topological class
 \be \label{pdg}
 D_i^{ab}(\Phi_k^{(n)})\bar A_i^{(n)b}=0~,
 \ee
 if  the phase $\Phi_0(\vec{x})$ satisfies the equation
 of the Gribov ambiguity
 \be \label{pgc}
 [D^2_i(\Phi^{(0)}_k)]^{ab} \Phi_0^{b}=0~.
 \ee
 In this case, the topological degeneration means the existence of
 the manifold of the Gribov copies of the  constraint-shell
 gauge~(\ref{pg}).
 One can show~~\cite{bpr} that the Gribov equation~(\ref{pgc}) together with
 \be\label{e1n}
 X[\Phi^{(n)}]= n
 \ee
 are compatible with the unique solution of classical equations.
 It is just the Wu-Yang monopole considered before.
 The nontrivial solution of the equation for the Gribov phase~(\ref{pgc})
 in this case is well-known
 \be \label{pgc0}
 \hat \Phi_0=-
 i \pi\frac{\tau^a x^a}{r} f_0^{BPS}(r)~,~~~~
 f_0^{BPS}(r)=\left[
 \frac{1}{\tanh(r/\epsilon)}-\frac{\epsilon}{r}\right]~,
 \ee
 it is the Bogomol'nyi-Prasad-Sommerfield (BPS)
 monopole~\cite{BPS}.

 Thus, instead of the topological degenerated classical vacuum
 for the instanton
 calculation (that is in the physically unattainable  region), we
 have the topological degenerated Wu-Yang monopole~(\ref{gnlp})
 \be
 \label{gnlf}
 \hat {\Phi}_i^{(n)}:=v^{(n)}(\vec{x})
 [\hat {\Phi}_i^{(0)} + \partial_i] v^{(n)}(\vec{x})^{-1}~,~~~~~~
 v^{(n)}(\vec{x})=\exp[n\Phi_0(\vec{x})]~,
 \ee
 and the topological degenerated multipoles
 \be
 \label{gnla}
 \hat {\bar A}_i^{(n)}:=v^{(n)}(\vec{x})
 \hat {\bar A}_i^{(0)}
  {v^{(n)}(\vec{x})}^{-1}~.
 \ee
 The Gribov copies are evidence of a zero mode in the left hand-side
 of both the constraints (\ref{ym01p}) and (\ref{ym01}).
 \be\label{has}
 \left[D_i^2\, (\Phi^{(0)})\right]^{ac} A^{c}_0=0~.
 \ee
 A nontrivial solution of this equation
 \be\label{ab0}
 A^c_0(t,\vec x)=\dot N (t)\Phi^{c}_0(\vec x)
 \ee
  can be removed  from the local equations of
  motion  by  the gauge transformation (a la Dirac of 1927)
  to convert the fields into the Dirac variables
 \be\label{D1}
 \hat A^{(N)}_i=\exp[N(t)\hat \Phi_0(\vec x)]
 [\hat A^{(0)}_i+\partial_i]\exp[-N(t)\hat \Phi_0(\vec x)]~.
 \ee
 But this solution~(\ref{ab0}) cannot be removed
 from the constraint-shell action $W^*=\int dt \dot N^2I/2+...$ and
 from the winding number $X[A^{(N)}]=N+X[A^{(0)}]$.
 Finally we obtained the Feynman path integral
 \be \label{F2}
 Z_F=\int DN \prod_{i,c}[D E_i^{c(0)}DA^{c(0)}_{i}] e^{iW^*}
 \ee
 that does not coincide with any artificial gauge (\ref{gf1}).

 We consider the derivation of this integral~(\ref{F2}) in details
 in the following.

 \subsection{ Topological dynamics}

 The repetition of the Dirac definition of the observable variables
 in QED allowed us to determine  the vacuum fields and phase of
 their topological degeneration in the form of the Gribov
 copies of the constraint-shell gauge.

 The degeneration of initial data is the evidence of the zero mode
 of the Gauss law constraint.
 In the lowest order of the considered perturbation theory, this
 constraint~(\ref{has}) has the solution~(\ref{ab0})
 with the nontrivial vacuum electric field
 \be\label{ab00}
  F^b_{i0}=\dot N(t)D^{bc}_i(\Phi_k^{(0)}) \Phi^c_0(\vec x)~.
 \ee
 We call the new variable $N(t)$ the winding number variable
 as the vacuum Chern-Simons functional is equal to a difference
 of the {\it in} and {\it out} values of this variable
  \be \label{pont}
  \nu[A_0,\Phi^{(0)}]=\frac{g^2}{16\pi^2}\int\limits_{t_{in} }^{t_{out} }dt
 \int d^3x F^a_{\mu\nu} \widetilde{F}^{a\mu \nu}=\frac{\alpha_s}{2\pi}
 \int d^3x F^b_{i0}B_i^b(\Phi^{(0)})[N(t_{out}) -N(t_{in})]
 \ee
 $$
 =N(t_{out}) -N(t_{in})~.
 $$
 The winding number functional
 admits its generalization to  noninteger degrees of a map~\cite{bpr}
 \be
 X[\Phi^{(N)}]=N, ~~~~N\not= n~, ~~~~~
 \left(\hat \Phi_i^{(N)}=e^{N\hat \Phi_0}
 [\hat \Phi_i^{(0)}+\partial_i ]e^{-N\hat \Phi_0}\right)~.
 \ee
 Thus, we can identify the global variable $N(t)$
 with the winding number degree of freedom in the Minkowskian space
 described by the action
 \be \label{ktg}
 W_{N}=\int d^4x \frac{1}{2} (F^c_{0i})^2=
 \int\limits_{ }^{ }dt\frac{{\dot N}^2 I}{2}~,
 \ee
 where  the functional
 \be I=\int\limits_{V}d^3x(D^{ac}_i(\Phi_k)\Phi^c_0)^2
 =\frac{4\pi^2}{\alpha^2_s} \frac{1}{V<B^2>}~.
 \ee
 does not contribute in local equations of motion.
 The topological degeneration of all fields
 converts into the degeneration of only one global topological
 variable $N(t)$ with respect to a shift of this variable on
 integers: $(N~\Rightarrow~ N+n,~ n=\pm 1,\pm 2,...)$.
 Thus, the topological variable is a free rotator with
 the instanton-type wave function~(\ref{twv}) in the Minkowskian
 space-time
 \be \label{mtvw}
 \Psi_N=\exp\left\{i P_N N\right\}~,~~~~~~P_N=\dot N I=2\pi k+\theta~,
 \ee
 where $k$ is a number of the Brilloin zone, and $\theta$ is
 the $\theta$-angle. In the contrast to the instanton wave function
 (\ref{twv}) the spectrum of the topological momentum is real and
 belongs to physical values.
 Finally,  equations~(\ref{ktg}) and~(\ref{mtvw}) determine the finite
 spectrum of the global electric tension~(\ref{ab00})
 \be \label{tef}
 F^b_{i0}=\dot N [D_i(\Phi^{(0)})A_0]^b=\alpha_s
 \left( \frac{\theta}{2\pi}+k\right)B^b_i(\Phi^{(0)})~.
 \ee
 It is analogue of the Coleman spectrum of the
 electric tension in the $QED_{(1+1)}$~\cite{col}.
 The application of the Dirac quantization to the 1-dimensional
 electrodynamics $QED_{(1+1)}$ in paper~\cite{gip}
 demonstrates the universality of the Dirac variables and their
 adequacy to the description of topological  dynamics with a
 nontrivial homotopy group.

 \subsection{ Zero mode of Gauss Law and Dirac variables}

 The constraint-shell theory is obtained by the explicit
 resolution of the Gauss law constraint
 \begin{equation}\label{173}
 \frac{\delta W}{\delta A^a_0}= 0 \quad \Longrightarrow \quad
 \left[D^2\, (A)\right]^{ac} A^c_0 = D^{ac}\,_i\, (A)
 \partial_0 A^c_i
 \end{equation}
 and next in dealing with the initial action  on surface of
 these solutions
 \begin{eqnarray}
  W^{*}\;=\; W [\;A_\mu\;]\;\Bigl\vert_{ \frac{\delta W}{\delta
 A^a_0}\;=\;0.}
 \end{eqnarray}
 The result of similar solution in QED was the electrostatics and
 the Coulomb - like atoms.
 In the non-Abelian case, the topological degeneration in the form
 of the Gribov copies means that a general solution of the Gauss law
 constraint~(\ref{173}) contains the zero mode ${\cal Z}$.
 A general solution of the inhomogeneous equation~(\ref{173}) is a sum of the
 zero-mode solution ${\cal Z}^a$  of the homogeneous equation
 \be\label{zm}
 (D^2(A))^{ab}{\cal Z}^b=0~,
 \ee
  and a particular solution ${\tilde A}_0^a$ of the
 inhomogeneous one
 \be \label{genl} A_0^a = {\cal
 Z}^a + {\tilde A}^a_0~.
 \ee

 The zero-mode ${\cal
 Z}^a$  at the spatial infinity has been represented in the form of
 sum of
 the product of a new topological variable $ \dot N(t)$ and the Gribov phase
 $\Phi_0(\vec{x})$ and weak multipole corrections
 \be \label{ass}
 \hat {\cal Z}(t,\vec{x})|_{\rm asymptotics}=\dot N(t)\hat
 \Phi_0(\vec{x})+O(\frac{1}{r^{1+l}})~, ~~~(l>0)
 \ee

 In this case, the single one-parametric variable $N(t)$
 reproduces the topological degeneration of all field variables,
 if the Dirac variables are
 defined by the gauge
 transformations
 \be \label{gt1} 0=U_{\cal Z}(\hat {\cal
 Z}+\partial_0)U_{\cal Z}^{-1}~,
 \ee
 $$
 {\hat A}^*_i=U_{\cal Z}({\hat A}^{(0)}_i+\partial_i)U_{\cal Z}^{-1},~~~
 A_i^{(0)}=\Phi_i^{(0)}+\bar A_i^{(0)}
 $$
 where the spatial asymptotic of $U_{\cal Z}$ is
 \be \label{UZ}
 U_{\cal Z}=T\exp[\int\limits^{t} dt' \hat {\cal
 Z}(t',\vec{x})]|_{\rm asymptotics} =\exp[N(t)\hat
 \Phi_0(\vec{x})]~.
 \ee
 The topological degeneration of all fields
 converts into the degeneration of only one global topological
 variable $N(t)$ with respect to a shift of this variable on
 integers: $(N~\Rightarrow~ N+n,~ n=\pm 1,\pm 2,...)$.

 \subsection{Constraining with the zero mode}

 Let us  formulate an equivalent unconstrained system for the YM theory
 in the monopole class of functions in the presence the zero
 mode ${\cal Z}^b$ of the Gauss law constraint
 \be
 \label{edg1} A_0^a={\cal Z}^a+\tilde A_0^a;~~~~~~ F_{0k}^a = -
 D_k^{ab}(A) {\cal Z}^b + {\tilde F_{0k}^a}~~~~~~
 (~(D^2(A))^{ab}{\cal Z}^b=0~)~.
 \ee
 To obtain the constraint-shell
 action
 \be \label{cs1} W_{YM}({\rm constraint}) =
 {\cal W}_{YM}[{\cal Z}] + \tilde W_{YM}[\tilde F]~,
  \ee
 we  use the evident decomposition \be
 F^2=(-D{\cal Z}+\tilde F)^2=
 (D{\cal Z})^2- 2\tilde F D{\cal Z} + (\tilde F)^2=
 \partial ({\cal Z}(D{\cal Z}))- 2\partial ({\cal Z}\tilde F)+(\tilde F)^2
 \ee
 and the Gauss Eqs. $D\tilde F=0$ and $D^2{\cal Z}=0$ which
 show that the zero mode part ${\cal W}_{YM}$ of the
 constraint-shell action~(\ref{cs1}) is the sum of two surface
 integrals
 \be \label{gp} {\cal W}_{YM}[{\cal Z}]=\int dt \int
 d^3x[\frac{1}{2}{\partial}_i({\cal Z}^a D_i^{ab}(A) {\cal Z}^b)-
 {\partial}_i(\tilde F_{0i}^a{\cal Z}^a)] ={\cal W}^0 + {\cal
 W}^{\prime}~,
 \ee
 where the first one ${\cal W}^0$ is the kinetic
 term and the second one ${\cal W}^{\prime}$ describes the coupling
 of the zero-mode to the local excitations. These surface terms are
 determined by the asymptotic of the fields $({\cal Z}^a,
 {A}_i^a)$ at spatial infinity (\ref{ass}), (\ref{ass1}) which we
 denoted by $(\dot N(t)\Phi_0^a(\vec{x}),
 {\Phi}_i^a(\vec{x}))$. The fluctuations $\tilde F_{0i}^a$ belongs
 to the class of multipoles. Since the surface integral over
 monopole-multipole couplings vanishes, the fluctuation part of the
 second term obviously drops out.
 The substitution of the solution
 with the asymptotic (\ref{ass}) into the first surface term
  Eq.~(\ref{gp}) leads to the zero-mode action~(\ref{ktg}).

 The action for the equivalent unconstrained system of the local
 excitations,
 \be \label{lym} \tilde W_{YM}[\tilde F]=\int d^4x
 \left\{E_k^a\cdot \dot{A}_k^{a(0)} - \frac{1}{2}\left\{E_k^{2}
 +B_k^2(A^{(0)})+[D_k^{ab}({\Phi}^{(0)})\tilde{\sigma}^b]^2\right\}
 \right\}~, \ee
 is obtained  in terms of variables with zero degree of map
 \be
 \label{evid} \hat {\tilde
 F}_{0k}=U_{\cal Z} \hat F^{(0)}_{0k}U_{\cal Z}^{-1}~,~~~~ \hat A_i =
 U_{\cal Z}(\hat A_i^{(0)}+\partial_i )U_{\cal Z}^{-1}~,~~~ \hat
 A_i^{(0)}(t,\vec x)=\hat \Phi^{(0)}_i(\vec x) + \hat {\tilde A}^{(0)}_i(t,\vec
 x)~.
 \ee
 by decomposing the electrical
 components of the field strength tensor $F_{0i}^{(0)}$ into transverse
 $E^a_i$ and longitudinal
 ${F^a_{0i}}^L=-D_i^{ab}({\Phi}^{(0)})\tilde \sigma^b$ parts, so that
 \be
 \label{longb2} F^{a(0)}_{0i}=E^a_i -D_i^{ab}({\Phi}^{(0)})\tilde
 \sigma^b;~~~ D_k^{ab}({\Phi}^{(0)}) E^b_k =0 ~.
 \ee
 Here the function
 $\tilde \sigma^b$ is determined from the Gauss equation
  \begin{equation}\label{gaub1}
 \left((D^2({\Phi}^{(0)}))^{ab} + g\epsilon^{adc}{\tilde A}^{d(0)}_i
 D_i^{cb}({\Phi}^{(0)})  \right) \tilde \sigma^b
 = - g\epsilon^{abc}{\tilde A}_i^{a(0)} E_i^{c}~.
 \end{equation}

 Due to gauge-invariance the dependence of the action for local
 ecxitations on the zero mode disappears, and we got
 the ordinary generalization of the Coulomb gauge~\cite{sch,f}
 in the presence of the Wu-Yang monopole.

 \subsection{Feynman path integral}

 The Feynman path integral over the independent variables includes
 the integration over the topological variable $N(t)$
  \be
 \label{fizm}
 Z_{\rm F}[J]=\int \prod\limits_{t}dN(t)
 \tilde Z [J^U]~, \ee where \be\label{schb} \tilde Z[J^U]=\int
 \prod\limits_{t,x}\left\{ \prod\limits_{a=1}^{3} \frac{[d^2
 A_a^{(0)}
 d^2 E_a^{(0)}]}{2\pi}\right\} \exp i\left\{{\cal W}_{YM}({\cal Z}) + \tilde
 W_{YM}( A_a^{(0)}) + S[J^U] \right\}~.
 \ee
 As we have seen above, the
 functionals ${\tilde W},S$ are given in terms of the variables
  which contain the nonperturbative phase factors $U=U_{\cal Z~}$
 (\ref{UZ}) of the topological degeneration of initial data. These
  factors disappear in the  action $\tilde W$, but not in  the
  source
  \be \label{sym} S[J^U]=\int d^4x J^a_i \bar A^a_i,~~~\hat
  {\bar A}_i=U(\hat A_a^{(0)} )U^{-1}
  \ee
  what reflects the fact of
  the topological degeneration of the physical fields.

 The constraint-shell formulation  distinguishes
  a bare "gluon" as a weak deviation of the monopole with the index $(n=0)$,
 and  an observable (physical) "gluon"
 averaged over the topological degeneration (i.e., Gribov's
 copies) \cite{p2}
 \be \label{photon}
 \bar A^{\rm Phys}=\lim\limits_{L \to \infty}\frac{1}{2L}
 \sum\limits_{n=-L }^{n=+L }\bar A^{(n)}(\vec x)\sim
 \delta_{r,0}~;
 \ee
 whereas, in QED the constraint-shell field is a transversal photon.

 \subsection{ Rising potential induced by monopole}

 We can calculate the instantaneous the Green function
 \begin{equation}\label{gaub01}
 (D^2(\Phi^{(0)}))^{ab}({\vec x})G^{bc}(\vec x,\vec
 y)=\delta^{ac}\delta^3(x-y)~.
 \end{equation}
 In the presence of the Wu-Yang monopole we have
 $$
 (D^2)^{ab}({\vec x})= \delta^{ab}\Delta-
 \frac{n^an^b+\delta^{ab}}{r^2}+
 2(\frac{n_a}{r}\partial_b-\frac{n_b}{r}\partial_a)~,
 $$
 and $n_a(x)={x_a}/{r};~ r=|{\vec x}|~$. Let us decompose $G^{ab}$
 into a complete set of orthogonal vectors in color space
 $$
 G^{ab}({\vec x},{\vec y})=  [n^a(x)n^b(y)V_0(z) +
 \sum\limits_{\alpha=1,2} e^a_{\alpha}(x)
 e^b_{\alpha}(y)V_1(z)];~~~(z=|{\vec x}-{\vec y}|)~.
 $$
 Substituting the latter into the first equation, we get
 $$
 \frac{d^2}{dz^2}V_n+ \frac{2}{z} \frac{d}{dz}V_n-
 \frac{n}{z^2}V_n=0 ~~~n = 0,~1~.
 $$
 The general solution for the last equation is
 \begin{equation} \label{pot}
  V_n(|{\vec x}-{\vec y}|)= d_n |{\vec x}-{\vec y}|^{l_1^n} +
 c_n |{\vec x}-{\vec y}|^{l_2^n}~,~~~~n=0,~1~,
 \end{equation}
 where $d_n$, $c_n$ are constants, and ${l_1^n},~{l_2^n}$ can be
 found as  roots of the equation $(l^n)^2 + l^n =n$~, i.e.
 \begin{equation} \label{l1}
 {l_1^n} =-\frac{1+\sqrt{1+4n}}{2};~~~ {l_2^n}
 =\frac{-1+\sqrt{1+4n}}{2}~.
 \end{equation}
 It is easy to see that for $n=0$ we get the Coulomb-type potential
 $d_0=-1/4\pi$,
 \begin{equation} \label{cfun}
 {l_1^0} =-\frac{1+\sqrt{1}}{2}= - 1~
 ;~~~{l_2^0} =\frac{-1+\sqrt{1}}{2}= 0~,
 \end{equation}
 and for $n=1$  the 'golden section' potential with
 \begin{equation} \label{fun}
 {l_1^1} =-\frac{1+\sqrt{5}}{2}\approx - 1.618~
 ;~~~{l_2^1} =\frac{-1+\sqrt{5}}{2}\approx 0.618~.
 \end{equation}
 The last potential (in the contrast with the Coulomb-type one)
 can leads to rearrangement of the naive perturbation series
 of the type of the spontaneous chiral symmetry breaking.
 This potential can be considered as the origin of
 the 'hadronization' of quarks and gluons
 in QCD \cite{a6}.

 \subsection{FP path integral}

 Thus, we can say that the Dirac variables with the topological
 degeneration of initial states in a non-Abelian theory
 determine the physical origin of hadronization and
 confinement as nonlocal monopole effects.
 The Dirac variables distinguish a unque gauge.
 In QED, it is the Coulomb gauge; whereas, in YM theory, it is
 covariant generalization the Coulomb gauge in the presence of a monopole.

 If we pass to another gauges of physical sources
 on the level of the FP integral
 in relativistic gauges, all the
 monopole effects of the degeneration and rising potential can be lost
 (as the Coulomb potential is lost in QED in relativistic invariant gauges).
 Recall that to prove the equivalence of the Feynman integral
 to the Faddeev-Popov integral in an arbitrary gauge, we
 change variables and  concentrate all monopole effects
 in the phase factors before the physical sources.
 The change of the sources removes all these effects.

 The change of the sources was possible in the Abelian theory
 only for the scattering amplitudes~\cite{f} when
 all particle-like excitations of the fields are on their mass-shell.
 However, for the cases of nonlocal bound states
 and  other  phenomena where these fields are off their mass-shell
 the Faddeev theorem of equivalence of different "gauges" is not valid.

 For the non-Abelian theory all states become nonlocal, and
 the range of validity of the equivalence theorem is equal to zero.

 \section{Dirac variables in  QCD: hadronization and confinement}

 \subsection{Action and first principles of QCD}

 In the previous Sections  we have shown that the Dirac variables
 of gauge theories is the best way  to describe both
 the topological degeneration of initial data and the relativistic
 covariant bound states.

 Here we would like to demonstrate the adequacy of the Dirac variables to
 the physical phenomena of hadronization and confinement in QCD.

 QCD was proposed~\cite{fgl} as the non-Abelian $SU_c(3)$
 theory with the action functional
 \be \label{5u}
 W=\int d^4x
 \left\{\frac{1}{2}({G^a_{0i}}^2- {B_i^a}^2)
 + \bar\psi[i\gamma^\mu(\partial _\mu+{\hat
 A_\mu})
 -m]\psi\right\}~,
 \ee
 where $\psi$ and $\bar \psi$ are the fermionic quark fields,
 \be \label{5v}
 G_{0i}^a = \partial_0 A^a_i - D_i^{ab}(A)A_0^b~,~~~~~~
 B_i^a=\epsilon_{ijk}\left(\partial_jA_k^a+
 \frac g 2f^{abc}A^b_jA_k^c\right)
 \ee
 are non-Abelian electric  and magnetic fields, and
 $D^{ab}_i(A):=\delta^{ab}\partial_i + gf^{acb} A_i^c$
 is the covariant derivative.

 The action (\ref{5u}) is invariant with respect to
 relativistic and gauge transformations
 $u(t,\vec x)$
 \be \label{gauge11}
 {\hat A}_{i}^u := u(t,\vec x)\left({\hat A}_{i} + \partial_i
 \right)u^{-1}(t,\vec x),~~~~~~
 \psi^u := u(t,\vec x)\psi~,
 \ee
 where ${\hat A_\mu}=g\frac{\lambda^a }{2i} A_\mu^a~$.

 \subsection{Topological degeneration of QCD}

 The group of stationary gauge transformations in QCD represents
 the group of three-dimensional paths  $u(t,\vec{x})=v(\vec{x})$
 lying  on the three-dimensional
 space of the $SU_c(3)$-manifold with the homotopy group
 $\pi_{(3)}(SU_c(3))=Z$.
 Thus, a difference between the Abelian and non-Abelian theories is
 the difference between the homotopy groups of maps of three-dimensional
 coordinate space onto the gauge manifolds $U(1)$, and $SU(3)_c$ respectively.

 The whole group of stationary gauge transformations is split into
 topological classes marked by the integer number $n$ (the degree of the map)
 which counts how many times a three-dimensional path turns around the
 $SU(3)$-manifold when the coordinate $x_i$ runs over the space where it is
 defined.
 The stationary transformations $v^n(\vec{x})$ with $n=0$ are called the small
 ones; and those with $n \neq 0$
 \be \label{gnl}
 {\hat A}_i^{(n)}:=v^{(n)}(\vec{x}){\hat A}_i(\vec{x})
 {v^{(n)}(\vec{x})}^{-1}
 +L^n_i~,~~~~L^n_i=v^{(n)}(\vec{x})\partial_i{v^{(n)}(\vec{x})}^{-1}~,
 \ee
 the large ones.

 The degree of a map
 \be \label{5gn2}
 {\cal N}[n]
 =-\frac {1}{24\pi^2}\int d^3x ~\epsilon^{ijk}~ Tr[L^n_iL^n_jL^n_k]=n~.
 \ee
 as the condition for
 normalization  means that the large transformations
 are given in the  class of functions with the spatial asymptotic
 ${\cal O} (1/r)$.
 Such a function $L^n_i$~(\ref{gnl}) is given by
 \be \label{5class0}
 v^{(n)}(\vec{x})=\exp(n \hat \Phi_0(\vec{x})),~~~~~
 \hat \Phi_0=- i \pi\frac{\lambda_A^a x^a}{r} f_0(r)~,
 \ee
 where the antisymmetric SU(3) matrices are denoted by
 $$\lambda_A^1:=\lambda^2,~\lambda_A^2:=\lambda^5,~\lambda_A^3:=\lambda^7~,$$
 and $r=|\vec x|$.
 The function $f_0(r)$ satisfies the boundary conditions~(\ref{bcf0})
 \be \label{5bcf0}
 f_0(0)=0,~~~~~~~~~~~~~~
 f_0(\infty)=1~.
 \ee
 The functions $L_i^n$ belong to monopole-type class of
 functions. The topological degeneration of QCD means
 that initial data of the transformed physical fields $A_i$
 in~(\ref{gnl}) should be also given in the same
 monopole class of functions.

 We have seen in the previous Section
 that, in the monopole class of functions in the sector
 with a zero value of the winding number functional,
 there is  the gauge Higgs effect. This effect
 leads to a stable  physical vacuum in the form of the $SU(3)$
 BPS monopole~\cite{BPS}
 \be \label{5bps}
 (\hat \Phi^{(0)}_i)^{BPS}=
 - i \frac{\lambda_A^a}{2}\epsilon_{iak}\frac{x^k}{r^2} f_1^{BPS},~~~~~~~
 f^{BPS}_{1}=
 \left[1 - \frac{r}{\epsilon \sinh(r/\epsilon)}\right]~,
 \ee
 where $\epsilon$ is defined by the monopole energy
 \be \label{energy}
 \int\limits_{ }^{ }d^3x [B^a_i(\Phi_k)]^2 =\frac{4\pi}{g^2
 \epsilon}~\Rightarrow~
 \epsilon=\frac{1}{\alpha_s V <B^2>}~,~~~~~\left(\alpha_s=
 \frac{g^2}{4\pi},~~V=\int d^3x\right)~.
 \ee
 The average vacuum magnetic tension
 $\hat {\Phi}^{(0)}_i(\vec{x})= O(\frac{1}{r})$ with a zero winding
 number and
 \be \label{magnet}
 <B^2>=\left[\frac{\int d^3x
 \left(B^a_k(\Phi^{(0)})\right)^2}{\int d^3x}\right]\not= 0
 \ee
 is the main difference of the {\it constraint-shell} formulation
 of non-Abelian theory in the Minkowski space-time from the
 instanton calculation \cite{ins} with functions interpolating between
 two classical vacua with zero energy
 that belong to nonphysical values of spectrum in the quantum theory.

 Any field is a sum
 \be \label{5bar}
 A^{(0)}_i(t,\vec{x}) = {\Phi}^{(0)}_i(\vec{x}) + \bar A^{(0)}_i(t,\vec{x})~,
 \ee
 where $\bar A^{(0)}_i$ is a weak perturbative part
 with the asymptotic at the spatial infinity
 \be \label{5ass1}
 \bar A^{(0)}_i(t,\vec{x})|_{\rm asymptotics} = O(\frac{1}{r^{1+l}})~~~~(l > 1)~.
 \ee
 The Dirac variables distinguish the constraint-shell gauge
 \be\label{gencou}
 D^{bc}_k(\Phi^{(0)}_i)\bar A^{c(0)}_k(t_0,\vec x)=0~.
 \ee
 The topological degeneration
 \be\label{su3}
 \hat A_k^{(n)}=\exp[n \hat \Phi_0(\vec x )]\left(\hat A^{(0)}_k+
 \partial_k \right)\exp[-n \hat \Phi_0(\vec x )]
 \ee
 keeps the constraint-shell gauge~(\ref{gencou}), if there is
 a nontrivial solution of the Gribov equation
 \be\label{}
 D_i^2(\Phi^{(0)}_k)\Phi_{0}=0~.
 \ee
 The nontrivial solution of the Gribov equation
  exists in the form of the $SU(3)$ BPS monopole~\cite{BPS}
 \be \label{5bps0}
 \hat \Phi_0^{BPS}=- i \pi\frac{\lambda_A^a x^a}{r} f^{BPS}_0(r)~,~~~~~~~
 f_0^{BPS}=\left[ \frac{1}{\tanh(r/\epsilon)}-\frac{\epsilon}{r}\right]~.
 \ee
 This solution  has the boundary condition~(\ref{5bcf0}) of a phase of the
 topological transformations~(\ref{5class0}).

 We get the vacuum $\Phi_k^{(0)}$ as an exact solution in a
 form of the Wu-Yang monopole
 in the limit of an infinite volume $\epsilon~\rightarrow~ 0$ at the
 end of the calculation of spectra and matrix elements
 \be\label{5wy}
 \lim\limits_{V\to \infty}f_1^{BPS}=1~.
 \ee

 \subsection{Superfluid dynamics of gluonic liquid}

 We have seen above (in Section 4) the constraint-shell action can contains
 the superfluid quantum motion of the gauge phases
 considered as  a zero mode (${\cal Z}$) of the Gauss law constraint
 \be \label{5gaussd}
 \frac{\delta W}{\delta A_0}=0~~~~~ \Rightarrow
 (D^2(A))^{ac} { A_0}^c = D_i^{ac}(A)\partial_0 A_i^c+ j_0^a~,
 \ee
 where $j_\mu^a=g\bar \psi \frac{\lambda^a}{2} \gamma_\mu\psi$
 is the quark current.
 In lowest order of the perturbation theory this constraint takes the form
 \be \label{5gausspt}
 (D_j^2(\Phi))^{ac} { A_0}^c = D_i^{ac}(\Phi)\partial_0 A_i^c~,
 \ee
  In the considered case of BPS-monopole (\ref{5bps}), the Gribov copies mean
  that  there is
 the zero mode of the covariant Laplace operator in the monopole field
 \be \label{5lap}
 (D^2)^{ab}({\Phi_k^{BPS}})({\Phi}_0^{BPS})^b(\vec{x})=0~.
 \ee

 This is a perturbative form of a zero mode of the Gauss law
 constraint~(\ref{5gaussd})~\cite{p2,vp1} as the solution
 ${\cal Z}^a$  of the homogeneous equation~(\ref{5gaussd})
 \be\label{5zm}
 (D^2(A))^{ab}{\cal Z}^b=0~,
 \ee
 with the asymptotics at the space infinity
 \be \label{5ass}
 \hat {\cal Z}(t,\vec{x})|_{\rm asymptotics}=\dot N(t)\hat \Phi_0(\vec{x})~,
 \ee
 where $\dot N(t)$ is the global variable of an excitation
 of the gluon system as a whole.

 From the mathematical point of view, the zero mode means that
 the general solution of the inhomogeneous equation~(\ref{5gaussd})
 for the time-like component $A_0$
 is a sum of the homogeneous equation~(\ref{5zm})
 and a particular solution
 ${\tilde A}_0^a$ of the inhomogeneous one~(\ref{5gaussd}):
 $A_0^a = {\cal Z}^a + {\tilde A}^a_0$~.

 The zero mode of the Gauss constraint and the
 topological variable $N(t)$ allow us to describe the topological
 degeneration of all fields by the non-Abelian generalization of
 the Dirac dressed variables in QED
 \be \label{5gt1}
 0=U_{\cal Z}(\hat {\cal Z}+\partial_0)U_{\cal Z}^{-1}~,~~~~
 {\hat A}^*_i=U_{\cal Z}({\hat A}^{(0)}+\partial_i)U_{\cal Z}^{-1},~~~
 \psi^*=U_{\cal Z}\psi^{(0)}~,
 \ee
 where the spatial asymptotic of $U_{\cal Z}$ is
 \be \label{5UZ}
 U_{\cal Z}=T\exp[\int\limits^{t} dt'
 \hat {\cal Z}(t',\vec{x})]|_{\rm asymptotics}
 =\exp[N(t)\hat \Phi_0(\vec{x})]=U_{as}^{(N)}~,
 \ee
 and $A^{(0)}=\Phi^{(0)}+\bar A^{(0)},\psi^{(0)}$ are the degeneration free variables.
 In this case, the topological degeneration of all color fields
 converts into the degeneration of only one global topological
 variable $N(t)$ with respect to a shift of this variable on integers:
 $(N~\Rightarrow~ N+n,~ n=\pm 1,\pm 2,...)$.

 One can check~\cite{bpr} that the Pontryagin index for
 the Dirac variables~(\ref{5gt1}) with the
 asymptotic~(\ref{5ass1}),~(\ref{5ass}),~(\ref{5UZ}) is determined
 by only the difference of the final and initial values of
 the topological variable
 \be \label{5pont}
 \nu[A^*]=\frac{g^2}{16\pi^2}\int\limits_{t_{in} }^{t_{out} }dt
 \int\limits_{ }^{ }d^3x G^a_{\mu\nu} {}^*G^{a\mu\nu}=N(t_{out}) -N(t_{in})~.
 \ee
 Thus, we can identify the global variable $N(t)$ with
 the winding number degree of freedom in the Minkowski space.

 The dynamics of physical variables including the topological one
 is determined by the constra\-int-shell action of an equivalent unconstrained
 system  as a sum
 of the zero mode part, and the monopole and perturbative ones
 \be \label{5csa}
 W^*_{l^{(0)}}=W_{Gauss-shell}=W_{\cal Z}[N]+W_{mon}[\Phi_i]+W_{loc}[\bar A]~.
 \ee
 The action for an equivalent unconstrained system~(\ref{5csa}) in the
 gauge
 \be \label{5qcdg}
 D_k^{ac}(\Phi)\bar A^c_k=0~.
 \ee
 with a monopole and a zero mode has been
 obtained  in the paper~\cite{bpr}.
 This action contains the dynamics of the topological variable in the
 form of a free rotator
 \be \label{5ktg}
 W_{\cal Z}=\int\limits_{ }^{ }dt\frac{{\dot N}^2 I}{2};~~~
 I=\int\limits_{V}d^3x(D^{ac}_i(\Phi_k)\Phi^c_0)^2=
 \frac{4\pi}{g^2}(2\pi)^2\epsilon~,
 \ee
 where $\epsilon$ is a size of the BPS monopole considered
 as a parameter of the infrared regularization which disappears
 in the infinite volume limit. The dependence of $\epsilon$ on
 volume can be chosen as $\epsilon \sim V^{-1}$, so that the density of energy
  was finite.

 The perturbation theory in the sector of local excitations $W_{loc}[\bar A]$
 is based  on the Green function in the form of the inverse
 differential operator of the Gauss law
 \be \label{5V}
 [D^2(\Phi)]^{ac}V^{cb}(x,y)=-\delta^3(x-y)\delta^{ab}
 \ee
 which is the non-Abelian generalization of the Coulomb potential.
 As it has been shown in the previous Section
 the non-Abelian Green function
 in the field of the Wu-Yang monopole
 is the sum of a Coulomb-type potential and a rising one~\cite{bpr}.
 This means that the instantaneous quark-quark interaction
 leads to spontaneous chiral symmetry breaking~\cite{yaf,fb},
 Goldstone mesonic bound states~\cite{yaf}, rearrangement of
  perturbation
 series for gluons,  glueballs~\cite{fb}, and
 the Gribov modification of the asymptotic freedom formula.
 If we choose a time-axis $l^{(0)}$
 along the total momentum of bound states~\cite{yaf}
 (this choice is compatible with the experience of QED in the description
 of instantaneous bound states), we get the bilocal generalization of
 the chiral Lagrangian-type mesonic interactions~\cite{yaf}.
 In this case, the U(1) anomalous interaction of $\eta_0$-meson with
 the topological variable lead to additional mass of this
 isoscalar meson~\cite{bpr}. All these effects can be described by
 the Feynman integral.

 \subsection{Feynman path integral}

 The Feynman path integral for the obtained unconstrained system in
 the class of functions of the topological transformations
 takes the form (see~\cite{bpr})
 \bea \label{qcdf}
 Z_F[l^{(0)},J^{a*}]&=&\int\limits_{ }^{ } DN(t)
 \int\limits_{ }^{ }\prod\limits_{c=1 }^{c=8 }
 [d^2A^{c*} d^2 E^{c*}]\prod [d\bar \psi^* d\psi^*]\nonumber\\
 &&\times\exp\left\{iW^*_{l^{(0)}} [N,A^*,E^*,\bar \psi^*,\psi^*]+iS^*
 \right\}~,
 \eea
 where  $W^*$ is given by Eq.~(\ref{5csa}),
 \be
 S^*=\int\limits_{ }^{ }d^4x
  [J^{c*}_{\mu} \cdot A^{c*}_{\mu}+\bar s^*\psi^*+\bar \psi^*s^*]~,
 \ee
 and $J^{c*},\bar s^*,s^*$ are physical sources.

 The action of the local variables does not depend on the Gribov phases
 \be \label{5tcsa}
 W^*_{l^{(0)}}(\bar A^*) + ...=
 W^*_{l^{(0)}}(\bar A^{(0)}) +...~.
 \ee
 Whereas the physical sources of the Dirac dressed variables  $A^*,\psi^*$
 are topologically degenerated. The nonperturbative
 phase factors of the topological degeneration can lead to
 a complete destructive interference of color amplitudes~\cite{p2,pn}
 due to averaging over all parameters of the degenerations.

 The perturbation theory is compatible with only one gauge.
 It is the Coulomb-type gauge in the monopole field~\cite{bpr}.

 The change of variables $A^*$
 concentrates all monopole physics in the form of
 the non-Abelian Dirac factor
 \be \label{dirqcd}
 U(A)=U_{\cal Z}\left\{1+\frac{1}{D^2(\Phi)} D_j(\Phi)\hat A_j+...\right\}
 \ee
 before the "physical" sources.
 The change of "physical" sources $A^*J^*~\Rightarrow~A J$ (that
 is called the transition to another gauge)
 removes all monopole physics, including confinement and hadronization,
 like the similar change in QED
 removes all electrostatic phenomena from the FP integral
 in the relativistic gauges.

 \subsection{A free rotator: topological confinement}

 We have seen in Section 4, that the topology can be the origin of
 color confinement as complete destructive interference of
 the phase factors of the topological degeneration of initial data.

 The mechanical analogue of the topological degeneration of initial data
 is the free rotator $N(t)$ with the action of free particle
 \be\label{rot}
 W(N_{out},N_{in}|t_1)=\int\limits_{0 }^{t_1 } dt \frac{\dot N^2}{2}I,
 ~~p=\dot N I,~~H_0=\frac{p^2}{2I}
 \ee
 given on a ring where the points $N(t)+n$
 ($n$ is integer) are physically equivalent.
 Instead of a initial date $N(t=0)=N_{in}$ in the mechanics in the
 space with the trivial topology, the observer of the rotator
 has the manifold of initial data $N^{(n)}(t=0)=N_{in}+n;~~n=0,\pm 1,\pm
 2,...$.

 An observer does not know where is the rotator. It can be at points
 $N_{in},N_{in}\pm 1,N_{in}\pm 2,N_{in}\pm 3,... $. Therefore, he
 should to average a wave function
 $$
 \Psi (N)=e^{ipN}~.
 $$
 over all values of the topological degeneration
 with the $\theta$-angle measure $\exp (i\theta n)$. In the result
 we obtain the wave function
 \be\label{wf}
 \Psi (N)_{\rm observable}=\lim\limits_{L \to \infty}\frac{1}{2L}
 \sum\limits_{n=-L }^{n=+L }
 e^{i\theta n}\Psi (N+n)=\exp\{i (2\pi k+\theta)N\}~,
 \ee
 where $k$ is integer. In the opposite case $p\not = 2\pi k + \theta$
 the corresponding wave function (i.e., the probability amplitude) disappears
  $\Psi(N)_{\rm observable}=0$ due to the complete destructive interference.

 The consequence of this topological degeneration is that
 a part of values of the momentum spectrum becomes unobservable in
 the comparison with the trivial topology.

 This fact can be treated as confinement of those values which do
 not coincide with the discreet ones
 \be\label{spc}
 p_k=2\pi k + \theta, ~~~~~ 0\leq \theta \leq \pi~.
 \ee
 The observable spectrum follows also
 from the constraint of the equivalence
 of the point $N$ and $N+1$
 \be\label{wf1}
 \Psi (N)=e^{i\theta}\Psi (N+1), ~~~~~~\Psi (N)=e^{ipN}~.
 \ee
 In the result we obtain the spectral decomposition
 of the Green function of the free rotator~(\ref{rot})
 (as the probability amplitude of transition from the point $N_{in}$
 to $N_{out}$) over the observable values of spectrum~(\ref{spc})
 \be \label{rotgf}
 G(N_{out},N_{in}|t_1)\equiv <N_{out}|\exp(-i\hat Ht_1)|N_{in}>
 = \frac{1}{2\pi}\sum\limits_{k=-\infty }^{k=+\infty }\exp\left[
 -i\frac{p_k^2}{2I}t_1+ip_k(N_{out}-N_{in})\right]~.
 \ee
 Using the connection with the Jacobian theta-functions~\cite{poll}
 $$
 \Theta_3 (Z|\tau)=\sum\limits_{k=-\infty }^{k=+\infty }\exp[i\pi k^2\tau
 +2ikZ]=(-i\tau)^{-1/2}\exp[\frac{Z^2}{i\pi\tau}]\Theta_3\left(
 \frac{Z}{\tau|-\frac{1}{\tau}}\right)
 $$
 we can represent expression~(\ref{rotgf}) in a form of the sum over
 all paths
 \be \label{rotgfp}
 G(N_{out},N_{in}|t_1)
 = \sqrt{\frac{I}{(i4\pi t_1)}}\sum\limits_{n=-\infty }^{n=+\infty }
 \exp[i\theta n]\exp\left[+iW(N_{out},N_{in}+n|t_1)\right]~,
 \ee
 where
 $$
 W(N_{out}+n,N_{in}|t_1)=\frac{(N_{out}+n-N_{in})^2I}{2t_1}
 $$
 is the rotator action~(\ref{rot}).

 \subsection{Confinement as destructive interference}

 The similar topological confinement
 as a complete destructive interference of phase factor of
 the topological degeneration (i.e., a pure quantum effect)
 can be in the "classical non-Abelian field theory".
 Recall that at the time of the first paper of Dirac~\cite{cj} (1927)
 the so called "classical relativistic field theory" were
 revealed in the papers of Schr\"odinger, Fock, Klein, Weyl \cite{fock,wh}
 as  a type of {\it relativistic quantum mechanics},
 i.e., the result of the primary quantization. The phase of the gauge
 transformations was introduced by Weyl~\cite{wh}  as a pure quantum quantity.

 The free rotator shows us that  the topological
 degeneracy can be removed, if all Green functions are averaged over values
 of the topological variable and all possible angles of orientation of
 the monopole unit vector ($\vec n=\vec x/r$)~(\ref{ass1}) in the
 group space (instead of the instanton averaging over interpolations
 between different vacua).

 Averaging over all parameters of the degenerations
 can lead to a complete destructive interference of all color
 amplitudes~\cite{in,p2,pn}. In this case, only colorless
 ("hadron") states have to form a complete set of physical states.
 Using the example of a free rotator,
 we have seen that the disappearance of a part of physical states
 due to the topological degeneration (confinement) does not violate
 the composition law for Green functions
  \be \label{compos}
 G_{ij}(t_1,t_3) =\sum\limits_{h }^{ } G_{ih}(t_1,t_2)G_{hj}(t_2,t_3)
 \ee
 defined as the amplitude of the probability to find a system with the
 Hamiltonian $H$
 in a state $j$ at the time $t_3$, if at the time $t_1$  this system was
 in a state $i$, where $(i,j)$ belongs to a complete set of all states
 $\{h\}$:
 $$G_{ij}(t_1,t_3)=<i|\exp(-i\int\limits_{t_1 }^{t_3 }H)|j>
 $$
 The particular case of this composition law~(\ref{compos}) is the unitarity
 of S-matrix
 $$
 SS^+=I~\Rightarrow~\sum\limits_{h }^{ }<i|S|h><h|S^+|j>=<i|j>~
 $$
 known as the law of probability conservation for S-matrix elements
 ($S=I+iT$)
 \be \label{qhd}
 \sum\limits_{h }^{ } <i|T|h> <h|T^*|j> =2 {\rm Im}<i|T|j>~.
 \ee
 The left side of this law is the analogue of the spectral
 series of the free rotator~(\ref{rotgf}).
 The destructive interference keeps only colorless "hadron" states.
 Whereas, the right side of this law
 far from resonances can presented  in a form of the perturbation
 series over the Feynman diagrams that follow from the Hamiltonian.
 Due to gauge invariance  $H[A^{(n)},q^{(n)}]=H[A^{(0)},q^{(0)}]$
 this Hamiltonian  does not depend on the
 Gribov phase factors and it contains the perturbation series in terms
 of only the zero - map  fields (i.e., in terms of constituent color
 particles) that can be identified with the Feynman partons.
 The Feynman path integral as the generation functional
 of this perturbation series is the analogue of the sum over
 all path of the free rotator~(\ref{rotgfp}).

 Therefore,  confinement in the spirit of the
 complete destructive interference of color
 amplitudes \cite{vp1,p2,pn} and
 the law of probability conservation for S-matrix elements~(\ref{qhd})
 leads to
 Feynman quark-hadron duality, that is foundation
 of all the parton model~\cite{feynman}
 and QCD application~\cite{efremov}. The quark-parton
 duality gives the method of direct experimental measurement
 of the quark and gluon quantum numbers from the deep inelastic
 scattering cross-section~\cite{feynman}.
 For example, according to Particle Data Group
 the ratio of the sum of the probabilities of $\tau$-decay
 hadron modes to the probability of $\tau$-decay muon mode is
 $$
 \frac{\sum\limits_{h }^{ }w_{\tau \to h}}{w_{\tau \to \mu}}=3.3\pm 0.3~.
 $$
 This is the left-hand side of Eq.~(\ref{qhd}) normalized to the value of the
 lepton mode probability of $\tau$-decay. On the right-hand side of
 Eq.~(\ref{qhd}) we have the ratio of the imaginary part of the sum of
 quark-gluon diagrams (in terms of constituent fields free from the Gribov
 phase factors) to the one of the lepton diagram. In the lowest order of QCD
 perturbation on the right-hand side we get the number of colors $N_c$ and,
 therefore,
 $$
 3.3\pm 0.3=N_c~.
 $$
 Thus in the constraint-shell QCD we can understand not only
 "why we do not see quarks", but also "why we can measure their
 quantum numbers".
 This mechanism
 of confinement due to the quantum interference of phase factors (revealed
 by the explicit resolving the Gauss law constraint~\cite{p2,pn,in})
 disappears after the change of "physical" sources
 $A^*J^*~\Rightarrow~A J$ that
 is called the transition to another gauge in the gauge-fixing method.

 \subsection{Relativistic equations for gluonic systems}


 The Wu-Yang monopole
 $$
 \partial_0^2  {\bar A}^c_i=[\delta_{ij}\left(D_k^2(\Phi)\right)^{cd} -
 2gG^a_{ij}(\Phi)f^{acd}] \bar A^d_j
 $$
 does not change qualitatively the spectrum of gluon excitations in
 the comparison with the naive perturbation theory
 $$
 \partial_0^2  {\bar A}^c_i=\partial_k^2 {\bar A}^c_i~.
 $$
 There are only the spin-color mixing~\cite{psm}.
 Here we omit this mixing
 and consider mainly the instantaneous current-current interaction
 described by the $QCD$  Hamiltonian \cite{fb,fb1}
 \begin{eqnarray} \label{5-1}
 H_{YM} &= &
 \int d{\bf x}
 {1\over 2}
 [ (E_{i}^{Ta})^{2}  + ( B_{i}^{a} )^{2} ]  + \nonumber \\
 &+& {1\over 2}
 \int d{\bf x} \int d{\bf y}
 f^{ b_{1}c_{1}d_{1}}
 E_{i}^{Tc_{1}}( {\bf x} )
 A_{i}^{Td_{1}}( {\bf x} )
 V^{b_{1}b_{2}} ( {\bf x}  - {\bf y} )
 f^{ b_{2}c_{2}d_{2}}
 E_{j}^{Tc_{2}} ( {\bf y} ) A_{j}^{Td_{2}}( {\bf y} ) + \nonumber
 \end{eqnarray}
 Here $ V( {\bf x} )$ denotes the sum of rising potentials induced by
 the Wu-Yang monopole.

 Let us represent the gluon fields as Bogolubov expansion in
 creation and annihilation operators
 \begin{eqnarray}  \label{5-2}
 E_{i}^{Tb}( {\bf x} ) & = & i
 \int { d{\bf p} \over (2\pi)^{3/2} }
 \sqrt{ { \phi( {\bf p} ) \over 2 } } [
 a_{\alpha}^{(+)b}({\bf p}) e_{i}^{\alpha} -
 a_{\alpha}^{(-)b}({\bf p}) e_{i}^{\alpha} ] e^{i {\bf px}}, \nonumber \\
 A_{i}^{Tb}( {\bf x} ) & = &
 \int { d{\bf p} \over (2\pi)^{3/2} }
 {1 \over { \sqrt{ 2 \phi({\bf p}) } }} [
 a_{\alpha}^{(+)b}({\bf p}) e_{i}^{\alpha} +
 a_{\alpha}^{(-)b}({\bf p}) e_{i}^{\alpha} ] e^{ - i {\bf px}},
 \end{eqnarray}
 where the function $ \phi({\bf p}) $ is to be calculated from the diagonalization
 condition of the Hamiltonian~(\ref{5-1}) with respect to the operators
 $ a^{(+)} , a^{(-)} $.

 In the normal form, the gluon Hamiltonian reads as
 \begin{eqnarray} \label{5-3}
 {\it H} = E_{0} + \int d{\bf k} [ &(&
 a_{\alpha_{1} }^{(+)}(-{\bf k}) a_{\alpha_{2} }^{(+)}( {\bf k}) +
 a_{\alpha_{1} }^{(-)}( {\bf k}) a_{\alpha_{2} }^{(-)}(-{\bf k})  )
 { {\it C}^{\alpha_{1}\alpha_{2}} (\phi ) \over 2 } + \nonumber \\ &+&
 a_{\alpha_{1} }^{(+)} a_{\alpha_{2} }^{(-)}
 \omega^{\alpha_{1} \alpha_{2} } (\phi ) + {\it O}( a^{4}) ] .
 \end{eqnarray}
 Here ${\it C}(\phi)$ and $\omega (\phi)$ are some defined functions
 the dependence on $\phi$ of which is determined from the diagonalization
 of the Hamiltonian~(\ref{5-3}).

 The diagonalization condition for the Hamiltonian~(\ref{5-3}) means
 that the coefficients ${\it C}(\phi)$ vanish
 \begin{eqnarray} \label{5-4}
 {\it C}(\phi)=0
 \end{eqnarray}
 For the solutions~(\ref{5-4}) the function $ \omega (\phi) $ defines
 the gluon energy spectrum in the same way as the Schwinger-Dyson equation
  defines the energy spectrum for the quarks.

 Let us illustrate this scheme by calculating the one-particle
 energy of the gluon and its bound states for the simplest example of
 the theory~(\ref{5-2})  where the operator for the potential is substituted
 by an effective potential. This means we consider the sum of the free
 Hamiltonian
 \begin{eqnarray*}
 H_{0} =
 {1\over 2}
 \int d{\bf x}
 [ (E_{i}^{T})^{2}  + ( \partial_{i}  A_{j}^{T} )^{2} ]  ,
 \end{eqnarray*}
 and the Hamiltonian of potential interaction between colour gluon
 currents
 \begin{eqnarray*}
 {\it H}_{I} &=& - {1\over 8} \int
 d{\bf p}_{1} d{\bf q}_{1} d{\bf p}_{2} d{\bf q}_{2}
 \delta(
 {\bf p}_{1}
 -{\bf p}_{2}+{\bf q}_{1}-{\bf q}_{2} )
 { \underline{V}({\bf p}_{1}-{\bf p}_{2}) \over (2\pi)^{3} }
 \sqrt{ {\phi( {\bf p}_{1} ) \phi( {\bf q}_{1} ) \over {\phi( {\bf p}_{2} )
 \phi( {\bf q}_{2} ) }} } \cdot   \\   &\cdot&
 f^{ab_{1}b_{2}}
 f^{ac_{1}c_{2}} [ a^{(+)}(-1)-a^{(-)}(-1) ]
        [ a^{(+)}(2) +a^{(-)}(-2) ] \cdot \\ & \cdot &
        [ a^{(+)}(-1^{\prime})-a^{(-)}(1^{\prime}) ]
        [ a^{(+)}(2^{\prime}) +a^{(-)}(-2^{\prime})].
 \end{eqnarray*}
 Here, the short - hand notation $
 (\pm {\bf p}_{1} , b_{1} , i_{1} ) = ( \pm 1 ),
 (\pm {\bf q}_{1} , c_{1} , j_{1} ) = ( \pm 2 ),        \\
 (\pm {\bf p}_{2} , b_{2} , i_{2} ) = ( \pm 1^{\prime} ),
 (\pm {\bf q}_{2} , c_{2} , j_{2} ) = ( \pm 2^{\prime} )
 $ has been used. In our case, the coefficients $ {\it C} $ and $ \omega $
 are given by the expressions
 \begin{eqnarray}
 {\it C}^{\alpha_{1} \alpha_{2}}(\phi ) &=&
 [ { {\bf k}^{2} \delta_{ij} + ( {\cal M}^{2}({\bf k}) )^{ij}
 \over {2\phi} } - { \phi \over 2} z^{ij} ]
 e^{\alpha_{1}}_{i}({\bf k}) e^{\alpha_{2}}_{j}(-{\bf k}) = 0 , \nonumber \\
 {\omega}^{\alpha_{1} \alpha_{2}}(\phi ) &=&
 [ { {\bf k}^{2} \delta_{ij} + (
 {\cal M}^{2}({\bf k}) )^{ij}
 \over {2\phi} } + { \phi \over 2} z^{ij} ]
 e^{\alpha_{1}}_{i}({\bf k})
 e^{\alpha_{2}}_{j}({\bf k})
 \end{eqnarray}
 with
 \begin{eqnarray}
 {\cal M}^{2}({\bf k}) )^{ij}
 &=& {N_{c} \over 2}
 \int { d{\bf q} \over (2\pi)^{3} } \underline{V}( {\bf k}-{\bf q})
 \phi ({\bf q})
 ( \delta_{ij} -q_{i} { 1 \over {\bf q}^{2} }q_{j} )  , \nonumber \\
 z^{ij}({\bf p}) & = & \delta_{ij} +
 {N_{c} \over 2}
 \int { d{\bf q} \over (2\pi)^{3} } \underline{V}( {\bf k}-{\bf q})
 { 1 \over {\phi ({\bf q})}}
 ( \delta_{ij} -q_{i} { 1 \over {\bf q}^{2} }q_{j} )  .
 \end{eqnarray}

 Since two gluons form the simplest bound state, by analogue
 with the mesons, we introduce a glueball creation operator
 \begin{eqnarray*}
 G^{+} =
 \sum_{b} \int
 d{\bf k}
 &[&
 X^{(++)}_{\gamma_{1} \gamma_{2} }({\bf k})
 e^{\gamma_{1}}_{i}({\bf k}) e^{\gamma_{2}}_{j}(-{\bf k})
 a^{(+)b}_{\gamma_{1}}({\bf k}) a^{(+)b}_{\gamma_{2}}(-{\bf k})  + \nonumber \\
 &+&
 X^{(--)}_{\gamma_{1} \gamma_{2} } ({\bf k})
 e^{\gamma_{1}}_{i}({\bf k}) e^{\gamma_{2}}_{j}(-{\bf k})
 a^{(-)b}_{\gamma_{1}}({\bf k}) a^{(-)b}_{\gamma_{2}}(-{\bf k}) ]
 \end{eqnarray*}
 and a "coherent" vacuum
 \begin{eqnarray*}
 \vert 0 >>_{\alpha} = \exp \{
 \sum_{c} &\int &
 d{\bf k}_{1} d{\bf k}_{2}
 \alpha({\bf k}_{1},{\bf k}_{2} )
 ( a^{(+)c}_{\gamma_{1}}({\bf k}_{1}) a^{(+)c}_{\gamma_{2}}({\bf k}_{2}) )
 \cdot \\ &\cdot&
 ( a^{(+)b}_{\gamma_{1}}({\bf k}_{1}) a^{(+)b}_{\gamma_{2}}({\bf k}_{2}) ) \}
 \vert 0 > .
 \end{eqnarray*}
 Then, the Schr\" odinger equation for eigenvalues of the Hamiltonian
 operator
 \begin{eqnarray}\label{5-60}
 { }_{\alpha}<< 0 \vert GHG^{+} \vert 0 >>_{\alpha} =
 M_{G} \,\, { }_{\alpha}<< 0 \vert GG^{+} \vert 0 >>_{\alpha}   ,
 \end{eqnarray}
 is equivalent  to the equation for the glueball wave function
 \begin{eqnarray} \label{5-61}
 ( 2 \omega ({\bf k}) &-& M_{G} )
 X^{(++)}_{ij}({\bf k})  \nonumber \\
 &=& {N_{c} \over 4}
 \hat{I}_{ {\bf kq}}
 \{
 ( W^{+}(
 {\bf q} \vert {\bf k}
 ) )^{2}
 X^{(++)}_{ij}({\bf q}) -
 ( W^{-}( {\bf q} \vert {\bf k}) )^{2}
 X^{(++)}_{ij}({\bf q})  \}          \nonumber \\      \\
 ( 2 \omega ({\bf k}) &+& M_{G} )
 X^{(--)}_{ij}({\bf k})  \nonumber \\
 &=& {N_{c} \over 4} \hat{I}_{ {\bf kq}}
 \{
 ( W^{+}( {\bf q} \vert {\bf k}) )^{2}
 X^{(--)}_{ij}({\bf q}) -
 ( W^{-}( {\bf q} \vert {\bf k}) )^{2}
 X^{(--)}_{ij}({\bf q})  \}          \nonumber \\
  \end{eqnarray}
 with
 \begin{eqnarray}   \label{5-63}
 W^{\pm} ( {\bf q} \vert {\bf k}) = [
 \sqrt { \phi ({\bf q}) \over \phi ({\bf k}) }
 \pm
 \sqrt { \phi ({\bf k}) \over \phi ({\bf q}) } ] , \nonumber \\
 \hat{I}_{ {\bf kq}} f({\bf q}) =
 \int {d{\bf q} \over (2\pi)^{3}} \underline {V} ({\bf k}-{\bf q}) f({\bf q}) .
 \end{eqnarray}
 Furthermore, one can write also the
 equation for the wave function of a antisymmetric three - gluon state
 $ \psi_{+++} ( \psi_{+--}, \psi_{-++} , \psi_{-+-} ) $ with
 eigenvalues $ M_{BG}$
 \begin{eqnarray}
 [
 \omega (1) &+& \omega (2) + \omega (3) - M_{BG} ]
 \psi_{+++}(1,2,3) =  \nonumber \\
 &=& {N_{c} \over 2} \{
 \hat{I}_{1, \underline{2} } (
 \omega^{+}_{1 \underline{1}} \omega^{+}_{2 \underline{2}}
 \psi_{+++}( \underline{1}, \underline{2},3) +
 \omega^{-}_{1 \underline{1}} \omega^{-}_{2 \underline{2}}
 \psi_{--+}( \underline{1}, \underline{2},3) ) + \,\, ... \,\, \}.
 \end{eqnarray}

 For an estimate of the solution to equations~(\ref{5-61}),~(\ref{5-63})
  we make use of the separable approximation for the potential
 \begin{eqnarray}
 \int
 { d{\bf q} \over (2\pi)^{3} }
 \underline{V}( {\bf k}-{\bf q})
 \phi ({\bf q})
 ( \delta_{ij} -q_{i} { 1 \over {\bf q}^{2} }q_{j} )  \simeq
 {2\over 3} { 1 \over { \mu^{2}_{QCD} } }
 \int^{L}_{0}
 { d{\bf q} \over (2\pi)^{3} }
 \phi ({\bf q}) \delta_{ij},
 \end{eqnarray}
 with the parameters  $L = 1.6
 GeV ,  \mu_{QCD} =  0.35  GeV $. Equation~(\ref{5-60}) take the form
 \begin{eqnarray}
 \sqrt {Z}
 m^{2}_{g}
 &=& {1 \over \mu^{2}_{QCD}}
 \int^{L}_{0}
 { d{\bf p} \over (2\pi)^{3} }
 \sqrt{ {\bf p}^{2}+m^{2}_{g}}, \nonumber \\
 Z &=& 1+{\sqrt{Z} \over \mu^{2}_{QCD}}
 \int^{L}_{0}
 { d{\bf p} \over (2\pi)^{3} }
 {1 \over {\sqrt{ {\bf p}^{2} + m^{2}_{g}}}},  \\
 \omega ({\bf p})  &=& \sqrt{Z}
 \sqrt{ {\bf p}^{2} - m^{2}_{g}} \nonumber
 \end{eqnarray}
 and have the solutions $ m_{g} \simeq 0.8 GeV , \sqrt{Z} \simeq 1.18 $.
 For the scalar glueball mass as an eigenvalue of equations~(\ref{5-61}),
 (\ref{5-63}) in the separable approximation
 \begin{eqnarray*}
 [ 2 \sqrt{Z} \sqrt{ {\bf k}^{2} + m_{g}^{2} } & - & M_{G} ] X =
 \nonumber \\
 &=&
 { N_{c} \over { 4 \mu^{2}_{QCD}}}
 \int^{L} { d {\bf q} \over (2\pi)^{3} }
 [ ( X - Y ) W ( {\bf k} \vert {\bf q} ) + 2 ( X + Y ) ],
 \end{eqnarray*}
 \begin{eqnarray}
  [ 2 \sqrt {Z} \sqrt { {\bf k}^{2} + m_{g}^{2} }  & + &  M_{G} ] X =  \nonumber \\
 &=&
 { N_{c} \over { 4 \mu^{2}_{QCD}}}
 \int^{L} { d {\bf q} \over (2\pi)^{3} }
 [ ( Y - X ) W ( {\bf k} \vert {\bf q} ) + 2 ( X + Y ) ],  \\
  W ( {\bf k} \vert {\bf q} ) & = &
 \sqrt {  {\bf k}^{2} + m_{g}^{2}  \over
 { {\bf q}^{2} + m_{g}^{2} } } +
 \sqrt {  {\bf q}^{2} + m_{g}^{2}  \over { {\bf k}^{2} + m_{g}^{2} } } ,
 \nonumber
 \end{eqnarray}
 one obtains the value $ M_{G} \simeq 1.6 GeV .$

 The Green function
 for the transversal gluon  corresponding to equations~(\ref{5-3})
 and~(\ref{5-4}) is given by
 \begin{eqnarray}
 {\cal D}_{ij}(q_{0},{\bf q})={\omega ({\bf q}) \over { \phi ({\bf q})}}
 {1 \over {q_{0}^{2}-\omega^{2}({\bf q})- i \epsilon }}
 ( \delta_{ij} -q_{i} { 1 \over {\bf q}^{2} }q_{j} ) .
 \end{eqnarray}
 From its meaning the quantity
 ${ \omega \over \phi }= Z({\bf q}) $ can be called the
 infrared renormalization constant of the wave function.

 The Green functions for the constituent quarks  and gluons
 are elements of a new quasiparticle perturbation theory in terms of
 which all matrix elements are calculated including the "running"
 coupling constant~\cite{fb}.

 The phenomenon of dimensional transmutation appearing in the
 "running" coupling constant should be investigated in accordance with
 the logic of quantum theory at the stage of defining the quark and gluon
 energy spectrum and their one-particle Green functions.
 We have seen before that in QCD like in QED the infrared parameters
 appear as nontrivial boundary conditions of the equations of motion.

 The appearance of constituent masses for quarks and gluons does
 of course influence the determination of the "running" coupling
 constant which in the new theory cannot have any singularities in the
 whole Euclidean region of the transferred momenta, among them also
 at $ {\bf q}^{2}=0 $ \cite{fb,a6,a31}. Thus, the rising potential
 rearranges the perturbation theory and makes it stable.

 \subsection{Hadronization  and chiral Lagrangian limit}

 One of our supposition was that the main role in the formation of
 instantaneous bound state in gauge theories
 plays the "instantaneous" interaction. In QCD this interaction
 \begin{eqnarray}  \label{5-38}
 { W}_{{Ins.}}
 &=& -\frac{1}{2} \int d^4x d^4y
 {\psi}^{j_2}_{\beta_2}(y)
 \bar{\psi}^{i_1}_{\alpha_1}(x)
 {\cal K}^{i_1 i_2;j_1 j_2 }_{ \alpha_1 \alpha_2 ; \beta_1 \beta_2 }
 (z \vert X)
 {\psi}^{j_1}_{\beta_1}(x)
 \bar{\psi}^{i_2}_{\alpha_2}(y) \\
 & \equiv& - \frac{1}{2} ( \psi \bar{\psi}, {\cal K} \psi \bar{\psi} )
 \,\,\, , \nonumber
 \end{eqnarray}
 is determined by the Green function~(\ref{5V}) as the inverse
 differential operator of the Gauss law constraint
 $G^{ab}$ in the field of the Wu-Yang monopole
 \begin{eqnarray}
 {\cal K}^{i_1 i_2;j_1 j_2 }_{ \alpha_1 \alpha_2 ; \beta_1 \beta_2  }
 ( z \vert X ) =
 (\frac{\lambda^a}{2})^{i_1 i_2}
 \rlap/\eta_{\alpha_1 \alpha_2}G^{ab}(z\vert X)
 \delta(z \cdot \eta) \delta({\bf z}^{\perp} )
 \rlap/\eta_{\beta_1 \beta_2}
 (\frac{\lambda^b}{2})^{j_1 j_2} \,\,\, ,
  \end{eqnarray}
 which is the non-Abelian generalization of the Coulomb potential.
 We have seen above in previous Section that
 the non-Abelian Green function
 in the field of the Wu-Yang monopole
 is the sum of a Coulomb-type potential and a rising one~\cite{bpr}.

 The procedure of the hadronization, as we have seen in Section 3
  consists in the application of the Legendre transformation with
 bilocal field ${\cal M}(x,y)={\cal M}(z \vert X)  $
 \begin{eqnarray}
 {1 \over 2} &\int &d^{4}x d^{4}y  ( \psi(y) \bar{\psi}(x) )
 {\cal K}(x,y )
 ( \psi(x) \bar{\psi}(y) )  = \nonumber \\  =
 -{1 \over 2} &\int &d^{4}x d^{4}y  {\cal M}(x,y)
 {\cal K}^{-1}(x,y) {\cal M}(x,y)
 + \\  +  &\int &d^{4}x d^{4}y ( \psi(x) \bar{\psi}(y) ) {\cal M}(x,y) \nonumber
 \end{eqnarray}
 where $ {\cal K}^{-1} $ is
 the inverse of the kernel~(\ref{5-38}).
 In the short- hand notation~(\ref{short})
 \begin{eqnarray*}
 \int d^{4}x \bar{\psi}(x) ( i
 \rlap/D(A^{\perp})
 - m^{0} ) \psi (x) &\equiv &
 \int d^{4}x d^{4}y \psi (y) \bar{\psi}(x) ( i
 \rlap/D(A^{\perp})  - m^{0} )
 \delta  (x-y)       \nonumber \\
 &=&  ( \psi \bar{\psi} , - G^{-1}(A^{\perp}) ) \, \, ,  \\
 \int d^{4}x d^{4}y
 ( \psi(x) \bar{\psi}(y) ) {\cal M}(x,y) &=& ( \psi \bar{\psi},
 {\cal M} )
 \end{eqnarray*}
 the QCD  action is represented in the form
 \begin{eqnarray}
 W_{{QCD}} = W_g(A^{\perp}) + W [\psi, {\cal M} \vert A^{\perp}] \,\,\, ,
 \end{eqnarray}
 where
 \begin{eqnarray}
 W [\psi, {\cal M} \vert A^{\perp}]  = ( \psi \bar{\psi},
 ( - G^{-1} (A^{\perp}) + {\cal M}) ) -
 { 1 \over 2} ( {\cal M}, {\cal K}^{-1} {\cal M} ) \,\,\,  ,
 \end{eqnarray}
 and $W_g(A^{\perp}) $ is the action for transverse gluons.
 After quantization ( or integration )
 over quark  fields and normal ordering, the
 action  takes the form
 \begin{eqnarray} \label{5-46}
 W [{\cal M}\vert A^{\perp}] = - {1 \over 2}
 ( {\cal M}, {\cal K}^{-1} {\cal M} )
 +
 i \mbox{Tr Ln }[ G_{\cal M}(A^{\perp}) ]
 \end{eqnarray}
 where the Green function $ G_{\cal M}(A^{\perp})  $ is defined by
 equation
 \begin{eqnarray}
 \bigl[
 G(A^{\perp})^{-1} - {\cal M}
 \bigr]
 G_{\cal M}(A^{\perp})        = I
 \end{eqnarray}
 or
 \begin{eqnarray}
 \int  d^{4}y  \Bigl[
 ( i \rlap/D (A^{\perp})  - m^{0} )
 \delta^4 (x-y) - {\cal M}(x,y)
 \Bigr]
 G_{\cal M}(y,z \vert A^{\perp})
  = \delta^4 (x-z)
 \end{eqnarray}

 As a result of such quantization, only
 the the contributions with inner fermionic lines (but no scattering
 and dissociation channel contribution) are included in the effective action
 since we are interested only in the bound states.

 Finally, we got the following result
 \begin{eqnarray}
 Z [ \ldots ] =
 < \int  d {\cal M} e^{ \{ i W_{\mbox{eff}}({\cal M}\vert A^{\perp})
 + \ldots \} }  >_{A^{\perp}}
 \end{eqnarray}
 here $< F (A^{\perp}) >_{A^{\perp}}$  is the Feynman integral~(\ref{qcdf})
 depended on the time- axis
 $\eta_\mu$ according to Markov- Yukawa prescription.

 The determination of a minimum of the action~(\ref{5-46}) leads
 to the SD equation
 \begin{eqnarray}
 { \delta W_{eff} ({\cal M}) \over { \delta {\cal M}} } =  0
 \rightarrow
 \Sigma(A^{\perp}) = i {\cal K} G_{\Sigma}(A^{\perp})~.
 \end{eqnarray}
 We denoted the corresponding classical solution for the bilocal field
 by $ \Sigma $.

 The next step is the expansion of the action~(\ref{5-46})
 around the point of minimum
 $ {\cal M} = \Sigma + {\cal M}^{\prime} $ ,
 \begin{eqnarray}
 W_{eff} ( \Sigma + {\cal M}^{\prime} ) &= &
 W_{eff}(\Sigma) + [ - {1\over2} {\cal M}^{\prime} {\cal K}^{-1} {\cal M}^{\prime}
 + { i \over 2} ( G_{\Sigma} {\cal M}^{\prime} )^{2} ] + \nonumber \\
 & + & i  \sum_{n=3}^{\infty} {1\over n} ( G_{\Sigma} {\cal M}^{\prime} )^{n},
 \, \, \,  ,
 \end{eqnarray}
 where
 \begin{eqnarray}
    G {\cal M}(x,y) &=& \int d^{4}z G (x,z) {\cal M}(z,y)
 =  \Phi (x,y)   , \nonumber \\
 \Phi^{2} &=& \mbox{tr}\int d^{4} x d^{4}y \Phi(x,y) \Phi(y,x) , \\
 \Phi^{3} &=& \mbox{tr}\int d^{4} x d^{4}y d^{4}z
 \Phi(x,y) \Phi(y,z) \Phi(z,x) \,\,  ,
 \ldots  \,\,\, . \nonumber
 \end{eqnarray}
 The small fluctuations $ {\cal M}^{\prime} $  we represent as
 a sum over the complete set of classical solutions $ \Gamma $,
 \begin{eqnarray}
 { \delta^{2}W_{Q} ( \Sigma + {\cal M}^{\prime} ) \over { \delta {\cal M}^{2}} }
 \vert_{ {\cal M}^{\prime} = 0 }  \cdot \Gamma = 0~.
 \end{eqnarray}

 The bound state function $\Gamma (x,y \vert A^{\perp})$ satisfies the
 Bethe- Salpeter equation in the form
 \begin{eqnarray}
 \Gamma (x,y \vert A^{\perp})
 = i {\cal K} (x,y)
 \int d^{4}z_{1} d^{4}z_{2} G_{\Sigma}
 ( x, z_1 \vert A^{\perp}  )
 \Gamma
 ( z_1 z_2  \vert A^{\perp}   ) G_{\Sigma}( z_2,y \vert A^{\perp} ) \,\,\, .
 \end{eqnarray}
 By the same way we can construct  any conventional vertex function
 \begin{eqnarray}
 \Gamma_C (x,y \vert A^{\perp})
 = i {\cal K} (x,y)
 \int d^{4}z_{1} d^{4}z_{2} G_{\Sigma}
 ( x, z_1 \vert A^{\perp}  )
 \gamma_C
 ( z_1 z_2  \vert A^{\perp}   ) G_{\Sigma}( z_2,y \vert A^{\perp} ) \,\,\, .
 \end{eqnarray}
 The $\gamma_C$ is one of the complete  set of  the Dirac  matrices.

 The obtained expression are the exact equations for the description
 heavy and light meson in QCD in which we may to try to separate
 the radiation corrections (retardations) from  bound state effects.
 All radiation corrections  are separated on two parts:
 the "retardation"  and the "instantaneous"
 (forming the instantaneous potential).

 We can consider the obtained expressions in the lowest order in the
 retardation for the colourless hadron
 \begin{eqnarray}
 Z  =
 \int  d {\cal M}
 e^{ \{ i W_{{eff}}({\cal M}) \} } \,\,\, ,
 \end{eqnarray}
 and
 \begin{eqnarray}
 W [{\cal M}] = - {1 \over 2} N_c ( {\cal M},
 { {\cal K}}^{-1} {\cal M} )
 + i N_c \sum_{n=1}^{\infty} {1 \over n} \underline{\Phi}^{n} \,\,\, ,
 \end{eqnarray}
 where $
 {\underline{\cal K}}$ and $\underline{\Phi} $ are the colourless
 operators
 \begin{eqnarray}
 \underline{{\cal K}} (z \vert X) =
 \frac{N_c^2-1}{2N_c}
 \rlap/\eta V_{\mbox{eff}} ({\bf z}^{\perp}) \delta(z \cdot \eta)
 \rlap/\eta  \,\,\, .
 \end{eqnarray}

 In this approach, we get the standard
 Schwinger- Dyson (SD) and Bethe- Salpeter (BS)
 equations:
 \begin{eqnarray}
 \Sigma(x-y) = m^{0} \delta^{(4)} (x-y) +
 i\underline{\cal K}(x,y) G_{\Sigma}(x-y),
 \end{eqnarray}
 \begin{eqnarray}
 \Gamma = i
 \underline{\cal K}(x,y) \int d^{4}z_{1} d^{4}z_{2} G_{\Sigma}(x-z_{1})
 \Gamma(z_{1},z_{2}) G_{\Sigma}(z_{2}-y)
 \end{eqnarray}
 They describe the spectrum of Dirac particles in bound states and the
 spectrum of the bound states themselves, respectively, like in
 Sections 3.3.-3-5..

 The relativistic bound state equations
 derived from the constraint-shell gauge theory with a rising potential
 contains the pure quantum effect of rearrangement of the ground state
 with spontaneous breakdown of chiral symmetry.
 This effect  leads to the low
 - energy theory of light mesons, in which the pion is considered
 in two different ways, as a quark - antiquark bound state and as a
 Goldstone particle.

 There is a number of paper (cf. \cite{yaf,a20,fb1,a6,a1,a2,puz,fin,a3,a4}
 and references therein) where SD and BS equations are used for
 the calculation of the mass spectrum of light mesons, the
 constituent quark masses and the meson decay constants. In the
 papers the potentials are determined from the spectroscopy of
 heavy quarkonia as sums of rising and Coulomb potentials, for
 instance \cite{a3}
 \begin{eqnarray}
 V(r) = { \alpha_{S} \over r} - V_{0} r^{2},~~~~~~ V_{0}^{1/3} \simeq 250
 MeV,~~~~~~~~~~~~ \alpha_{S} \simeq 0.3 \,\,  .
 \end{eqnarray}
 Thereby, the heavy quarkonia ($ m^{0} >> 250 MeV $) themselves
 are described by Schr{\"o}\-dinger equation  which, as has
 been shown, can be derived from BS equation in the limit of large
 masses. In this limit, the effect of spontaneous breakdown of
 chiral symmetry also disappears, and the constituent quark masses
 are identical to the current ones \cite{yaf,puz}.

 The advantage of such a potential approach compared with all other
 ones consists in the first constructive connection between the
 fundamental parameters for physics at short distances (the
 parameters of rising and Coulomb potentials and the current quark
 masses) with those of hadron physics for long distances (the pion
 mass and its weak decay constant $ F_{\pi} $).

 The shortcomings of this approach were the following: the
 nonrelativistic formulation (in the rest frame) of the bound
 state, the absence of an relativistic meson interaction
 Hamiltonian, and the open problem of the status of radiative $ QCD
 $ corrections. The first two disadvantages are absent in the new
 relativistic potential model \cite{yaf,fb1} considered here
 on the basis of the Dirac variables and the Markov-Yukawa
 prescription. This model
 represents a logical interpretation of relativistic atomic
 physics, i.e. an interpretation of the "atomization" of $ QED $.

 From this point of view the "hadronization" of $ QCD $
 qualitatively differs only by the short - range property of the
 quark - antiquark interaction potential for light quarkonia.
 Furthermore, the effective action for light mesons must be an
 action for a chiral Lagrangian~\cite{b4}. The proof of the fact that the
 effective bilocal action
 leads to a chiral Lagrangian has been performed in \cite{a20} with
 the help of the separable approximation which can be used just for
 short - range potentials. For low orbital momenta such potentials
 can be represented with good accuracy as a product of two factors
 \begin{eqnarray*}
 < l=0 \vert V( { p}-{ q})^{\perp} \vert l=0 > = f( p^{\perp})
 f(q^{\perp}) .
 \end{eqnarray*}
 The underlying bilocal model  becomes equivalent to one of versions of
 the Nambu - Jona - Lasinio model \cite {a22,a21} with explicit
 indication of the form factor $ f^{2}(p) $ for the ultraviolet
 regularization. It is well known \cite{a22,a21}, that this model
 leads to chiral Lagrangians.

 The validity of the separable approximation for short - range
 potentials explains the fact of the weak dependence of  low -
 energy physics for light mesons on the form of the potential.
 Therefore, there exists a number of models yielding  a
 satisfactory description of the experimental data.

 Here, one should mention also papers dealing with the derivation
 of nonlinear chiral Lagrangians from $ QCD $ (cf.ref. in
 \cite{a22}). The essence of those proofs consists in a formal
 derivation of this Lagrangian by means of chiral
 transformation which are parameterized by the meson field. Thereby,
 in many cases no derivation of the equation for the meson spectrum
 is  given to say nothing of its solution. The main aim of these
 papers is to find the coefficients in higher order terms in the
 expansion of chiral Lagrangians in meson momenta and to establish
 the description of baryons in the form of "skyrmions" \cite{b31}.
 All these papers concerning the justification of chiral Lagrangian
 from $QCD$ are not devoted to the determination of essential
 parameters of the low - energy physics ( $ F_{\pi}, F_{K},
 m_{\pi}, \,\, ... $) from $QCD$.

 The relativistic bilocal model for atoms and hadrons  as
 compared with the above - mentioned popular non - relativistic
 \cite{a1,a3} and nonlinear \cite{a22} approaches unifies aspects of
 both approaches and gives a constructive generalization of chiral
 Lagrangians to heavy quarkonium physics, i.e. it allows to
 describe decays of heavy quarkonia into light ones in the
 framework of the bilocal action  with a minimal
 number of parameters, defined in the short - range region where
 the perturbation theory begins to work.

 The construction of such a quantum relativistic hadron theory  has
 been given in papers \cite{yaf,a6,puz}.

 \subsection{The low energy theorems}

 Let us check the existence of low energy theorems,
 in the bilocal approach~\cite{yaf,a20,fb1,a6,walter}.
 The low energy theorems of hadron physics are independent of any model of
 hadrons. Consequently, they play a role of a pure tool to test the
 QCD-inspired
 models. These models  (for example, the QCD sum rules ~\cite{a1},
 Nambu - Jona - Lasinio (NJL) model~\cite{a21}, QCD lattice
 calculations~\cite{a3}, nonrelativistic quark potential
 models~\cite{fin,a4,puz}) relate the PCAC in hadron physics,
 induced by the spontaneous chiral symmetry
 breakdown, to the quark condensate. It should be noted that in all these
 quark models the definition of the local quark condensate is used.

 In the present section we discuss the validity of the notion of the local
 condensate in the quark potential model within the low energy theorems.
 Since a wide class
 of nonlocal potentials used in the model lead to equations that are
 unsolvable by an analytical method,
 we investigate the problem by means of the separable approximation.
 We calculate the amplitudes of leptonic and anomalous decays of mesons
 and consider the possibility of the generalization to
 heavy quark limit.

 As it was shown above the "constituent" quarks and their bound states
 are described by the following SD and BS coupled equations:
 $$
 \left\{\begin{array}{c}
 E(p)\cos{2\vartheta(p)}  = m^0 + \frac{1}{2}\hat{I}_q
 \left[\cos{2\vartheta(p)}\right] \qquad (1a) \\
 E(p)\sin{2\vartheta(p)} = (\mid {\bf p}\mid) + \frac{1}{2}\hat{I}_q
 \left[\hat{p}\hat{q}\sin{2\vartheta(p)}\right], \qquad (2a) \end{array}\right.
 $$
 \begin{equation}\left\{\begin{array}{c}
 M_{H}L_2(p) = E_t(p)L_1(p)-\hat{I}_q\left\{\left[c^{(-)}_pc_q^{(-)}
 +\hat{p}\cdot\hat{q}s_p^{(-)}s_q^{(-)}\right]L_1(q)\right\} \\
 M_{H}L_1(p) = E_t(p)L_2(p)-\hat{I}_q\left\{\left[c^{(+)}_pc_q^{(+)}
  +\hat{p}\cdot\hat{q}s_p^{(+)}s_q^{(+)}\right]L_2(q)\right\}.
 \end{array}\right.
 \end{equation}
 Here $m^{0}$ is the "current" quark mass, $ E_t=E_1+E_2, \quad E_1 $ and
 $ E_2 $ are the energies, respectively, of the quark and antiquark,
 and $\vartheta(p)$ is the solution to the SD
 equation, $L_{1},\,\, L_{2}\,$\, and \,$\,M_{H}$  are the eigenfunctions
 and eigenvalue of the BS equation, respectively, identified by the
 wave functions and mass of pseudoscalar meson (H), and the following
 abbreviations are used:
 $$
 c_p^{(\pm)}\equiv\cos{\left[\vartheta_1(p) \pm \vartheta_2(p)\right]}, \,\,\,
 s_p^{(\pm)} \equiv\sin{\left[\vartheta_1(p) \pm \vartheta_2(p)\right]},\,\,\,
 \hat{p}\cdot\hat{q} \equiv\frac{{\bf p\cdot q} }{pq},\,\,\, p=\left|{\bf p}\right|,
 $$
 $$
 \hat{I}_qF\equiv \int \frac{d{\bf q}}{(2\pi)^3}V(\left|{\bf p -q}\right|)F
 $$
 where $ V(\mid {\bf p}-{\bf q}\mid) $ is the potential.
  We define the mass function as
 \begin{eqnarray}
 m(p) = E(p)cos{2(p)}.\,\,\,
 \end{eqnarray}

 The solutions of the BS equation satisfy the normalization condition
 \begin{eqnarray}\label{c4}
 \frac{4 N_c}{M_{H}}
 \int \frac{d{\bf q}}{(2\pi)^3} L_1(q)L_2(q) = 1 .\,\,\,
 \end{eqnarray}

 In the chiral limit, these equations admit the solutions satisfying the Goldstone
 theorem. In other words, if a solution to the SD equation for the massless
 quark ($m^{0}=0$) exists, then the same
 solution satisfies the BS equation for a massless ($M_{\pi} = 0$)
 pseudoscalar meson too:
 \begin{eqnarray} \label{c5}
 L^{G}(q) = L_1(q) = \frac{cos{2\vartheta (p)}}{F_{\pi}} \,\,\, ,
 \end{eqnarray}
 where $F_{\pi}$ is the pion leptonic decay constant having the
 experimental value 132 MeV.

 The latter equation is obtained from the amplitude of the decay
 $\pi$ $\to$ $\mu$ $\nu$ in the chiral limit and normalization
 condition.
 In the potential model, the pion leptonic decay constant is defined
 as
 \begin{eqnarray}\label{c7}
 F_{\pi} = \frac{4 N_c}{M_\pi}
 \int \frac{d{\bf q}}{(2\pi)^3}
 \cos{2\overline{\vartheta}(p)}L_2(q)
 \end{eqnarray}
 where ${\overline{\vartheta}(p) =
 \frac{{\vartheta}_{1}(p)+{\vartheta}_{2}(p)}
 {2}}$,\,\,\, $M_{\pi}$ is the pion mass.

 Notice that in the limit of heavy quarks ($\frac{p}{m^{0}}<<1$)
 the BS equation reduces to the Schr\"odinger equation
 \begin{eqnarray*}
 (M_{H} - m_{1}^{0} - m_{2}^{0} - \frac {p^{2}}{2\mu})\Psi (p) &
 = & - \hat I_{q}\cdot \Psi (p) \,\,\, ,  \end{eqnarray*}
 where
 \begin{eqnarray*}
 \Psi(p) &=& 2\sqrt {\frac {N_{c}}{M_{H}}}L_{1}(p) \approx
 2\sqrt{\frac {N_{c}}{M_{H}}}L_{2}(p) \,\,\, ,\\
 \mu & = & \frac {m_{1}^{0}m_{2}^{0}}{m_{1}^{0} + m_{2}^{0}} \,\,\,,
 \end{eqnarray*}
 whereas equation~(\ref{c7}) coincides with the nonrelativistic definition
 of the meson decay constant
 $$
 F_{H}^{NR}=\sqrt{\frac{N_c}{\mu}}\int \frac{d{\bf
 q}}{(2\pi)^3}\Psi(q)=\sqrt{\frac{N_c}{\mu}}\Psi(0) \,\,\,.
 $$

 Thus, we can see that the quark potential model reflects the properties
 of the QCD at long and short distances.


 The separable approximation to the potential consists in the following
 factorization of the potential:
 \begin{eqnarray}\label{c8}
 V({\bf p}-{\bf q})= \frac{e^2}{L^2}
 f(\frac{p}{L}) \cdot f(\frac{q}{L}) \,\,\, , \,\,\, f(0)=1 \,\,\, ,
 \end{eqnarray}
 where $e$ is the coupling constant, $f$ is the form factor, and $L$ is the
 cut-off parameter.

 In this approximation one neglects the dependence of the spectra
 on the angular momentum.
 As a result, the SD equation takes the following simple form:
 \begin{eqnarray}
 m(p) = m^0 + m f(\frac{p}{L})
 \end{eqnarray}
 or

 \begin{eqnarray} \label{c10}
 1- \frac{m^0_R}{m}  = \frac{e^2}{L^2}  <\frac{f^2}{2E}>  \,\,\, , \,\,\,
 m^0_R= m^0 \frac{e^2}{L^2} < \frac{f}{2E} > \,\,\, ,
 \end{eqnarray}
 where the following abbreviations are used:

 \begin{eqnarray*}
 m \equiv m(0),\,\,\,
 <F> = \int \frac{d{\bf q}}{(2\pi)^3} F \,\,\, .
 \end{eqnarray*}

 Let us consider the solution of this equation in the low energy limit,
 more exactly, when $m^{0}/L \to 0$. In this limit, equation~(\ref{c10})
 can be solved under the coupling constant
 \begin{eqnarray}
 e^2 =
 4 \pi^2 \Biggl(
 \int_0^{\infty} dx \frac{x^2 f^2(x)}{\sqrt{ x^2+f^2(x) \gamma^2  }}
 \Biggr)^{-1} \sim
 4\pi
  \Biggl(
 \int_0^{\infty} dx {x f^2(x)}
 \Biggr)^{-1} \,\,\, ,
 \end{eqnarray}
 where the approximation for the integral is valid as the dependence on $\gamma \equiv
 m/L$ is weak.

 We define the "averaged" quark Green function $G(q)$ with
 the density function $f$ as follows:
 \begin{eqnarray} \label{c12}
 << q \bar{q} >>  = 4N_c i \mbox{tr}
 \int \frac{d q}{(2\pi)^4} ( f(q) G(q) ) =
 -4 N_c <\frac{m f^2}{2E}> = - 4N_c \frac{L^2}{e^2} m .
 \end{eqnarray}

 In the separable approximation, the BS equation also takes
 the form of the usual algebraic one
      \begin{eqnarray}
 \frac{m^0_R}{m} = \frac{M_\pi^2}{4 m^2}
 \Biggl[
 {\cal D}^{(2)}
 +
 \frac{{\cal D}^{(3) 2} }{1-{\cal D}^{(4)} }
 \Biggr] \,\,\, ; \,\,\,
 {\cal D}^{(k)} = \frac{e^2}{L^2} < \frac{f^{k}}{2 E^3}  >\,\,\,.
 \end{eqnarray}
 In the low energy limit, $M_{\pi}<<4E^{2}(0)$, one obtains the
 following solutions to the equation:
 \begin{eqnarray}
 L_1(p) &=& \frac{1}{F_{\pi}} \frac{m(p)}{E(p)}
 \Biggl(  1+ \frac{M_\pi^2}{4E^2 } \frac{{\cal D}^{(3)} }{1-{\cal D}^{(4)} }
 \Biggr)
 \sim \frac{m(p)}{F_{\pi} E(p)}  \,\,\, ,  \\ \nonumber \\
 L_2(p) & =& \frac{M_\pi}{2E(p)}
 \Biggl(  1+ f \frac{{\cal D}^{(3)} }{1-{\cal D}^{(4)} }
 \Biggr) L_1(p) \,\,\, .
 \end{eqnarray}

 Now using eqs.~(\ref{c4}),~(\ref{c10}), and~(\ref{c12})  we arrive
 at the following relation:
 \begin{eqnarray}
 -4 m^0_R
 << q \bar{q}>> = F_{\pi}^2 M_\pi^2\,\,.
 \end{eqnarray}
 We see that this equation turns out into the well known low energy
 theorem of the
  local theory if one replaces the averaged Green function,
 $<<q \bar{q}>>$, with the quark condensate, $<q \bar{q}>$.
 The quantities included in this relation can be estimated if the form of the
 potential and the current quark mass are given.

 To study the relation of the averaged Green function of the quark to
 the usual quark condensate, we consider the local NJL model type potential,
 $f(x)=\theta(1-x)$,
 and use the conventional values for the free parameters, namely,
 \begin{equation}
 L =950 MeV,\quad m^{0}_{R} =4 MeV,
 \end{equation}
 then we obtain the following numerical estimations
 \begin{eqnarray}    \label{c18}
 M_\pi&=&140 \mbox{MeV}   \,\,\, ; \,\,\,
 F_{ \pi} = 132 \mbox{MeV}
 \,\,\, ; \,\,\,  \nonumber \\ \\
 << q \bar{q}>> &=&  (-250 \mbox{MeV} )^3
 \,\,\, ; \,\,\,
 {\cal D}^{(k)}= {\cal D} =
  \frac{1}{6} \,\,\, ;
 \,\,\, \gamma^2 = \frac{1}{12} \,\,\, . \nonumber
 \end{eqnarray}

 So, we see that these estimations are in agreement with the conventional
 values
 for the mass and decay constant of the pion as well as the quark condensate.
 We can conclude that equation~(\ref{c12}) defines the quark condensate
 formed of the quark - antiquark pair due to the nonlocal interaction in the
 separable approximation.

 Let us now consider the mode of the pion decay, $\pi \to \gamma \gamma$,
 induced by vector current. The amplitude of this process
 in the potential model is written as~\cite{a20}
 \begin{eqnarray}
 M ( \pi^0 \longrightarrow  \gamma \gamma ) &=&
 <\gamma \gamma \vert W_{\mbox{eff.}} \vert \pi^0>
 = \frac{(2\pi)^4 \delta ({\cal P}-k_1-k_2)}{\sqrt{ (2\pi)^9 8
 {\cal P}_0 k_{10} k_{20} }} \cdot \nonumber \\ \\
 &\cdot&  2 N_c
 (\epsilon^\alpha_\mu(k_1) \epsilon^\beta_\nu(k_2) \frac{1}{\sqrt{2}}
 (e_u^2 - e_d^2) \epsilon_{\mu \nu \alpha \beta} k_1^\alpha k_2^\beta
 I(k_1,k_2)
 \,\,\, , \nonumber
 \end{eqnarray}
 where $\epsilon^{\alpha}_{\mu}(k_{i})$ is the photon polarization tensor
 $(i=1,2)$,
 $e_{q}$ are the quark charges, and $N_{c}$ is the number of colors.
 The integral of this equation in the limit of zero photon momenta has the form
 \begin{eqnarray} \label{c20}
 I(k_1=0,k_2=0) = \int \frac{d{\bf q}}{(2\pi)^3}
 \frac{m(q)}{E^2(q)}
 \frac{1}{E^2(q)- \frac{M_\pi^2}{16}}
 \frac{1}{2}
 \bigl[ L_1(q) + \frac{4E(q)}{M_\pi} L_2(q) \bigr] \,\,\, ,
 \end{eqnarray}
 where the approximation of equal masses of the quarks is used,
 $E(q) = \sqrt{q^{2}+m(q)^{2}}$ is the quark
 energy, {\it P} and $k_{1},k_{2}$ are the 4 - momenta of the initial and final
 particles, respectively.

 In the separable approximation the above integral takes the form
 \begin{eqnarray}
 I(k_1=0,k_2=0) &=& \frac{J}{4 \pi^{2}F_{\pi}} \,\,\,\,\,,
 \nonumber \\  \\
 J &=&
 \Biggl( \frac{3-{\cal D}}{1- {\cal D}} \Biggr)
 \int_0^{\infty}  dx
 \frac{ f^2(\gamma x) }{ \biggl( \sqrt{ x^2 + f^2(\gamma x) } \biggr)^5  } =
 \nonumber \\  \\
  &=&
 \Biggl( \frac{3-{\cal D}}{1- {\cal D}} \Biggr)
 \Biggl[ \frac{1}{3 (1+ \gamma^2)^{3/2}}
 \Biggr]
 =1.02\,\,\,,               \nonumber
 \end{eqnarray}
 where for the numerical estimation the values of the
 parameters~(\ref{c18}) are used.

 So, the separable approximation for the integral $J$ leads to the the
 Bell - Jackiw's  theoretical estimation~\cite{ABJ}
 (${J}^{BJ} = 1$) and the experimental result
 (${J}^{exp} \approx 1.04$).

 Let us investigate whether one can self-consistently reproduce the low energy
 theorems for the pion decays using the conventional definition of the local
 quark condensate, namely,
 \begin{eqnarray} \label{c22}
 < q \bar{q} > = iN_c \mbox{tr}
 \int \frac{dq}{(2\pi)^4} G(q) = - 2 N_c
 \int \frac{d{\bf q}}{(2\pi)^3}\cos(\vartheta(\mid {\bf p}\mid))
 \,\,\,,
 \end{eqnarray}
 where $G(q)$ is the quark Green function.
 It is easy to see that the low energy theorem can be obtained from
 equations~(\ref{c4}),~(\ref{c5}), and~(\ref{c22})
 if the "small" component of the bound wave function ($L_{2}(p)$) in this energy
 scale has the behavior
 \begin{eqnarray}
 L_2 = \frac{m^0_R}{2 F_{ \pi} M_\pi} = \mbox{const} \,\,\, .
 \end{eqnarray}

 However, this asymptotic function contradicts the Bell - Jackiw estimation.
 Indeed, substitution of
 \begin{eqnarray}\label{c24}
 L_1(p=0) \sim \frac{1}{F_\pi}\,\,\, , \,\,\,
 L_2 =  0 \,\,\, ,
 \end{eqnarray}

 into~(\ref{c20}) leads to
 \begin{eqnarray}
 I(k=0) &=& \frac{1}{4\pi^2 F_\pi} J \,\,\, , \,\,\,
 J = \int_0^{\infty} dx \frac{x^2}{(1+x^2)^2} = \frac{\pi}{4} \,\,\, ,
 \end{eqnarray}

 Notice that by using
 \begin{eqnarray*}
 L_1(q)=\frac{1}{F_\pi}  \frac{m}{\sqrt{m^2 +{\bf q}^2}}
 \end{eqnarray*}
 instead of the value~(\ref{c20}) one would obtain $J = \frac{1}{3}$.

 Thus, a straightforward application of the local quark condensate in the
 nonlocal potential model does not provide a self-consistent reproduction of
  the low energy theorems for the pion decays.

 The main result of the section is the check  of the low  energy theorems
 on the level of the effective bilocal meson theory
 and the generalization to heavy quark limit.
 This effective approach  may be a reasonable
 for the detailed  study  of amplitudes and the Isgur--Wise function for
 heavy quarkonia decays (for example, $B \longrightarrow
 D^{*}(\rho) l\nu$)~\cite{a1}.

 \subsection{U(1) - problem}

 We have seen that the bilocal linearization of the four fermion interaction
 leads to an effective bilocal field $\eta_M$ action \cite{pre,a22} in both
 QED~(\ref{s}) and QCD.

 This meson action includes Abelian anomalies
 in the pseudoscalar isosinglet channel \cite{2,1,3,bl}
 (positronium $\eta_M=\eta_P$, in QED; and $\eta_M=\eta_0$-meson, in QCD).

 We have chosen the total-motion variable $\eta_M(t)$
 so that
 the effective action for the
 total motion of the the pseudoscalar bound state with anomaly term
 have the universal form for all gauge theories
 ($QED_{(3+1)},QED_{(1+1)}, QCD_{(3+1)}$) in terms of physical variables
 similar to~(\ref{s})
 \be \label{sm}
 W_{eff}=\int dt \left\{\frac 1 2 \left({\dot\eta_M}^2-M_P^2{\eta_M}^2\right)
 V +
 C_M\eta_P \dot X[A^{(N)}] \right\}~,
 \ee
 where in 3-dimensional $QED_{(3+1)}$  the constant
 $C_M$ is given by Eq.~(\ref{t}),
 $$
 C_M=C_P=\frac{\sqrt{2}}{m_e}8{\pi}^2
 \left(\frac{\underline{\psi}_{Sch}(0)}{m_e^{3/2}}\right)~,
 $$
 in 1-dimensional $QED_{(1+1)}$~\cite{gip},  $C_M=2\sqrt{\pi}$; and
 in 3-dimensional $QCD_{(3+1)}$,
 $$
 C_M=C_{\eta}=\frac{N_f}{F_\pi}\sqrt{\frac 2 \pi } , ~~~(N_f=3)~.
 $$
 $X[A]$ is the "winding number" functional.
 In $QED_{(3+1)}$  the "winding number" functional defined
 by Eq.~(\ref{wnqed}) describes the two $\gamma$ decay of  a positronium.
 In $QED_{(1+1)}$ and  $QCD_{(3+1)}$ these  winding number functionals
 $$
 \dot X_{\rm QED}(A^{(N)})=\frac{e}{4\pi}
 \int\limits_{-V/2 }^{V/2 }dx F_{\mu\nu}
 \epsilon^{\mu\nu}= \dot N(t)~\Rightarrow~~F_{01}=\frac{2\pi\dot N}{eV}~,
 $$
 \be \label{nor}
 \dot X_{\rm QCD}[A^{(N)}]= \frac{g^2}{16 \pi^2}
 \int d^3x G^a_{\mu \nu}{}^*G^a_{\mu \nu}=  \dot N(t)+\dot X[A^{(0)}]
 \ee
 contain the independent topological variable.
 Recall that in QCD with the Wu-Yang monopole
 we obtained the normalizable zero-mode
 $$
 G^a_{0i}=\dot N D_i^{ab}({\Phi})\Phi^b_0 = \dot N B_i^a({\Phi})
 \frac{2\pi}{\alpha_s V<B^2>}~,
 $$
 so that
 $$
 \frac{g^2}{8 \pi^2}\int d^3D_i^{ab}({\Phi})\Phi^b_0 B_i^a({\Phi})=1~.
 $$
 In $QED_{(1+1)}$ and  $QCD_{(3+1)} $
 the effective action should be
 added by the topological dynamics of the zero mode with the actions
 $$
  W_{\rm QED} =\frac{1}{2}\int dt  \int\limits_{-V/2 }^{V/2 }dx F^2_{01}=
 \int dt\frac{\dot N^2 I_{\rm QED}}{2}~,
 $$

 $$
  W_{\rm QCD} =\frac{1}{2}\int dt  \int\limits_{V }^{ }d^3x G^2_{0i}=
 \int dt \frac{\dot N^2 I_{\rm QCD}}{2}~,
 $$
 where
 $$
 I_{\rm QED}=\left(\frac{2\pi}{e}\right)^2\frac{1}{V}~,
 $$
 $$
 I_{\rm QCD}=\left(\frac{2\pi}{\alpha_s}\right)^2\frac{1}{V<B^2>}~.
 $$
 It is easy to show that the diagonalization of the total Lagrangian
 of the type of
 $$
 L=[\frac{\dot N^2I}{2}+C_M\eta_M \dot N] =
 [\frac{(\dot N+C_M\eta_M/I)^2I}{2}- \frac{C_M^2}{2IV} \eta_M^2V ]
 $$
 leads to additional mass
 of the pseudoscalar bound state in both $QED_{(1+1)}$ and  $QCD_{(3+1)}$
 $$
  \triangle {M}^2=\frac{C_{M}^2}{IV}~.
 $$
 In $QED_{(1+1)}$ this formula describes the well-known Schwinger mass
 $$
  \triangle {M}^2=\frac{C_{M}^2}{I_{\rm QED}V}=\frac{e^2}{\pi}~;
 $$
  whereas in $QCD_{(3+1)}$ we obtain  the additional mass of $\eta_0$ meson
 \be  \label{dl}
  L_{eff}= \frac{1}{2}[{\dot\eta_0}^2-\eta_0^2(t)(m_0^2
 +\triangle m_{\eta}^2)]V
 \ee
 \be  \label{mm}
  \triangle {m_\eta}^2=\frac{C_{\eta}^2}{I_{\rm QCD}V}
 = \frac{N_f^2}{F_{\pi}^2}\frac{\alpha_s^2<B^2>}{2\pi^2}.
 \ee
 This result allows us to estimate the value of the vacuum magnetic
 field (\ref{magnet}) in QCD
 $$
 <B^2>=\frac{}{}\frac{2\pi^3F_{\pi}^2\triangle
 {m_\eta}^2}{N_f^2\alpha_s^2}=\frac{0.06 GeV^4}{\alpha_s^2}~.
 $$
 After calculation
 we can remove infrared  regularization $V \to \infty $.

 Thus, the Dirac constraint-shell formulation of gauge theories~\cite{cj}
 allows us to describe on  equal footing the set
 of well-known results on anomalous interactions of pseudoscalar
 bound states in gauge theories including the anomalous
 decay of a positronium in $QED_{(D=1+3)}$,
 the Schwinger mass in $QED_{(D=1+1)}$, and U(1)-problem
 in $QCD_{(D=1+3)}$.
 These results
 include also the zero probability of the two gluon decay of a pseudoscalar
 meson due to confinement as the destructive interference
 of the Gribov copies for gauge theories with the homotopy
 group $\pi_{(D-1)}(G)=Z$.

 \section{ Conclusion}

 Why does the non-Abelian gauge theory lead to the spontaneous chiral
 symmetry breaking, quark-hadron duality, rising potential, and
 color confinement?

 The formulation of the theory of strong interactions is completed,
 if the theory can explain and unify all working phenomenological schemes
 (chiral Lagrangians with the Goldstone mesons,
 the parton model based of quark-parton duality,
 the Schr\"odinger equation with  rising potentials for heavy quarkonia, etc.)
 from the first principles (dynamic, quantum, and symmetric).

 The pure water of the theory of strong interactions
 can clean the muddy swamp of phenomenology if there are constructive
 explanations of  the paradoxes:

 i) How to combine the observation of
 nonlocal gauge-invariant variables with
 the variational principles formulated for local fields?

 ii) How to combine the dependence of the Hamiltonian approach
 to quantization on the time axis with the relativistic
 covariance of S-matrix elements?

 iii) How to reconcile the presence of poles of  color particle
 Green functions
 (that needs to ground the parton model and
 to obtain the Shr\"odinger equation) with nonobservability
 of  physical states corresponding to these poles?

 The first two questions belong also to the consistent scheme of
 the description of the nonlocal bound-state sector
 (spectrum and interactions) in QED.
 Nevertheless, these questions become essentially actual in QCD
 where observables are only colorless bound states.
 In this case, the
 theorem of equivalence of different gauges~\cite{f,fs,ft} on the
 mass-shell of elementary particles is not adequate to the
 physical situation of the description of observable nonlocal objects
 where local elementary particles are off mass-shell.
 For nonlocal bound states, even in QED, the dependence on the time-axis
 and gauge exists. All peculiarities of bound states (including time
 initial data, spatial boundary conditions, normalization
 of wave functions, time evolution)
 reflect the choice of their rest frame of reference
 distinguished by the axis of time chosen to lie along the
 total momentum of any bound state in order
 to obtain the relativistic - covariant dispersion
 law and invariant mass spectrum.

 Thus, the answers to the first two questions (i,ii)
 are the {\it equivalent
 unconstrained system} obtained by  resolving
 the Gauss law constraint and the Markov-Yukawa prescription
 of the choice of the time axis parallel to the eigenvector
 of the bound-state total momentum operator.
 In this case, we can combine the relativistic and quantum principles
 of the dynamic description of gauge fields.

 The last question iii) belongs only to QCD. The answer is the
 topological degeneration of initial data in the non-Abelian
 theory that leads to the pure gauge Higgs effect
 of appearance of a physical vacuum in the form of the Wu-Yang monopole.
 This vacuum determines  the effects of hadronization and confinement
 in QCD omitted by the conventional gauge-fixing method.

 We brought out that the physical vacuum leads to rising potentials,
 chiral Lagrangians, and an additional mass of the $\eta_0$-meson
 (as a consequence of the rearrangement of the perturbation series).
 The averaging over the Gribov copies of the topological
 degeneration leads to the color confinement and quark-hadron duality
 as a consequence of the destructive interference of the Gribov
 phase factors. This confinement is one of the consequences
 of the Dirac definition of measurable  variables that keeps
 quantum principles. This confinement (omitted by all other
 methods of quantization) appears as pure quantum effect that gives
 the theoretical basis
 of the quark-hadron duality as the experimental method of measurement of
 the quark-gluon quantum numbers.

  Here we can recall the words by J. C. Maxwell in the Introduction of his
 {\it A Treatise on Electricity and Magnetism}  (Oxfrord, 1873):
 "The most important aspect of any phenomenon from mathematical point of
 view is that of a measurable quantity. I shall therefore
 consider electrical phenomena chiefly with a view to their measurement,
 describing the methods of measurement, and defining the standards
 on which they depend."

\section*{Acknowledgments}

\medskip

I am gratefull to   B. M. Barbashov, D. Blaschke, A. V. Efremov, G.
A. Gogilidze, N. Ilieva, W. Kallies, A. M. Khvedelidze, E. A. Kuraev, M.
Lavelle, D. McMullan, J. Polonyi, G. R\"oepke and W. Thirring for interesting
and critical discussions.

 \end{document}